\documentclass[lettersize,journal]{IEEEtran}
\usepackage[defaultcolor=red]{changes}
\usepackage{amsmath,amsfonts}
\usepackage{algorithmic}
\usepackage{algorithm}
\usepackage{array}
\usepackage[caption=false,font=footnotesize,labelfont=rm,textfont=rm]{subfig}
\usepackage{textcomp}
\usepackage{stfloats}
\usepackage{url}
\usepackage{verbatim}
\usepackage{graphicx}
\usepackage{cite}
\usepackage{xcolor}

\usepackage[table,xcdraw]{xcolor}
\usepackage{threeparttable}
\usepackage{xcolor}
\usepackage{pdfpages}
\usepackage{changes}
\usepackage{amsthm,verbatim,amssymb,amsfonts,amscd}
\usepackage{array}
\usepackage{bm}
\usepackage{multicol}
\usepackage{multirow}
\usepackage{graphicx}
\usepackage{url}
\graphicspath{{figure/}}
\usepackage{epstopdf}
\usepackage{makecell}
\usepackage{soul, color, xcolor}
\usepackage{subfloat}
\begin{document}

\title{A Tutorial on 3GPP Rel-19 Channel Modeling for 6G FR3 (7–24 GHz): From Standard Specification to Simulation Implementation
}

\author{Pan Tang, Huixin Xu, Jianhua Zhang,~\IEEEmembership{Fellow,~IEEE}, Ximan Liu, Enrui Liu, Haiyang Miao,\\ Xiaodong Sun, Wei Jiang, and Guangyi Liu
\thanks{Pan Tang, Huixin Xu, Jianhua Zhang, Ximan Liu, and Enrui Liu are with the State Key Lab of Networking and Switching Technology, Beijing University of Posts and Telecommunications, Beijing 100876, China (e-mail: tangpan27@bupt.edu.cn; xuhuixin@bupt.edu.cn; jhzhang@bupt.edu.cn; liuxm2020@bupt.edu.cn; liuenrui@bupt.edu.cn).}
\thanks{Haiyang Miao is with Tsinghua University, Beijing, 100084, China (e-mail: miaohy@tsinghua.edu.cn).}
\thanks{Xiaodong Sun and Wei Jiang are with vivo Mobile Communication Co., Ltd., Dongguan 523850, China (e-mail: sunxiaodong@vivo.com; wei.jiang@vivo.com).}
\thanks{Guangyi Liu is with China Mobile Research Institute, Beijing 100053, China (e-mail: liuguangyi@chinamobile.com).}
}

\markboth{Journal of \LaTeX\ Class Files,~Vol.~14, No.~8, August~2021}%
{Shell \MakeLowercase{\textit{et al.}}: A Sample Article Using IEEEtran.cls for IEEE Journals}

\bstctlcite{IEEEexample:BSTcontrol}

\maketitle

\begin{abstract}
The upper-mid band (7–24 GHz), designated as Frequency Range 3 (FR3), has emerged as a definitive ``golden band" for 6G networks, strategically balancing the wide coverage of sub-6 GHz with the high capacity of mmWave. To compensate for the severe path loss inherent to this band, the deployment of Extremely Large Aperture Arrays (ELAA) is indispensable. However, the legacy 3GPP TR 38.901 channel model faces critical validity challenges when applied to 6G FR3, stemming from both the distinct propagation characteristics of this frequency band and the fundamental physical paradigm shift induced by ELAA. Specifically, due to a scarcity of empirical data, channel parameters in the legacy model were primarily derived via linear interpolation between sub-6 GHz and mmWave bands, thereby failing to accurately capture the specific frequency-dependent characteristics. Compounding this, the massive aperture of ELAA renders the traditional far-field and wide-sense stationary assumptions invalid, necessitating the explicit modeling of near-field spherical wavefront effects and spatial non-stationarity. In response, 3GPP Release 19 (Rel-19) has validated the model through extensive new measurements and introduced significant enhancements. This tutorial provides a comprehensive guide to the Rel-19 channel model for 6G FR3, bridging the gap between standardization specifications and practical simulation implementation. First, we provide a high-level overview of the fundamental principles of the 3GPP channel modeling framework. Second, we detail the specific enhancements and modifications introduced in Rel-19, including the rationale behind the new Suburban Macro (SMa) scenario, the mathematical modeling of ELAA-driven features such as near-field and spatial non-stationarity, and the recalibration of large-scale parameters. Third, this article presents a step-by-step implementation guideline for the updated Geometry-Based Stochastic Model (GBSM) procedure, demonstrating how to integrate these new features into simulation platforms supported by benchmark results. Finally, we discuss remaining open issues and outline potential directions for future research. Overall, this tutorial serves as an essential guide for researchers and engineers to master the latest 3GPP channel modeling methodology, laying a solid foundation for the accurate design and performance evaluation of future 6G FR3 networks.
\end{abstract}

\begin{IEEEkeywords}
6G, FR3 (7–24 GHz), ELAA, 3GPP Release 19, Standardized Channel Modeling, Near-Field Propagation, Spatial Non-Stationarity, Simulation Implementation.
\end{IEEEkeywords}

\section{Introduction}
\subsection{Standardization Background}
Operating as the United Nations specialized agency mandated to regulate the global radio-frequency spectrum and harmonize international telecommunication standards, the International Telecommunication Union Radiocommunication Sector (ITU-R) has formalized the vision for the sixth generation (6G) of mobile networks through the approval of the IMT-2030 Framework \cite{C1_6G_survey,series2023imt}. This recommendation substantially broadens the horizons of 5G by delineating six interconnected usage scenarios: Immersive Communication, Massive Communication, Hyper-reliable and Low-latency Communication, Ubiquitous Connectivity, Artificial Intelligence (AI) and Communication, and Integrated Sensing and Communication (ISAC). To meet the stringent Key Performance Indicators (KPIs) associated with these scenarios—such as peak data rates surpassing 50–200 Gbit/s and centimeter-level localization accuracy—transformative physical layer technologies are imperative. Prominent enablers identified by the global research community include ISAC \cite{liu2022integrated}, Reconfigurable Intelligent Surfaces (RIS) \cite{di2020smart}, AI-native air interfaces \cite{letaief2019roadmap}, and ELAA \cite{ye2024extremely,Tang2024XL-MIMOChannelSurvey}. In parallel, the identification of new spectrum resources to support these technologies is indispensable. While the sub-6 GHz band (FR1) is congested and the millimeter-wave band (FR2) faces severe coverage limitations, the 7–24 GHz range—often referred to as the ``upper-mid band" or FR3—has emerged as the strategic ``golden band" for 6G \cite{zhang2024new}. This spectral focus is strongly underpinned by the regulatory outcomes of the World Radiocommunication Conference 2023 (WRC-23), which identified specific frequency bands within this range (e.g., 7–8.5 GHz and 14.8–15.35 GHz) as candidate bands for future IMT identification studies \cite{ITU_WRC23}. The effective utilization of this promising spectral range will demand a holistic evolution across the entire system chain, from the radio propagation environment to terminal hardware and network architecture.

Functioning as the premier technical specification group for mobile networks, the 3rd Generation Partnership Project (3GPP) is tasked with translating ITU-R’s high-level requirements into detailed global standards. In essence, 3GPP acts as the technology proponent, developing and submitting specifications for IMT certification. Throughout the 5G era, 3GPP has systematically advanced the ecosystem: solidifying the New Radio (NR) foundation in Release 15 and 16, diversifying into vertical domains in Release 17, and inaugurating the 5G-Advanced era with Release 18 \cite{3gpp_tr21918}. Currently, the consortium has embarked on Release 19 (Rel-19), a pivotal phase functioning as the technological bridge toward 6G. While academia proposes a plethora of potential 6G enablers—such as RIS and sub-THz communications—3GPP strategically prioritizes technologies based on industrial readiness, deployment feasibility, and regulatory alignment (e.g., WRC-23 outcomes). Consequently, within the physical layer (RAN1) of Rel-19, significant attention is converged on two exploratory frontiers: unlocking the potential of the upper-mid band (7–24 GHz) via ELAA, and evaluating the feasibility of Integrated Sensing and Communication (ISAC) \cite{3gpp_rp232906}. The former addresses the urgent industry need for a capacity-coverage balance, while the latter represents a tangible step toward functional expansion beyond mere connectivity. However, the successful standardization and performance evaluation of these specific technologies are predicated on a fundamental prerequisite: the availability of accurate, validated, and standard-compliant channel models \cite{Gong20256GUnifiedChannelModel}.
\subsection{Motivation and Updates}
Standardized channel models constitute the bedrock for the unbiased evaluation and benchmarking of physical layer technologies across different companies and organizations. The widely adopted 3GPP TR 38.901 \cite{3gpp_tr38901_rel17}, developed throughout the 5G era, provides a unified GBSM framework nominally covering frequencies from 0.5 to 100 GHz. However, its direct application to the emerging Rel-19 frontiers is challenged by discrepancies in two distinct dimensions: new spectrum characteristics and new sensing capabilities. Regarding the 7–24 GHz band and ELAA systems, the legacy model suffers from a scarcity of empirical data (relying on linear interpolation) and the breakdown of far-field/stationary assumptions, leading to inaccurate representations of delay spreads and beamforming gains. Simultaneously, regarding ISAC, the legacy communication-centric framework is fundamentally incapable of characterizing the scattering properties of sensing targets (e.g., Radar Cross Section) and the specific propagation paths required for monostatic or bistatic sensing topologies. To bridge these gaps, 3GPP RAN1 launched two parallel Study Items (SIs) in Release 19: one dedicated to ``Channel Modeling Enhancements for 7–24 GHz" \cite{3gpp_rp234018} and another for ``Channel Modeling for ISAC" \cite{3gpp_rp234069}. In light of the breadth and complexity of these topics, and to provide a sufficiently deep technical analysis, this tutorial explicitly targets the 7–24 GHz channel modeling enhancements.

The initiation of the specific work on 7–24 GHz channel modeling was precipitated by the recognition that the legacy model's validity in this band was mathematically extrapolated rather than empirically proven. Specifically, channel parameters in the legacy TR 38.901 were largely derived via simple linear interpolation between sub-6 GHz and millimeter-wave data points. However, comprehensive academic surveys and measurement campaigns \cite{miao20256g,miao2023sub,Fan2025MidbandELAANearField,Miao2024SPAWC1115GHzNearField} have revealed that propagation characteristics in the upper-mid band—such as scattering richness and material penetration loss—exhibit complex non-linear frequency dependencies, thereby compromising the fidelity of link budget analysis based on interpolation. Compounding this data scarcity, the introduction of ELAA technologies precipitates a fundamental breakdown in physical consistency. As the antenna aperture size increases to compensate for path loss, the radiative near-field (Fresnel) region expands significantly, extending to hundreds of meters in typical deployment scenarios \cite{cui2024near}. This expansion renders the plane-wave assumption inherent in the legacy model. Furthermore, measurements utilizing large arrays have demonstrated that the received signal strength can fluctuate by over 10 dB across the array aperture due to partial blockage or cluster visibility \cite{Xu2026THzXL-MIMOIndoorChannel,Xu2024ICC132GHzTHzXL-MIMO}, a phenomenon of spatial non-stationarity that violates the Wide-Sense Stationary (WSS) assumption of the legacy GBSM framework.

To address these gaps, 3GPP RAN1 conducted a rigorous study spanning nearly two years, culminating in the significant updates embodied in TR 38.901 V19.1.0 \cite{tang2025preliminary}. This standardization effort involved active contributions from over 20 leading companies, who submitted and analyzed massive datasets to derive statistically robust parameters using a weighted approach across frequency bands. Consequently, the updated standard incorporates a comprehensive set of enhancements beyond the baseline model. First, it introduces the SMa scenario with recalibrated large-scale parameters (LSPs) and updated O2I penetration models to address the specific coverage challenges in suburban environments. Second, to support ELAA, it integrates the Near-field channel model with spherical wavefront corrections and Spatial Non-Stationarity (SNS) models to capture element-wise power variations. Furthermore, to better reflect the realistic stochastic nature of the channel, particularly in scenarios with sparse scattering, the standard introduces a mechanism for number of cluster variability, allowing the model to dynamically reproduce the fluctuation in multipath richness observed in real-world measurements. This tutorial aims to dissect these complex standard specifications, providing a comprehensive bridge from theoretical principles to practical simulation implementation.

\subsection{Related Work and Contributions}
While the research community has extensively investigated radio propagation for future networks, existing literature can be broadly classified into three categories: research papers on specific channel characteristics, comprehensive surveys, and tutorials on legacy standards. However, a gap remains in addressing the practical implementation of the specific 3GPP Rel-19 standard for the 7–24 GHz band.

The first category focuses on specific research contributions regarding measurements and theoretical modeling. For the 7–24 GHz band, pioneering measurement campaigns \cite{rappaport2019wireless, ju2021millimeter} have provided valuable empirical data on path loss exponents and material penetration losses, highlighting the non-linear frequency dependency compared to sub-6 GHz bands. Regarding ELAA, theoretical studies \cite{cui2023near, zhang2023nearfield_tutorial} have rigorously derived near-field propagation characteristics using analytical electromagnetic approaches (e.g., spherical wave expansion) and analyzed the impact of spatial non-stationarity. While these works provide the physical basis and empirical evidence for 6G channels, they are often deterministic or site-specific. They do not explain how to approximate these complex physical phenomena within the stochastic GBSM framework standardized by 3GPP for system-level simulations.

The second category comprises comprehensive surveys that map out the landscape of 6G channels. Works such as \cite{wang2023paving} and \cite{zhang20206g} provide extensive reviews of channel characteristics across sub-6 GHz, millimeter-wave, and sub-THz bands. These surveys are instrumental in identifying key challenges, such as the sparsity of multipath components in higher frequency bands and the necessity of non-stationary modeling. However, due to their broad scope, they function more as a literature review rather than a technical specification. They typically lack the granular details of the finalized Rel-19 parameter tables (e.g., for the new SMa scenario) and the specific algorithmic procedures required to build a compliant simulator.

The third category consists of tutorials and guides dedicated to legacy 3GPP standards (Rel-15/16/17). Classical references such as [18] and [19] successfully demystified the baseline 5G channel models and their implementation (e.g., the QuaDRiGa extension). However, these articles are inherently obsolete regarding the 7–24 GHz band. They fail to cover the newly defined SMa scenario and do not address the mandatory modeling components for spatial non-stationarity and near-field phase corrections introduced in Rel-19, leaving a significant gap for engineers aiming to upgrade their simulation platforms.

To bridge these gaps, this tutorial distinguishes itself as the first comprehensive guide dedicated to the implementation of the 3GPP Rel-19 channel modeling standard for the 6G FR3 band. The primary contributions of this article are summarized as follows:
\begin{itemize}
\item We analyze the physical deficiencies of legacy models in the 7–24 GHz band and explain the rationale behind Rel-19's specific modeling choices, specifically the geometrical definition of the new SMa scenario and the recalibration of large-scale parameters.
\item We demystify how the standard mathematically incorporates ELAA physics into the stochastic GBSM framework. Specifically, we detail the modeling formulations for spherical wavefronts and spatial non-stationarity, clarifying how these deterministic features are grafted onto the stochastic cluster generation process.
\item We provide a step-by-step procedure for implementing the Rel-19 channel generation flow. Unlike general surveys, this guide highlights specific modifications required in the coefficient generation stage to support the new frequency band and antenna types.
\item We critically analyze the remaining limitations of the current Rel-19 model, such as computational complexity and the interface with deterministic tools, and identify key directions for future research in Rel-20 and beyond.
\end{itemize}
\subsection{Organization of This Tutorial}
The remainder of this tutorial is organized as follows. Section II provides a high-level overview of the fundamental 3GPP GBSM framework. Section III details the specific standardization enhancements introduced in Rel-19 for the 7–24 GHz band, including the rationale behind the new SMa scenario, recalibrated propagation parameters, and the theoretical modeling of ELAA features. The practical implementation guide is divided into two parts: Section IV presents the step-by-step procedure for the core channel model (based on TR 38.901 Clause 7.5), validated by baseline simulation results; Section V focuses on the implementation of additional modeling components (based on TR 38.901 Clause 7.6), specifically detailing how to integrate near-field corrections and spatial non-stationarity into the simulation loop. Section VI discusses the remaining open issues and outlines potential directions for future research. Finally, Section VII draws the conclusions.

\begin{figure}[htbp]
	\centering
	\includegraphics[width=9cm]{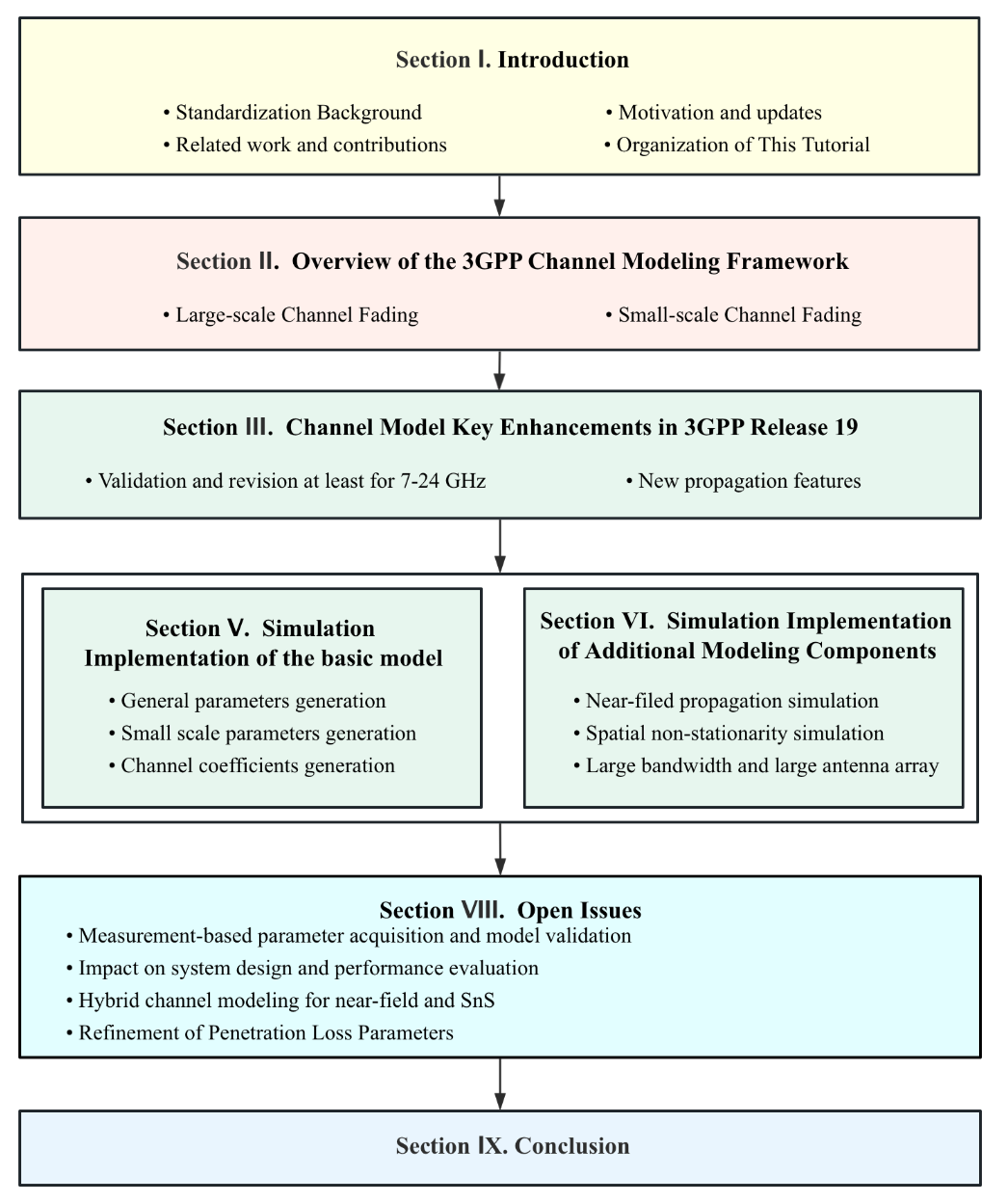}
	\caption{Logical structure and technical scope of the tutorial on the 3GPP Rel-19 channel model.}
    \label{Struct}
\end{figure}

\section{Overview of the 3GPP Channel Modeling Framework}

The 3GPP Rel-19 channel model serves as the fundamental benchmark for evaluating candidate 6G technologies, providing a realistic representation of radio propagation conditions across diverse environments. Built upon the widely adopted GBSM framework—akin to the WINNER \uppercase\expandafter{\romannumeral2} \cite{winnerII} and ITU-R 5G channel models \cite{M2412}—it characterizes the channel through ray directions rather than explicit scatterer locations. This geometry-based approach is pivotal as it enables the decoupling of propagation parameters from specific antenna configurations. Radio propagation is inherently subject to fading, which is categorized into large-scale fading (encompassing path loss and shadowing) and small-scale fading (representing multipath-induced rapid fluctuations). The following subsections detail the modeling principles within the 3GPP Rel-19 framework for these two categories, respectively.

\subsection{Modeling Principles for Large-Scale Fading}

\begin{figure}[htbp]
	\centering
	\includegraphics[width=8.4cm]{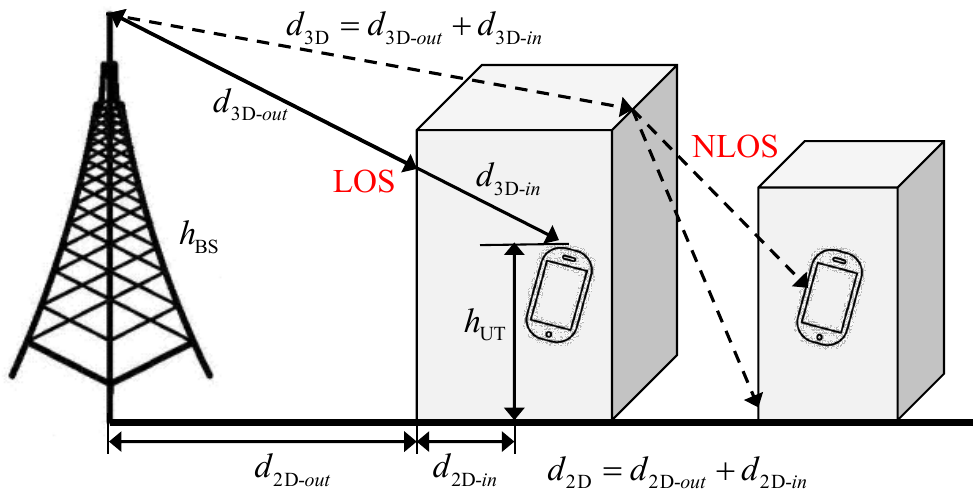}
	\caption{Geometrical relationships underlying large-scale fading components, including path loss, LOS probability, and O2I penetration.}
 \label{fig_LSP}
\end{figure}

Large-scale fading characterizes average signal attenuation over distance and specific environmental contexts, playing a fundamental role in determining coverage, cell planning, and interference management. In the 3GPP framework, large-scale fading is jointly described by three complementary components: path loss, line-of-sight (LOS) probability, and penetration loss. These components respectively capture distance- and frequency-dependent attenuation, blockage-induced visibility conditions, and additional losses caused by obstructions such as buildings or vehicles. Fig.~\ref{fig_LSP} conceptually illustrates the interaction among these elements: path loss establishes the baseline attenuation, LOS probability determines the stochastic classification of a link as LOS or non-LOS (NLOS), and penetration loss introduces additional attenuation for obstructed links (e.g., outdoor-to-indoor (O2I) scenarios).

\subsubsection{Path Loss}
In 3GPP modeling, path loss is formulated as a log-distance model with explicit dependencies on distance and frequency. The link distance is typically defined by the 3D separation, \(d_{3\mathrm{D}}\). Generally, the path loss model can be abstracted as:
\begin{equation}
\begin{aligned}
 \mathrm{PL}(f_c,d_{3\mathrm{D}})= &A_{\mathrm{PL}} + B_{\mathrm{PL}}\log_{10}(d_{3\mathrm{D}})+\\
 &C_{\mathrm{PL}}\log_{10}(f_c)+ G_{\mathrm{PL}} + X_\sigma ,
\end{aligned}
\end{equation}
where \(f_c\) denotes the carrier frequency (in GHz) \cite{3GPP_38901}. The coefficients \(A_{\mathrm{PL}}\), \(B_{\mathrm{PL}}\), and \(C_{\mathrm{PL}}\) represent the reference path loss offset, the path loss exponent governing distance attenuation, and the frequency-dependent loss factor, respectively; these are determined empirically for specific deployment scenarios. \(G_{\mathrm{PL}}\) represents an additional geometry-dependent correction term, capturing effects beyond simple distance/frequency scaling, such as average building height, street width, or breakpoint distances in dual-slope models. Notably, 3GPP models often employ a breakpoint structure where the distance dependence shifts when the 2D distance \(d_{2\mathrm{D}}\) exceeds a breakpoint \(d_{BP}\), reflecting a transition in propagation mechanisms.

The shadow fading term, \(X_\sigma\), accounts for large-scale power fluctuations caused by slow-varying blockage and environmental features (e.g., buildings, foliage). It is modeled as a zero-mean Gaussian random variable in the logarithmic (dB) domain (i.e., log-normal in the linear domain). Its standard deviation, \(\sigma\), is scenario-dependent (e.g., LOS, NLOS, O2I), and realizations are generated per link to reflect location-specific variations around the mean path loss.

\subsubsection{LOS Probability}
The LOS probability, \(\mathrm{Pr}_{\mathrm{LOS}}\), models the likelihood of an unobstructed path between the transmitter (TX) and receiver (RX), statistically capturing the impact of environmental blockage. Physically, at short TX–RX separations, the probability of blockage is negligible, implying a high confidence in LOS conditions. As the distance increases, blockage probability rises, causing \(\mathrm{Pr}_{\mathrm{LOS}}\) to decay. Accordingly, the LOS probability is typically modeled as a piecewise distance-dependent function of the horizontal separation \(d_{2\mathrm{D}}\):
\begin{equation}
\begin{aligned}
    \mathrm{Pr}_{\mathrm{LOS}}(d_{2\mathrm{D}})=
\begin{cases}
1, & d_{2\mathrm{D}} \le d_c,\\
 g_{\mathrm{LOS}}(d_{2\mathrm{D}}), & d_{2\mathrm{D}} > d_c,
\end{cases}
\end{aligned}
\end{equation}
where \(d_c\) is a scenario-dependent critical distance within which blockage is negligible. The function \(g_{\mathrm{LOS}}(d_{2\mathrm{D}})\) is a non-negative decaying function representing the random blocking process, often modeled exponentially beyond the critical region. In system-level simulations, \(\mathrm{Pr}_{\mathrm{LOS}}\) is utilized to stochastically classify each link, which subsequently determines the applicable path loss model, shadow fading statistics, and other large-scale parameters.

\subsubsection{Penetration Loss}
Penetration loss accounts for the attenuation incurred when radio waves traverse obstructing materials, such as building walls or vehicle chassis. The total large-scale attenuation, incorporating penetration effects, is expressed as:
\begin{equation}
 \mathrm{PL}_{\mathrm{tot}} = \mathrm{PL}_{\mathrm{out}} + L_{\mathrm{pen}} + L_{\mathrm{in}}+N(0,\sigma_P^2),
 \label{PL_INDOOR_OUEDOOR}
\end{equation}
where \(\mathrm{PL}_{\mathrm{out}}\) is the outdoor path loss, \(L_{\mathrm{pen}}\) denotes the penetration loss through external materials, \(L_{\mathrm{in}}\) represents additional indoor or in-vehicle losses, and \(\sigma_P\) is the standard deviation of the penetration loss. In 3GPP models, penetration loss is treated as a random variable with material- and frequency-dependent statistics to reflect the diversity of construction materials and incidence angles. These losses are generated on a per-link basis in the logarithmic domain, enabling flexible modeling of O2I and in-vehicle scenarios without altering the underlying outdoor path loss formulation.

\subsection{Modeling Principles for Small-Scale Fading}

In the 5G era, small-scale fading modeling was primarily confined to far-field and spatially stationary channels, where the assumptions of planar wavefronts and array-invariant path parameters were typically sufficient. In this paradigm, a cluster is conceptually represented as a scattering region comprising $M$ rays, with a total of $N$ clusters assumed. The system comprises $S$ transmit antennas and $U$ receive antennas.

\begin{figure*}[htbp]
	\centering
	\includegraphics[width=16cm]{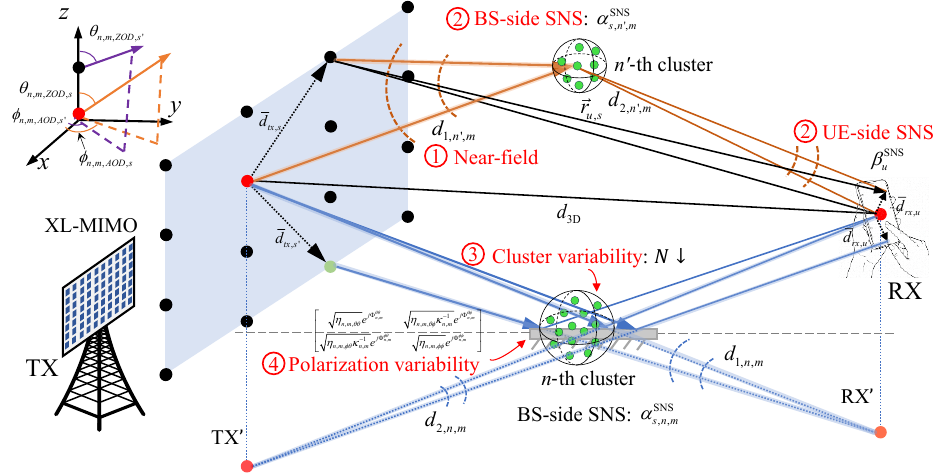}\\
	\caption{Conceptual illustration of the 6G FR3 channel model highlighting four key features: near-field propagation, SNS, cluster number variability, and polarization variability \cite{3GPP_38901}.}\label{6Gchannelmodel}
\end{figure*}

However, as antenna array apertures expand for 6G, channel models must evolve to capture near-field (NF) propagation and spatial non-stationarity. Fig.~\ref{6Gchannelmodel} illustrates the conceptual framework of the FR3 channel model for 6G. Unlike far-field modeling, 6G models allow ray-level small-scale parameters—such as azimuth/elevation angles of departure (AOD/ZOD) and arrival (AOA/ZOA)—to vary across antenna elements, acknowledging that widely spaced elements observe distinct local geometries. Near-field effects are modeled by introducing spherical-wave source distances (e.g., $d_1$ and $d_2$), which define the effective wavefront curvature. Consequently, phase evolution across the array is governed by true geometric distances rather than planar approximations. To incorporate SNS, the framework introduces element-wise power weighting factors applied to each ray. This mechanism mathematically captures physical phenomena such as partial blockage and incomplete scattering across the massive Base Station (BS) aperture, as well as element-dependent self-blockage (e.g., hand/head grip effects) at the User Equipment (UE). Furthermore, to enhance stochastic realism, the model integrates two novel variability features: cluster number variability, where the number of active clusters is drawn from a scenario-dependent distribution rather than being fixed; and polarization power variability, which applies ray-specific random weights to cross-polarization coefficients. These extensions constitute a unified 6G small-scale fading model capable of consistently reproducing spherical-wave propagation, spatial non-stationarity, and realistic polarization dynamics.

The channel impulse response (CIR), denoted as $H_{u,s}(\tau, t)$ between the $u$-th receive (RX) and $s$-th transmit (TX) antenna element, is modeled as a superposition of line-of-sight (LOS) and non-line-of-sight (NLOS) components. As formulated in (\ref{equ_CIR_LOS}), these components are scaled by the Ricean K-factor:
\begin{equation}
\begin{aligned}
H_{u,s}^{{\rm{LOS}}}\left( {\tau ,t} \right) =& \sqrt {\frac{1}{{{K_R} + 1}}} H_{u,s}^{{\rm{NLOS}}}\left( {\tau ,t} \right) + \\
&\sqrt {\frac{{{K_R}}}{{{K_R} + 1}}} H_{u,s,1}^{{\rm{LOS}}}\left( t \right)\delta \left( {\tau - {\tau _1}} \right),
\end{aligned}
\label{equ_CIR_LOS}
\end{equation}
where \(K_R\) is the linear Ricean K-factor (with $K_R = 0$ indicating purely NLOS conditions). The NLOS component consists of \(N\) clusters, each containing \(M\) rays, and is expressed as:
\begin{equation}
\begin{aligned}
H_{u,s}^{{\rm{NLOS}}}\left( {\tau ,t} \right) = &\sum\limits_{n = 1}^2 {\sum\limits_{i = 1}^3 {\sum\limits_{m \in {R_i}} {H_{u,s,n,m}^{{\rm{NLOS}}}(t)\delta \left( {\tau  - {\tau _{n,i}}} \right) } } } \\
&+ \sum\limits_{n = 3}^N {\sum\limits_{m = 1}^M {H_{u,s,n,m}^{{\rm{NLOS}}}(t)\delta \left( {\tau - {\tau _n}} \right)}},
\end{aligned}
\label{equ_CIR_NLOS}
\end{equation}
where rays in the two strongest clusters ($n=1,2$) are spread in delay into three sub-clusters indexed by $i$. 

To accurately capture 6G-specific propagation features, the explicit channel coefficients for the LOS component, $H_{u,s,1}^{\mathrm{LOS}}(t)$, and the NLOS rays, $H_{u,s,n,m}^{\mathrm{NLOS}}(t)$, are mathematically expanded in (\ref{equ_H_LOS_6G}) and (\ref{equ_H_NLOS_6G}), respectively \cite{Xu2026FR3XL-MIMOChannelModel}. Crucially, these formulations incorporate two transformative modifications:
\begin{itemize}
    \item SNS Power Scaling: The terms $\beta_{u}^{\rm{SNS}}$ and $\alpha_{s,n,m}^{\rm{SNS}}$ represent the power attenuation at the UE and BS sides, respectively, enabling element-wise power variations.
    \item Spherical Phase Evolution: The phase terms are governed by the precise geometric distances ($d_{1,n,m}$, $d_{2,n,m}$) between antenna elements and spherical-wave sources, replacing the linear phase approximation of far-field models. 
\end{itemize}
The coefficients involve the polarization matrix $\mathbf{\Phi}_{n,m}$, given by 
$\mathbf{\Phi}_{n,m} =
\begin{bmatrix}
\sqrt{\eta_{n,m,\theta \theta}} e^{j \Phi_{n,m}^{\theta\theta}} & \sqrt{\eta_{n,m,\theta \phi}\kappa_{n,m}^{-1}} e^{j \Phi_{n,m}^{\theta\phi}} \\
\sqrt{\eta_{n,m,\phi \theta}\kappa_{n,m}^{-1}} e^{j \Phi_{n,m}^{\phi\theta}} & \sqrt{\eta_{n,m,\phi \phi}}e^{j \Phi_{n,m}^{\phi\phi}}
\end{bmatrix}$, 
where $\kappa_{n,m}$ denotes the cross-polarization power ratio (XPR), and $\Phi$ represents independent random phases uniformly distributed over $[0, 2\pi)$. 
For clarity, Table~\ref{Tab_Par} provides a comprehensive definition of the key geometric and physical parameters used in these equations, including the spherical unit vectors, field patterns, and element-wise spherical-wave distance variables.

\par Finally, to derive the complete channel realization, large-scale fading effects (path loss and shadowing) are applied to these normalized small-scale fading coefficients.

\begin{figure*}
\begin{equation}
\begin{aligned}
H_{u,s,1}^{\rm{LOS}}&\left( t \right) = \sqrt{\beta_{u}^{\rm{SNS}} \alpha_{s,1}^{\rm{SNS}}} \left[ {\begin{array}{*{20}{c}}
{{F_{rx,u,\theta }}\left( {{\theta _{{\rm{LOS,ZOA}},u,s}},{\phi _{{\rm{LOS,AOA}},u,s}}} \right)}\\
{{F_{rx,u,\phi }}\left( {{\theta _{{\rm{LOS,ZOA}},u,s}},{\phi _{{\rm{LOS,AOA}},u,s}}} \right)}
\end{array}} \right]\left[ {\begin{array}{*{20}{c}}
1&0\\
0&{ - 1}
\end{array}} \right] \\
&\cdot \left[ {\begin{array}{*{20}{c}}
{{F_{tx,s,\theta }}\left( {{\theta _{{\rm{LOS,ZOD}},u,s}},{\phi _{{\rm{LOS,AOD}},u,s}}} \right)}\\
{{F_{tx,s,\phi }}\left( {{\theta _{{\rm{LOS,ZOD}},u,s}},{\phi _{{\rm{LOS,AOD}},u,s}}} \right)}
\end{array}} \right] \exp \left( { - j2\pi \frac{{{d_{{\rm{3D}}}}}}{{{\lambda _0}}}} \right)\exp \left( { - j2\pi \frac{{\left| {{{\vec r}_{u,s}}} \right| - {d_{{\rm{3D}}}}}}{{{\lambda _0}}}} \right)\exp \left( {j2\pi \frac{{\hat r_{rx,{\rm{LOS}}}^T\bar v}}{{{\lambda _0}}}t} \right).
\end{aligned}
\label{equ_H_LOS_6G}
\end{equation}
\end{figure*}

\begin{figure*}
\begin{equation}
\begin{aligned}
H&_{u,s,n,m}^{\rm{NLOS}}\left( t \right) = \sqrt{\beta_{u}^{\rm{SNS}} \alpha_{s,n,m}^{\rm{SNS}}} \sqrt {\frac{{{P_n}}}{M}} \left[ {\begin{array}{*{20}{c}}
{{F_{rx,u,\theta }}\left( {{\theta _{n,m,{\rm{ZOA}},u}},{\phi _{n,m,{\rm{AOA}},u}}} \right)}\\
{{F_{rx,u,\phi }}\left( {{\theta _{n,m,{\rm{ZOA}},u}},{\phi _{n,m,{\rm{AOA}},u}}} \right)}
\end{array}} \right]\mathbf{\Phi}_{n,m} \left[ {\begin{array}{*{20}{c}}
{{F_{tx,s,\theta }}\left( {{\theta _{n,m,{\rm{ZOD}},s}},{\phi _{n,m,{\rm{AOD}},s}}} \right)}\\
{{F_{tx,s,\phi }}\left( {{\theta _{n,m,{\rm{ZOD}},s}},{\phi _{n,m,{\rm{AOD}},s}}} \right)}
\end{array}} \right] \\
& \cdot\exp \left( {j2\pi \frac{{{d_{2,n,m}} - \left\| {{d_{2,n,m}} \cdot {{\hat r}_{rx,n,m}} - {{\bar d}_{rx,u}}} \right\|}}{{{\lambda _0}}}} \right)\exp \left( {j2\pi \frac{{{d_{1,n,m}} - \left\| {{d_{1,n,m}} \cdot {{\hat r}_{tx,n,m}} - {{\bar d}_{tx,s}}} \right\|}}{{{\lambda _0}}}} \right)\exp \left( {j2\pi \frac{{\hat r_{rx,n,m}^T\bar v}}{{{\lambda _0}}}t} \right).
\end{aligned}
\label{equ_H_NLOS_6G}
\end{equation}
\end{figure*}

\begin{table}[]
\renewcommand\arraystretch{1.2}
 \caption{Definitions of key parameters in channel model}
 \centering
 \begin{tabular}{m{3.8cm}<{\centering}|m{4cm}<{\centering}}
 \hline
 Channel Model Symbols & Definitions \\
\hline
\(\phi_{{\rm{LOS},AOD,}u,s}\), \(\phi_{{\rm{LOS},AOA,}u,s}\), \(\theta_{{\rm{LOS},ZOD,}u,s}\), \(\theta_{{\rm{LOS,ZOA}},u,s}\) & Near-field: AOD, AOA, ZOD, and ZOA of the TX–RX \((s, u)\) pair\\
\hline
\(\phi_{n,m,{\rm{AOD}},s}\), \(\phi_{n,m,{\rm{AOA}},u}\), \(\theta_{n,m,{\rm{ZOD}},s}\), \(\theta_{n,m,{\rm{ZOA}},u}\) &{Near-field: AOD, ZOD (TX element \(s\)) and AOA, ZOA (RX element \(u\)) for ray \(m\) in cluster \(n\)}\\
 \hline
 \raisebox{-1ex}{\({{{\vec r}_{u,s}}}\)} & { Near-field: Vector from TX element \(s\) to RX element \(u\)} \\
 \hline
\(d_{1,n,m}\) & Distance from BS to the spherical-wave source for ray \(m\) in cluster \(n\) \\
 \hline
\(d_{2,n,m}\) & Distance from UE to the spherical-wave source for ray \(m\) in cluster \(n\) \\
 \hline
\(\alpha_{s,n,m}\) & Power attenuation factor for ray \(m\) in cluster \(n\) at TX antenna element \(s\)\\
 \hline
 \(\beta_{u}\) & Power attenuation factor at RX antenna element \(u\) \\
 \hline
 {\({F_{rx,u,\theta }}\), \({F_{rx,u,\phi }}\) }&Field patterns of RX element \(u\) along spherical basis vectors \(\hat{\theta}\) and \(\hat{\phi}\), respectively \\
\hline
{\({F_{tx,s,\theta }}\), \({F_{tx,s,\phi }}\) }& Field patterns of TX element \(s\) along spherical basis vectors \(\hat{\theta}\) and \(\hat{\phi}\), respectively \\
\hline
{\({\hat{r}_{rx,n,m }}\), \({\hat{r}_{tx,n,m }}\)}& Spherical unit vectors of ray \(m\) in cluster \(n\) at the RX and TX, respectively \\
 \hline
{\({\bar{d}_{rx,u}}\), \({\bar{d}_{tx,s}}\) }& Location vectors of RX element \(u\) and TX element \(s\), respectively\\
 \hline
{\({{d}_{\rm{3D}}}\) }& 3D distance between the reference points of the BS and the UE \\
 \hline
\end{tabular}
 \label{Tab_Par}
\end{table}

\section{Channel Model Key Enhancements in 3GPP Release 19}

To accommodate the requirements of emerging 6G technologies, 3GPP Release 19 (Rel-19) introduces substantial enhancements to the baseline channel model. The initiative was formalized in December 2023, when the 3GPP Technical Specification Group (TSG) for Radio Access Network (RAN) approved a new study item titled ``Channel Modeling Enhancements for 7–24 GHz'' at the RAN \#102 meeting. Executed by the RAN1 working group, the rigorous standardization process spanned multiple technical meetings (from RAN1 \#116 to \#121), involving intensive deliberations and contributions. These efforts culminated in an updated channel model that integrates new measurement data, refined UE antenna models, and, crucially, native support for near-field propagation and spatial non-stationarity in the TR 38.901 V19.1.0 specification. The specific modifications can be categorized as follows:

\begin{itemize}
\item \textbf{Validation and Revision for the 7-24 GHz Band:}
\begin{itemize}
\item Recalibration of channel parameters, including delay spread, angular spreads (AOA/AOD/ZOA), cluster AOD spread, and angle generation scaling factors.
 \item Updated penetration loss modeling for plywood.
 \item Introduction of a new deployment scenario: SMa.
\item New antenna models for handheld and consumer premise equipment (CPE) UE.
 \item Absolute time-of-arrival modeling for Indoor hotspot (InH), Urban microcell (UMi), Urban macrocell (UMa), Rural macrocell (RMa), and SMa scenarios.
\item Flexibility in minimum intra-cluster ray counts for large bandwidth/array models.
\item Introduction of cluster number variability and polarization power variability.
\end{itemize}
\item \textbf{New Propagation Features:}
\begin{itemize}
 \item \textit{Near-field channel propagation}: Modeling spherical wavefront characteristics.
\item \textit{Spatial non-stationarity modeling}: Characterizing antenna element-wise power variations for both BS and UE.
\end{itemize}
\end{itemize}

\subsection{Validation and Revision at least for 7-24 GHz}
As operational frequencies scale up, existing channel models exhibit significant deviations from channel measurement results, leading to inaccuracies in system performance evaluation. Therefore, 3GPP has rigorously validated and refined the channel models for the 7-24 GHz frequency range. This includes updates to scenarios, characteristics, parameters, and additional modeling components, specially: the introduction of the SMa deployment scenario; enhanced modeling of channel characteristics such as plywood penetration loss and new antenna models; revisions of key channel model parameters including delay spread, AOA/AOD/ZOA spreads, cluster AOD spread, and angle generation scaling factors; and additional modeling procedures encompassing absolute time of arrival, the minimum number of intra-cluster rays for large bandwidth and large antenna array models, variability in the number of clusters, and polarization power.

\subsubsection{SMa Scenarios}

\begin{figure}[htbp]
	\centering
	\includegraphics[scale=0.1]{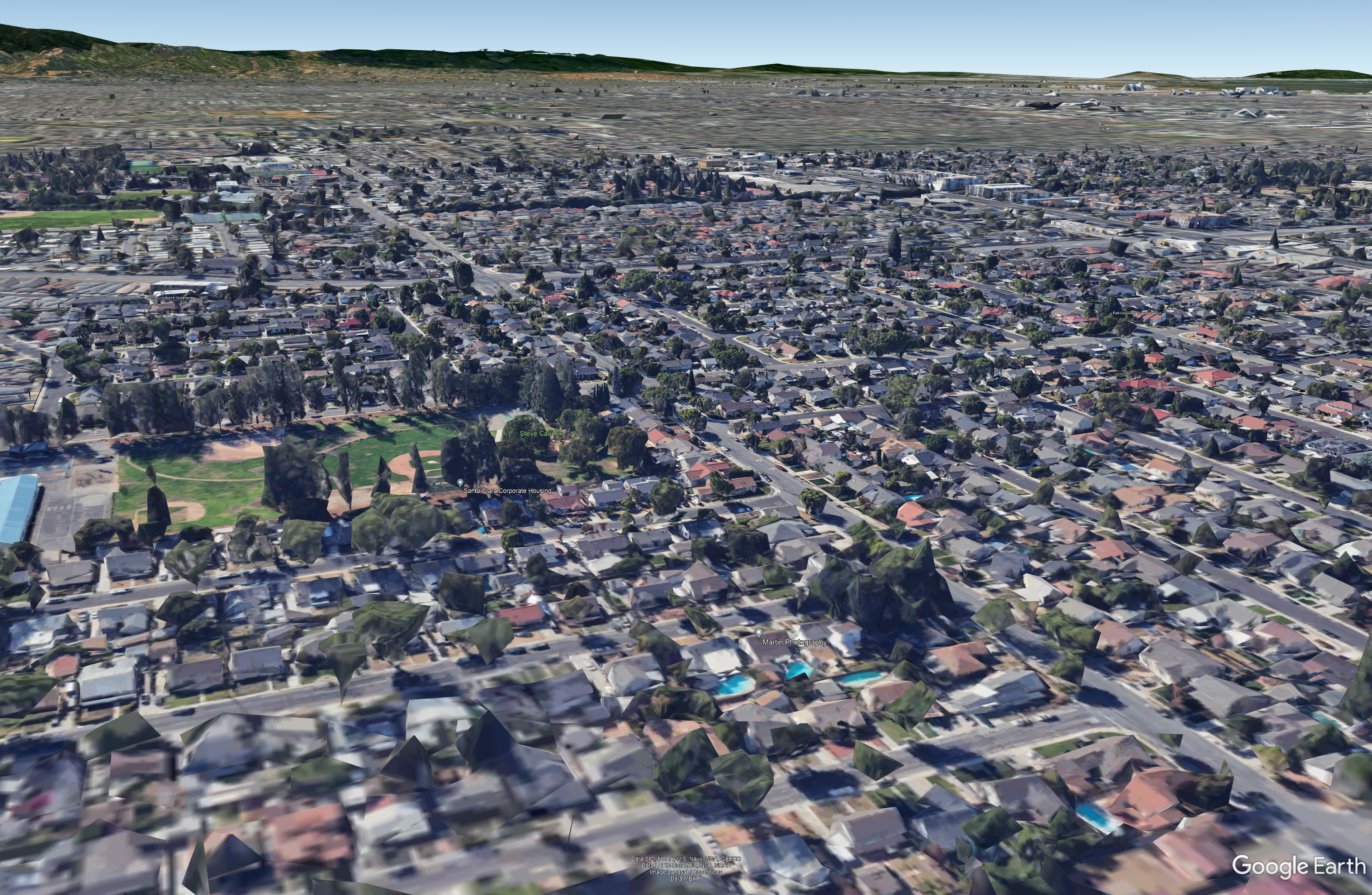}
	\caption{ Illustration of the SMa deployment scenario adopted in 3GPP Rel-19 \cite{r1-2403991}.}
    \label{SMa}
\end{figure}

\begin{figure}[htbp]
	\centering
	\includegraphics[scale=0.8]{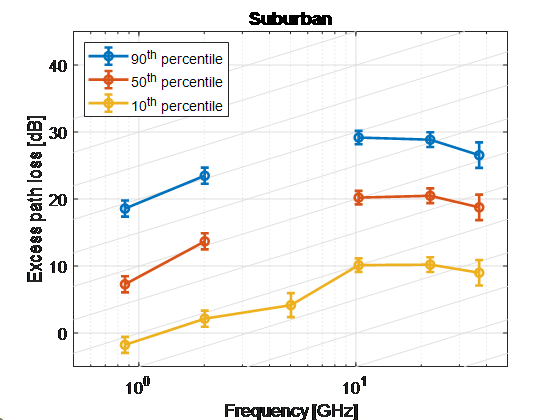}
	\caption{Measured excess path loss in the SMa scenario, with error bars denoting the measurement uncertainty. Note that some percentiles are missing for 5 GHz due to an equipment misconfiguration that affected the sensitivity \cite{r1-2402613}.}
    \label{SMaPL}
\end{figure}

The SMa scenario is defined as follows: In suburban macro-cells, base stations are placed above the surrounding environment to provide wide-area coverage, while mobile stations are located outdoors at street level as well as inside commercial and residential buildings, as illustrated in Fig. \ref{SMa}. Detailed evaluation parameters are provided in Table. \ref{SMa_para}. The focus on the SMa scenario arises from its prevalence in regions such as Europe and the United States, where a significant number of user terminals operate within this type of scenario. This new scenario is introduced because the frequency-dependent path loss models of the existing scenarios (e.g., UMa, UMi) could not accurately predict path loss in SMa scenarios, as illustrated in \ref{SMaPL}. Consequently, based on extensive channel measurement results in SMa, Ericsson pioneered the proposal at the 3GPP RAN 1 116-bis meeting to add SMa scenarios and the specific channel parameters to the TR 38.901 standard \cite{r1-2402613}.

\begin{table*}[t]
\centering
\renewcommand{\arraystretch}{1.15}
\label{SMa_para}
\caption{Evaluation parameters for SMa scenarios.}
\begin{tabular}{p{3.2cm}|p{3.2cm}|p{8.4cm}}
\hline
\multicolumn{2}{c|}{Parameters} & SMa \\
\hline

\multicolumn{2}{p{6.4cm}|}{Cell layout} &
Hexagonal grid, 19 macro sites, 3 sectors per site \newline
(ISD = 1299 m and 1732 m, see note 1 and 2) \newline
Up to two floors for residential buildings, up to five floors for commercial buildings. \newline
Building distribution are 90\% residential and 10\% commercial buildings \\
\hline

\multicolumn{2}{p{6.4cm}|}{BS antenna height $h_{\mathrm{BS}}$} & 35 m \\
\hline

\multirow{3}{*}{UE\newline location} & Outdoor/indoor & Outdoor and indoor \\
\cline{2-3}
& LOS/NLOS & LOS and NLOS \\
\cline{2-3}
& Height $h_{\mathrm{UE}}$ &
1.5 m for outdoor, \newline
1.5 or 4.5 m for residential buildings, \newline
1.5/4.5/7.5/10.5/13.5 m for commercial buildings \\
\hline

\multicolumn{2}{p{6.4cm}|}{Indoor UE ratio} & 80\% \\
\hline

\multicolumn{2}{p{6.4cm}|}{UE mobility (horizontal plane only)} &
Indoor UEs: 3 km/h \newline
Outdoor UEs (in car): 40 km/h \\
\hline

\multicolumn{2}{p{6.4cm}|}{Min.\ BS -- UE distance (2D)} & 35 m \\
\hline

\multicolumn{2}{p{6.4cm}|}{UE distribution (horizontal)} & Uniform \\
\hline

\multicolumn{2}{p{6.4cm}|}{UE distribution (vertical)} &
Uniform distribution across all floors for a building type \\
\hline

NOTE 1: & \multicolumn{2}{p{11.8cm}}{For evaluation in study items/work items, if needed, admission control policies may be considered in conjunction with the  Inter-Site Distance (ISD) choice to address out of coverage UEs} \\
\hline

NOTE 2: & \multicolumn{2}{p{11.8cm}}{SMa scenarios with ISDs between 1200--1800 m can be used for evaluations.} \\
\hline

\end{tabular}
\end{table*}

At subsequent 3GPP 118 and 118-bis meetings, a layout for the SMa scenario in TR 38.901 was tabled, taking the WINNER II \cite{winner2} as references. Further updates—covering path loss, LOS probability, angular spreads, and other parameters—were presented at subsequent meetings. The SMa model parameters defined in IMT-Advanced are used as references in combination with actual measurements to determine the starting point for specifying additional large-scale parameters in TR 38.901, such as correlation distances, spatial consistency, and blockage-region characteristics. Eventually, 3GPP approved the inclusion of the SMa scenario and its corresponding modeling methodology in the channel modeling standard. For example, the CR of the 3GPP meeting defines the path loss model in the SMa scenario as formulated as:

\begin{equation}
PL_{\text{SMa-LOS}} =
\begin{cases}
  PL_1, & 10\,\text{m} \le d_{2\mathrm{D}} < d_{\mathrm{BP}} \\[4pt]
  PL_2, & d_{\mathrm{BP}} \le d_{2\mathrm{D}} \le 5\,\text{km}
\end{cases}
\label{sma_pl}
\end{equation}

with

\begin{equation}
\begin{aligned}
PL_1 =& 20\log_{10}\left(\frac{40\pi d_{3\mathrm{D}} f_c}{3}\right)
       + \min\bigl(0.03h^{1.72},10\bigr)\cdot\\
       &\log_{10}(d_{3\mathrm{D}}) - \min\bigl(0.044\,h^{1.72},14.77\bigr)\\
       &+ 0.002\log_{10}(h)\,d_{3\mathrm{D}}
\end{aligned}
\end{equation}

\begin{equation}
PL_2 = PL_1(d_{\mathrm{BP}}) + 40\log_{10}\!\left(\frac{d_{3\mathrm{D}}}{d_{\mathrm{BP}}}\right)
\end{equation}

\noindent where $d_{\mathrm{BP}}$ is defined as $d_{\mathrm{BP}} =2\pi h_{\mathrm{BS}} h_{\mathrm{UE}} f_c /c$. $f_c$ is the center frequency in Hz, $c = 3.0 \times 10^8$ m/s is the propagation velocity in free space, and $h_{\mathrm{BS}}$ and $h_{\mathrm{UE}}$ are the antenna heights at the BS and UE, respectively.

\subsubsection{UE Pattern}
\begin{figure}[htbp]
	\centering
	\includegraphics[width=3 in]{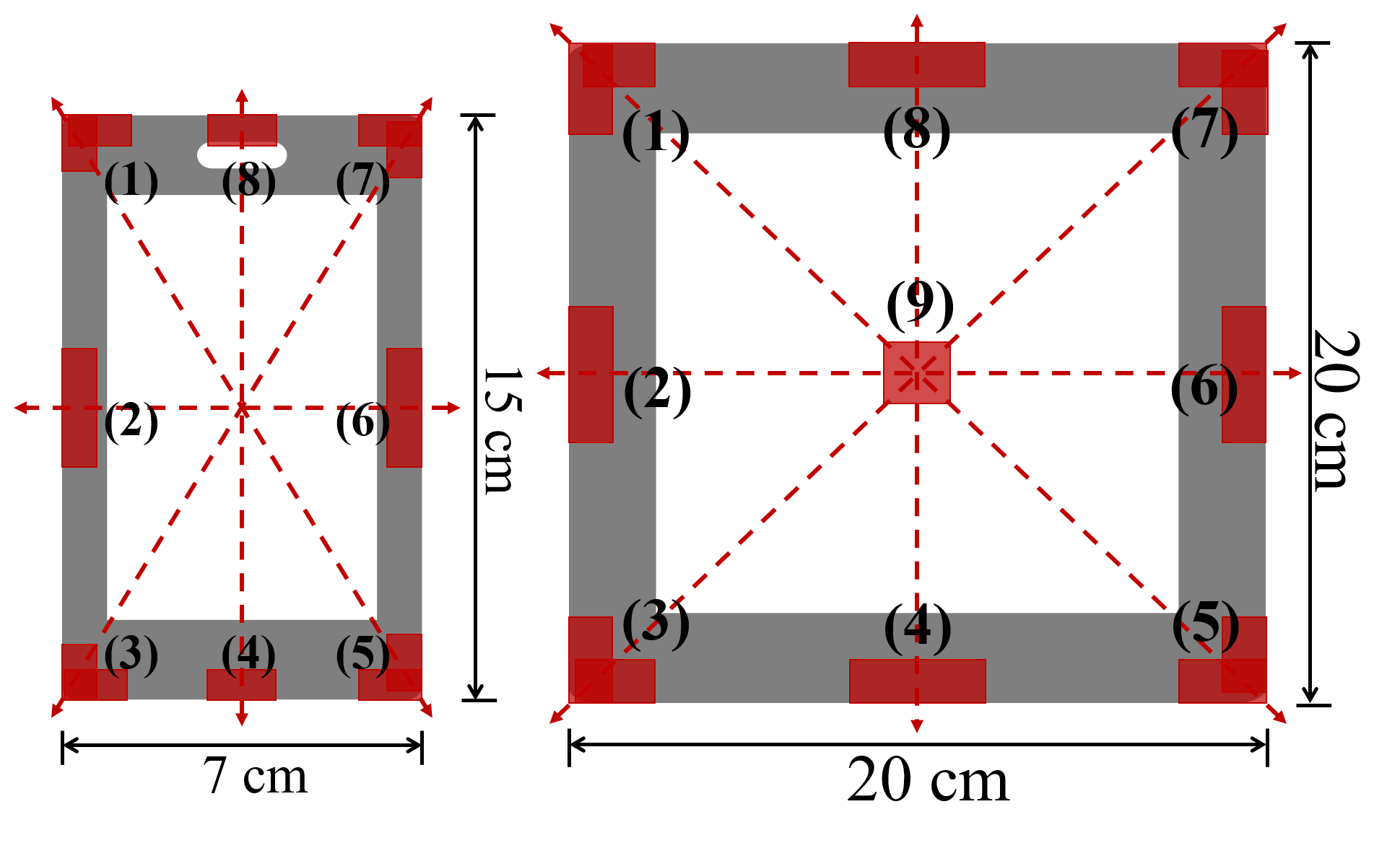}
	\caption{Handheld (top down view) and CPE (side view) UE antenna placement candidate locations relative to center of device \cite{3GPP_38901}.}\label{handheld_UE}
\end{figure}

As communication systems evolve with ever-increasing demands for bandwidth and data rates, the existing antenna patterns of user terminals (UEs) are often insufficient to meet the requirements of new-generation devices. This necessitates the development of new UE antenna pattern models. The UE antenna can be modeled by placing the antenna element or array at the placement candidate locations relative to the center of the UE. The overall device size can be defined in terms of depth, width, and height as (X cm, Y cm, Z cm). For handheld UE devices, potential antenna placement sites are typically chosen at the four corners and the midpoints of the edges of the rectangular form factor. This configuration captures the key geometric features of the UE, as illustrated in Fig. \ref{handheld_UE}. Each antenna is assumed to be oriented along the direction determined by the vector connecting the center of the rectangle to the antenna location. In such modeling, the device dimensions, i.e., the length, width, and thickness, for the UE antenna are generally specified as (15 cm, 7 cm, 0 cm). The representation of the handheld UE orientation will be the same between the global coordinate system (GCS) and UE local coordinate system (LCS) for $\Omega _{UE,\alpha} = 0^\circ $, $\Omega _{UE,\beta } = 0^\circ$, $\Omega _{UE,\gamma } = 0^\circ$. For consumer premise equipment (CPE) devices, 9 candidate antenna/array positions are typically arranged on a vertical square plane: these include the four corners, the midpoints of the four edges, and the center of the square, as illustrated in Fig. \ref{handheld_UE}. The device dimensions used for UE antenna modeling in this configuration are (0 cm, 20 cm, 20 cm). 

\subsubsection{Penetration Loss}
Wall penetration loss—defined as the difference between the median signal level on one side of a wall and the signal level on the opposite side at the same height above ground \cite{M2412}—is a critical factor in wireless system planning. It directly impacts coverage, capacity, and quality of service (QoS) \cite{Ryan2017}, \cite{Maccartney2015}. Penetration loss is strongly frequency-dependent, and its magnitude can vary significantly across different frequency bands. This is especially relevant for FR1 and FR3, which together cover a wide frequency range from approximately 450 MHz to 24 GHz. To improve the accuracy of large-scale channel modeling, recent 3GPP channel models have introduced material-specific penetration loss parameters for plywood. The penetration loss $L_{\text{plywood}}$ (in dB) can be expressed as:
\begin{equation}
L_{\text{plywood}}= 1.03+0.17f_c
\end{equation}
where $f_c$ is the frequency in GHz. Incorporating this material-specific parameter enables more precise prediction of signal attenuation through plywood partitions, thereby enhancing system-level simulations and network planning in mid-band and FR3

\subsubsection{Large-Scale Parameters}
\begin{figure}[htbp]
	\centering
	\includegraphics[scale=0.5]{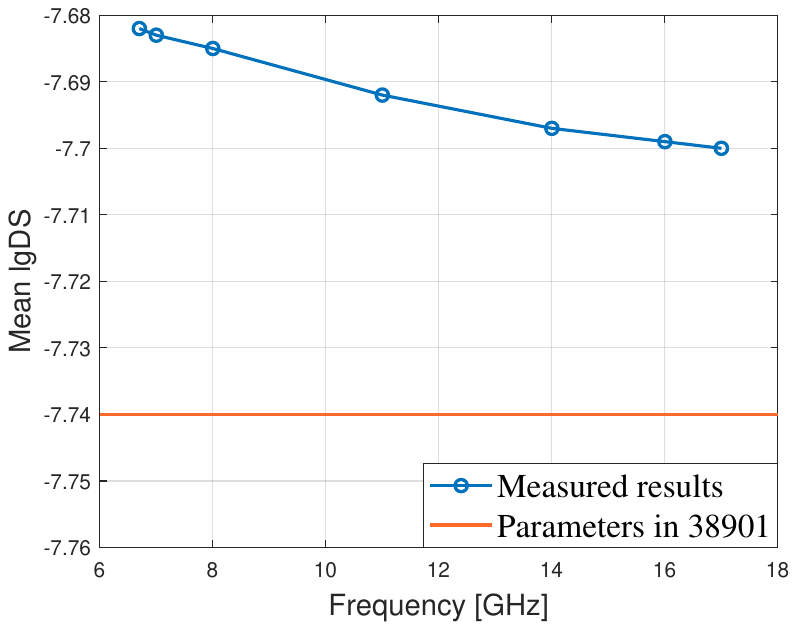}
	\caption{Frequency dependence of DS in the indoor factory scenario: measurements versus TR 38.901 reference values.}
    \label{DS_UMi}
\end{figure}

WRC-23 \cite{ITU_WRC23} has made a resolution about sharing and compatibility studies and development of technical conditions for the use of IMT in the frequency bands 4400-4800 MHz, 7125-8400 MHz, and 14.8-15.35 GHz for the terrestrial component of IMT. Also, it is expected to identify the exact frequency bands in WRC-27. The existing 3GPP channel model, while nominally covering 0.5 to 100 GHz, relied on parameters derived primarily from data below 6 GHz and above 25 GHz. To obtain accurate frequency-dependent characteristics for the newly assigned spectrum, robust channel measurements were deemed indispensable in the 7–24 GHz range. A consortium of 15 companies carried out measurement campaigns spanning 7–24 GHz across five scenarios: UMa, UMi, Indoor, SMa, and O2I. Based on these collective findings, 3GPP updated the existing large-scale parameter table. Fig. \ref{DS_UMi} shows the delay spread (DS) measurements in the indoor-factory scenario, revealing sizable discrepancies from the legacy reference values. It is found that the DS decreases with increasing frequency. The value of DS at 6.7 GHz is 20.8 ns. As the frequency increases, the DS decreases gradually. At 17 GHz, the delay spread is 19.95 ns. In addition, the measured results are generally greater than those in the standard.

\subsubsection{Cluster Number}
\begin{figure}[htbp]
	\centering
	\includegraphics[scale=0.6]{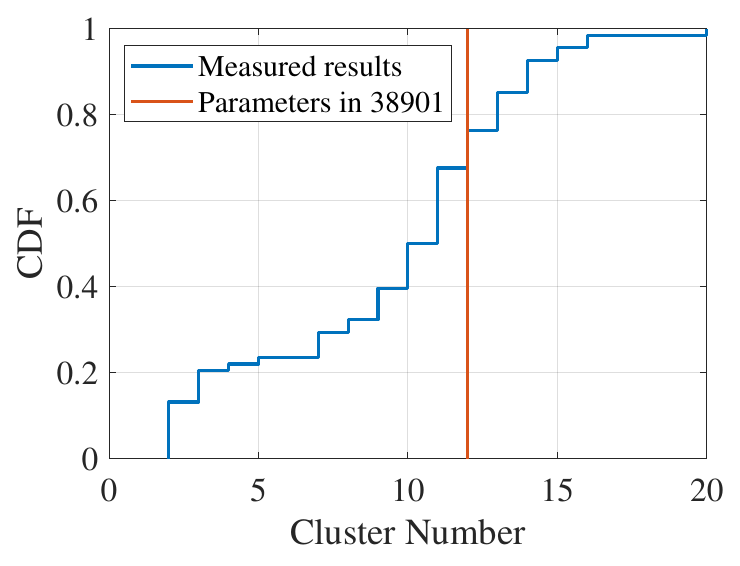}
	\caption{The CDF of the cluster number in the UMa scenario.}
    \label{cluster_number}
\end{figure}

The number of clusters observed in real-world deployments exhibits significant variability due to factors such as the propagation environment, carrier frequency, bandwidth, and spatial resolution. Moreover, with increasing frequency, the channel becomes spatially sparser, resulting in a reduced number of clusters. Contributions to RAN1 highlighted this trend using KPowerMeans clustering on UMa measurement data, showing that the mean number of clusters at 13 GHz is significantly lower than the reference value assumed in the original model (Fig. \ref{cluster_number}). Consequently, 3GPP approved a ``cluster number variability" model, wherein the number of clusters for each link is randomly chosen between a scenario-dependent closed range $[D_1, D_2]$. This approach allows the model to dynamically reproduce the fluctuation in multipath richness observed in measurements.

\begin{table}[htbp]
    \centering
    \caption{Range of number of clusters}
    \begin{tabular}{|l|c|c|c|}
        \hline
        \textbf{Scenarios} &
        \textbf{LOS} &
        \textbf{NLOS} &
        \textbf{O2I} \\
        \hline
        UMi &
        \begin{tabular}[c]{@{}c@{}}$D_1: 6$ \\ $D_2: 12$\end{tabular} &
        \begin{tabular}[c]{@{}c@{}}$D_1: 6$ \\ $D_2: 19$\end{tabular} &
        \begin{tabular}[c]{@{}c@{}}$D_1: 6$ \\ $D_2: 12$\end{tabular} \\
        \hline
        UMa &
        \begin{tabular}[c]{@{}c@{}}$D_1 : 10$ \\ $D_2 : 12$\end{tabular} &
        \begin{tabular}[c]{@{}c@{}}$D_1 : 15$ \\ $D_2 : 20$\end{tabular} &
        \begin{tabular}[c]{@{}c@{}}$D_1 : 10$ \\ $D_2 : 12$\end{tabular} \\
        \hline
        Indoor-office &
        \begin{tabular}[c]{@{}c@{}}$D_1 : 7$ \\ $D_2 : 15$\end{tabular} &
        \begin{tabular}[c]{@{}c@{}}$D_1 : 6$ \\ $D_2 : 19$\end{tabular} &
        N/A \\
        \hline
    \end{tabular}
\end{table}

\subsubsection{Large Bandwidth and Large Antenna Array}
\begin{figure}[htbp]
	\centering
	\includegraphics[scale=0.3]{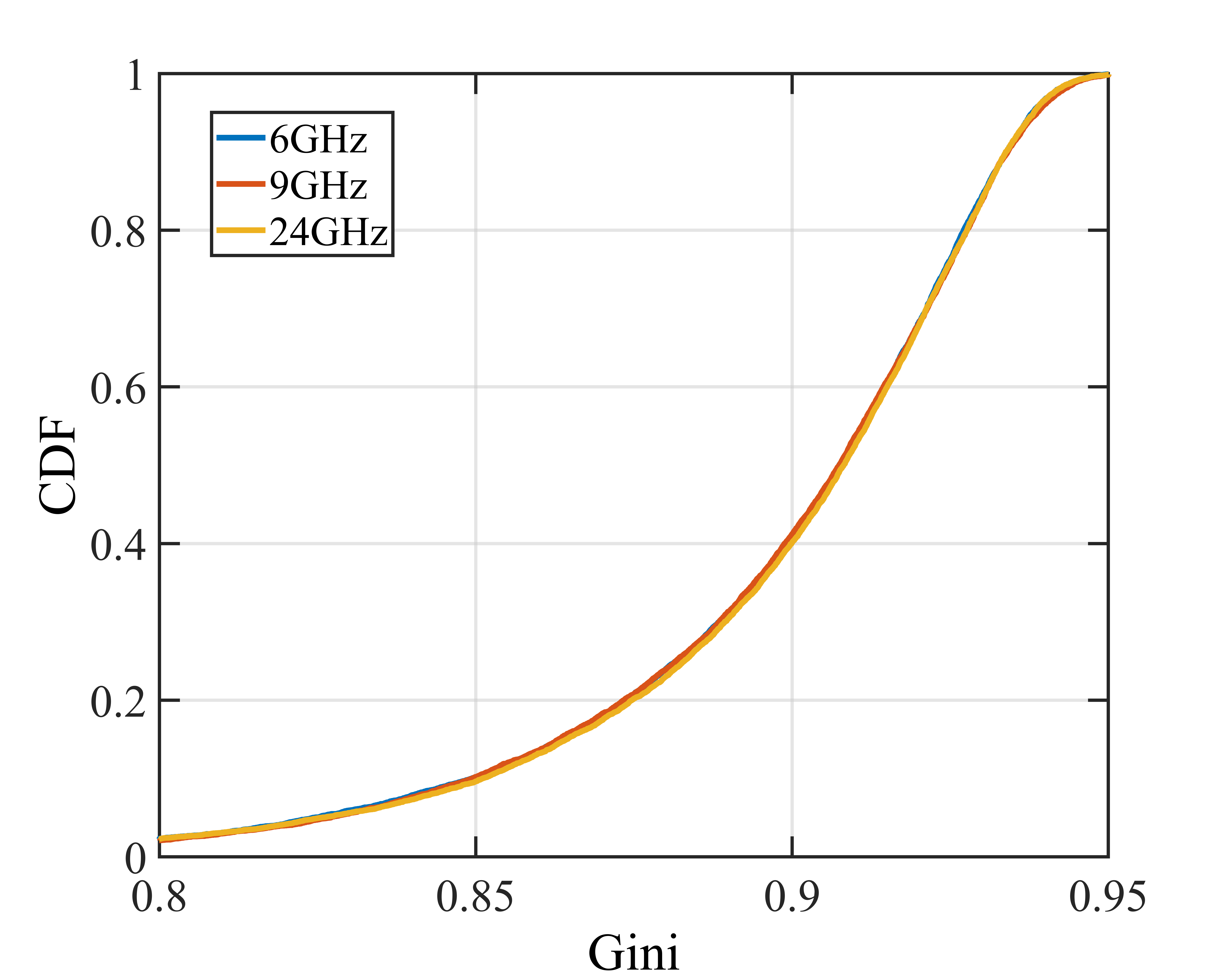}
	\caption{The CDF of the Gini index obtained by the original model in UMa scenarios.}
    \label{Gini_20}
\end{figure}

To accurately characterize frequency-dependent channel characteristics, model parameters must be allowed to vary with frequency. However, the legacy ray-number scaling equation failed to generate frequency-dependent channel impulse responses because the product of resolvable rays $(M_tM_{AOD}M_{ZOD})$ hit a fixed floor of 20 across the frequency range, which limits the capability to characterize the sparse properties across different frequency bands, as illustrated in Fig. \ref{Gini_20}. Simulation analysis identified that the calculated product for 6–24 GHz is consistently below 20, rendering the frequency-dependent scaling ineffective. To rectify this, Rel-19 revises the minimum number of multipath components $(M_{min})$ to explicitly capture channel sparsity. And the number of rays per cluster shall be calculated as follows:

\begin{equation}
M = \min\left( \max(M_t\cdot M_{AOD}\cdot M_{ZOD}, M_{\min}), M_{\max} \right)
\end{equation}

\noindent with 

\begin{equation}
M_t = \left \lceil 4kc_{DS}B \right \rceil
\end{equation}

\begin{equation}
M_{AOD} =  \left \lceil 4kc_{ASD} \frac{\pi \cdot D_h}{180 \cdot \lambda}  \right \rceil 
\end{equation}

\begin{equation}
M_{ZOD} = \left \lceil 4kc_{ZSD} \frac{\pi \cdot D_v}{180 \cdot \lambda} \right \rceil  
\end{equation}

\noindent where $M_{\min}$ can be chosen based on the mean of dominant rays across all clusters for a specific frequency and deployment scenario. For example, the number of dominant rays in a cluster is defined as the minimum number of rays that contain $95\%$ of the total cluster power, when the rays in a cluster are sorted in descending order of power. The default value of $M_{\min}=20$ is assumed. $M_{\max}$ is the upper limit of $M$, and it should be selected by the user of channel model based on the trade-off between simulation complexity and accuracy.
$D_h$ and $D_v$ are the array size in m in horizontal and vertical dimensions, $B$ is the bandwidth in Hz, $c_{ASD}$ and $c_{ZSD}$ are the cluster spreads in degrees, and $\lambda$ is the wavelength. $k$ is a ``sparseness" parameter with value 0.5.

\subsubsection{Polarization}
In the legacy TR 38.901 model, the two co-polar components are assumed to have equal power, as are the two cross-polar components. However, empirical evidence indicates a slow, independent variability around the mean co-polar and cross-polar power levels \cite{r1-2402613}. 


To support the simulation of this phenomenon, polarization variability factors $\eta_{n,m,\theta\theta}$, $\eta_{n,m,\theta\phi}$, $\eta_{n,m,\phi\theta}$ and $\eta_{n,m,\phi\phi}$ for each ray $m$ of each cluster $n$ are generated and applied to NLOS channel coefficients to enable polarization power variability. 
\begin{figure*}
\begin{align*}
H_{u,s,n}^{\text{NLOS}}(t) = & \sqrt{\frac{P_n}{M}} \sum_{m=1}^{M} 
 \begin{bmatrix} 
    F_{rx,u,\theta} \big( \theta_{n,m,\text{ZOA}}, \phi_{n,m,\text{AOA}} \big) \\
    F_{rx,u,\phi} \big( \theta_{n,m,\text{ZOA}}, \phi_{n,m,\text{AOA}} \big) 
\end{bmatrix}^T 
\begin{bmatrix} 
    \sqrt{\eta_{n,m,\theta\theta}} \exp(j\Phi_{n,m}^{\theta\theta}) & 
    \sqrt{\eta_{n,m,\theta\phi} \kappa _{n,m}^{-1}} \exp(j\Phi_{n,m}^{\theta\phi}) \\ 
    \sqrt{\eta_{n,m,\phi\theta} \kappa _{n,m}^{-1}} \exp(j\Phi_{n,m}^{\phi\theta}) & 
    \sqrt{\eta_{n,m,\phi\phi}} \exp(j\Phi_{n,m}^{\phi\phi}) 
\end{bmatrix} \\
& \begin{bmatrix}
    F_{tx,s,\theta} \big( \theta_{n,m,\text{ZOD}}, \phi_{n,m,\text{AOD}} \big) \\
    F_{tx,s,\phi} \big( \theta_{n,m,\text{ZOD}}, \phi_{n,m,\text{AOD}} \big)
\end{bmatrix}  \exp \left( \frac{j2\pi (\hat{r} ^T_{rx,n,m} \cdot \bar{d}_{rx,u})}{\lambda_0} \right)
  \exp \left( \frac{j2\pi (\hat{r}^T_{tx,n,m} \cdot \bar{d}_{tx,s})}{\lambda_0} \right)  \exp \left( j2\pi \frac{\hat{r}^T_{rx,n,m} \bar{v}}{\lambda_0} t \right),
\end{align*}
\end{figure*}
where $Q_{n,m,\left \{ \theta \theta ,\theta \phi ,\phi \theta ,\phi \phi  \right \} }$  is a Normal distributed random variable with mean 0 and variance $3^2$, i.e., $N(0,3^2 )$, independently drawn for each ray, cluster, and polarization component, and used to generate polarization variability powers $\eta_{n,m,\theta\theta}$, $\eta_{n,m,\theta\phi}$, $\eta_{n,m,\phi\theta}$ and $\eta_{n,m,\phi\phi}$ based on following equations: 

\begin{equation}
\eta '_{n,m,\left \{ \theta\theta,\theta\phi,\phi\theta,\phi\phi \right \} } = 10^{\frac{Q_{n,m,\left \{ \theta\theta,\theta\phi,\phi\theta,\phi\phi \right \} }}{10}} 
\end{equation}

\begin{align*}
&\eta_{n,m,\left \{ \theta\theta,\theta\phi,\phi\theta,\phi\phi \right \} }=\eta '_{n,m,\left \{ \theta\theta,\theta\phi,\phi\theta,\phi\phi \right \} } \\
& \frac{2+2\kappa _{n,m}^{-1} }{\eta '_{n,m,\theta\theta }+\eta '_{n,m,\phi\phi }+\eta '_{n,m,\theta\phi }\kappa _{n,m}^{-1}+\eta '_{n,m,\phi\theta }\kappa _{n,m}^{-1}} 
\end{align*}

\subsubsection{Absolute Delay}

To account for the positioning and sensing in which absolute time of arrival is important, the propagation time delay due to the total path length is considered in step 11 of the fast fading model \cite{3GPP_38901}. The impulse response in NLOS is expressed as

\begin{equation}
H_{u,s,n}^{\text{NLOS}}(\tau,t) = H_{u,s,n}^{\text{NLOS}}(t) \delta \left(\tau-\tau_{n}-\frac{d_{3D}}{c}-\Delta\tau\right)
\end{equation}



\noindent where $c$ is the speed of light. $\Delta\tau$ is generated from a lognormal distribution with parameters according to Table \ref{absolute}. $\Delta\tau$ is generated independently for links between the same UE and different BS sites. 

For InF and InH scenarios, the excess delay in NLOS, $\Delta\tau$, should further be upper bounded by $\frac{2L}{c} $, where $L$ is the largest dimension of the factory hall and office room. For UMi-Street Canyon, UMa, SMa, and RMa scenarios, the excess delay is not upper bounded.

\begin{table*}[htbp]
    \centering
    \caption{Parameters for the absolute time of arrival model in NLOS}
    \label{absolute}
    \begin{tabular}{|c|c|c|c|c|c|c|c|c|}
        \hline
        \multicolumn{2}{|c|}{Scenarios}
        & InH & \begin{tabular}{@{}c@{}}InF-SL, \\ InF-DL\end{tabular} & \begin{tabular}{@{}c@{}}InF-SH, \\ InF-DH\end{tabular} & UMi & UMa & RMa & SMa \\
        \hline
        \multirow{2}{*}{\( \lg \Delta \tau = \log_{10}(\Delta \tau / 1\,\text{s}) \)} 
            & \( \mu_{\lg \Delta \tau} \) & -8.6 & -7.5 & -7.5 & -7.5 & -7.4 & -8.33 & -7.702 \\
        \cline{2-9}
            & \( \sigma_{\lg \Delta \tau} \) & 0.1 & 0.4 & 0.4 & 0.5 & 0.2 & 0.26 & 0.4 \\
        \hline
        \multicolumn{2}{|c|}{\begin{tabular}{@{}c@{}}Correlation distance in \\ the horizontal plane [m]\end{tabular} } 
            &10 & 6& 11 & 15 &50& 50 & 50 \\
        \hline
    \end{tabular}
\end{table*}

\subsection{New Propagation Features}
With the integration of ELAA, the planar-wave assumptions inherent in legacy models become invalid when the link operates in the radiating near field. In this region, wavefront curvature becomes non-negligible, resulting in spherical propagation, element-dependent delays/angles, and non-linear phase variations across the aperture. Concurrently, partial blockage and incomplete scattering cause different array partitions to observe distinct multipath clusters, inducing SNS. Consequently, accurately modeling near-field and SNS characteristics is essential for representing physical propagation mechanisms in future systems operating in the 7–24 GHz range and beyond.

\begin{table*}[]
\renewcommand\arraystretch{1.2}
    \caption{A brief review of current near-field propagation and SNS channel modeling approaches}
    \centering
    \begin{tabular}{m{0.8cm}<{\centering}|m{3.4cm}<{\centering}|m{1.5cm}<{\centering}|m{4cm}<{\centering}|m{5.7cm}<{\centering}}
    \hline
    Features & Models  & Complexity & Pros & Cons  \\
    \hline
    \multirow{5}{*}[-1.5em]{\parbox{2em}{Near-field}} & Ray-tracing \cite{Yuan_RT,C1_Hybrid_HC} & High & Accurate near-field spherical-wave modeling & Very high computational cost; incompatible with 3GPP framework  \\
    \cline{2-5}
    & Quadriga model \cite{QuaDRiGa,C1_NF_HC} & Moderate & Extra-parameter-free & Optimization may diverge or yield outliers \\
    \cline{2-5}
    &Random first/last bounce scatterer distance model \cite{3GPP_BUPT_hefei} & Low & Simple and easy to implement & Incapable of modeling some interactions (e.g., specular reflection) \\
    \cline{2-5}
    & Based on spatial consistency model \cite{3GPP_ZTE_hefei} & Low & Minor-modification 3GPP model & Beyond the valid range of TR 38.901 spatial consistency \\
    \cline{2-5}
    & Adopted model \cite{3GPP_38901} & Low & Simple and consistent with propagation mechanism & - \\
    \hline
    \multirow{6}{*}[-4em]{SNS} & Ray-tracing \cite{Yuan_RT} & High & Accurate in array power variation & Very high computational cost; incompatible with 3GPP framework  \\
    \cline{2-5}
    & Fading channel model \cite{SNS_fading} & Low & Suitable for link-level simulation & Not applicable to 3GPP system-level simulation \\
    \cline{2-5}
    & BS-VR model (e.g., COST 2100) \cite{C1_VR_COST,C1_VR_ljz} & Moderate & Consistent across antenna elements & Lack of consistency among clusters; hard to obtain scenario parameters; sharp power jumps between visible/invisible regions \\
    \cline{2-5}
    &Birth–death process \cite{C1_cluster_LR,C1_BD_CX} & Moderate & Suitable for spatial-correlation analysis & Cluster evolution fully random; inconsistent across antennas; Intractable birth–death probabilities \\
    \cline{2-5}
    & Adopted stochastic based model \cite{3GPP_38901} & Moderate & Computationally efficient, reflecting SNS from incomplete scattering & - \\
    \cline{2-5}
    & Adopted physical blocker-based model \cite{3GPP_38901} & Relatively high &Physically accurate and consistent, reflecting SNS from partial blockage & - \\
    \hline
    \end{tabular}
    \label{Tab_NF_SNS}
\end{table*}

\subsubsection{Near-Field Propagation}
The near-field channel is modeled by explicitly introducing the geometric distances from the BS and UE to ``spherical-wave sources". These distances characterize the element-wise path parameter variations across the large antenna array. This spherical-wave formulation accurately captures wavefront curvature, thereby overcoming the limitations of the conventional far-field planar-wave assumption. To provide a comprehensive overview of this modeling framework, the following subsections discuss three key aspects: (i) the definition of the boundary between near-field and far-field regions, (ii) the modeling methodology, and (iii) the derivation of antenna element-wise channel parameters.

\paragraph{Near-Field and Far-Field Boundary}

The distinction between near- and far-field propagation has long been a subject of academic debate. Rather than simply replacing the far-field assumption with a near-field one, recent studies have emphasized two intertwined aspects: the transition region between near- and far-fields (cross-field \cite{Han2024,Liu2023,Wang2025}) and the coexistence of both components (hybrid-field \cite{Wei2022,WangHongwei2025}) within the same link. To ensure the model flexibly supports diverse transmission technologies, 3GPP conducted extensive discussions on this issue during multiple RAN1 meetings (from \#116-bis to \#118-bis).

Despite intensive deliberations, no unified or physically unambiguous criterion was established for defining a hard boundary between near- and far-field regions. For direct paths, a simplified assumption was adopted to determine the field condition based on the 3D geometric distance between the BS and the UE. For non-direct paths (clusters), several alternatives were considered, including aligning with the direct-path condition, using the distance between the array and the cluster, or defining a probabilistic near-field condition. However, existing metrics were deemed insufficient; for instance, the widely used Rayleigh distance is derived under a strict $\pi/8$ phase-error criterion, while other metrics such as scaled Rayleigh distances, Fraunhofer-based approximations, or uniform-power-distance definitions lack universality. Consequently, a consensus emerged that the primary objective of channel modeling is to faithfully reproduce physical propagation phenomena rather than to anchor the boundary to a specific performance metric.

\begin{figure}[htbp]
	\centering
	\includegraphics[width=7cm]{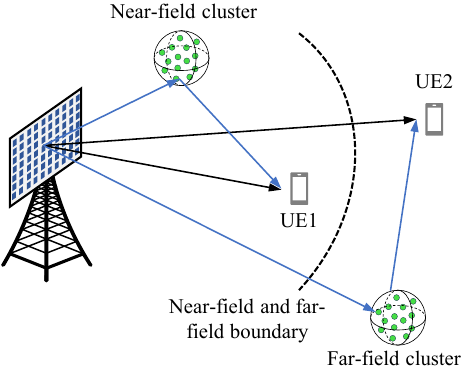}
	\caption{Illustration of channel discontinuity caused by separate near- and far-field modeling at the boundary.}
 \label{fig_NF_Boundary}
\end{figure}

In practice, the transition between near- and far-field regions is continuous. Based on this physical reality, 3GPP adopted a unified modeling framework that eschews an explicit hard boundary or switching mechanism. Instead, the model relies on a spherical-wave representation that asymptotically degenerates into the planar-wave case as the propagation distance increases. This approach ensures physical consistency and maintains fairness in system-level evaluations. Explicitly partitioning the simulation into near- and far-field regimes could introduce artificial discontinuities in channel realizations, leading to non-comparable results for users located near the switching boundary. As shown in Fig.~\ref{fig_NF_Boundary}, two UEs are close to each other, but the channels corresponding to two UE-links are modeled by different implementations of the channel models. This may incur a certain dramatic discontinuity between the channels. By applying the near-field model uniformly, the framework guarantees modeling continuity and robustness, with evaluations indicating only marginal computational overhead (less than 20\%)  \cite{3GPP_vivo_helan}.

\paragraph{Modeling Methodology}

Various approaches have been explored in academia to capture the near-field propagation characteristics. These include deterministic ray-tracing (RT) methods, optimization-based models (e.g., the Quadriga framework), random first/last-bounce scatterer distance models, and spatial-consistency-based extensions to the existing 3GPP model, as shown in Table~\ref{Tab_NF_SNS}. Specifically, ray-tracing-based methods explicitly resolve geometric propagation paths and can accurately capture spherical-wave effects, but their high computational complexity and lack of compatibility with the stochastic structure of 3GPP TR 38.901 limit their applicability for standardized system-level simulations \cite{Yuan_RT,C1_Hybrid_HC}. Optimization-based approaches, such as those built upon the Quadriga framework, infer effective near-field propagation distances by reconstructing multi-bounce geometries from existing large-scale parameters \cite{QuaDRiGa,GaoTianyang2023}; while parameter-efficient and formally compatible with 3GPP, they often suffer from convergence issues, unrealistic outliers, and the absence of closed-form expressions, which complicates standardization. Random first/last-bounce distance models follow the stochastic philosophy by statistically generating near-field distances from distributions fitted to ray-tracing data \cite{3GPP_BUPT_hefei}; although simple and easy to implement, they lack the ability to represent specific physical interactions. Finally, spatial-consistency-based extensions attempt to reuse the TR 38.901 spatial consistency mechanism to emulate near-field behavior through smooth parameter evolution \cite{3GPP_ZTE_hefei}, but evaluations show that this approach cannot reliably reproduce spherical-wave phase curvature and may violate physical consistency when applied beyond its validated range. While each approach offers specific advantages regarding physical interpretability or implementation efficiency, none fully achieves an optimal trade-off between modeling accuracy and computational complexity. 

\begin{figure*}[h]
\centering
\subfloat[]
{\includegraphics[width=8cm]{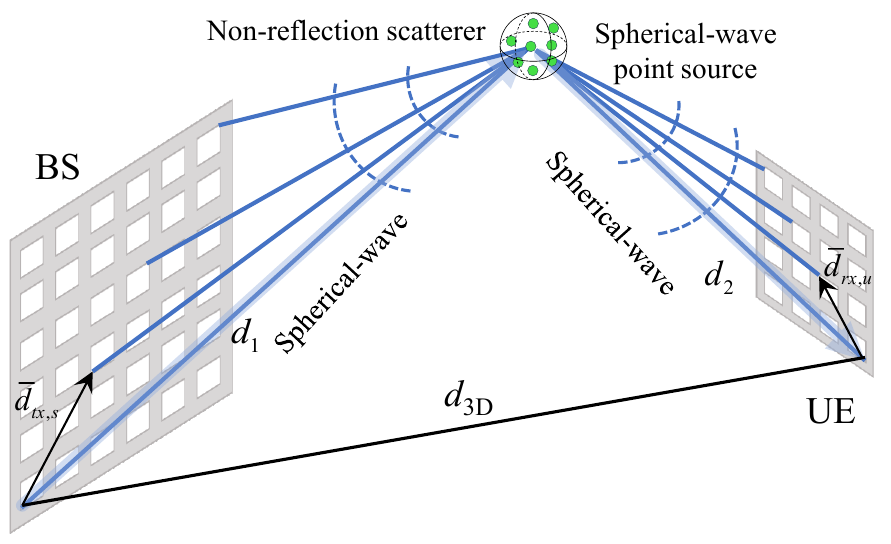}\label{fig_NF_model_a}}
\subfloat[]
{\includegraphics[width=7.5cm]{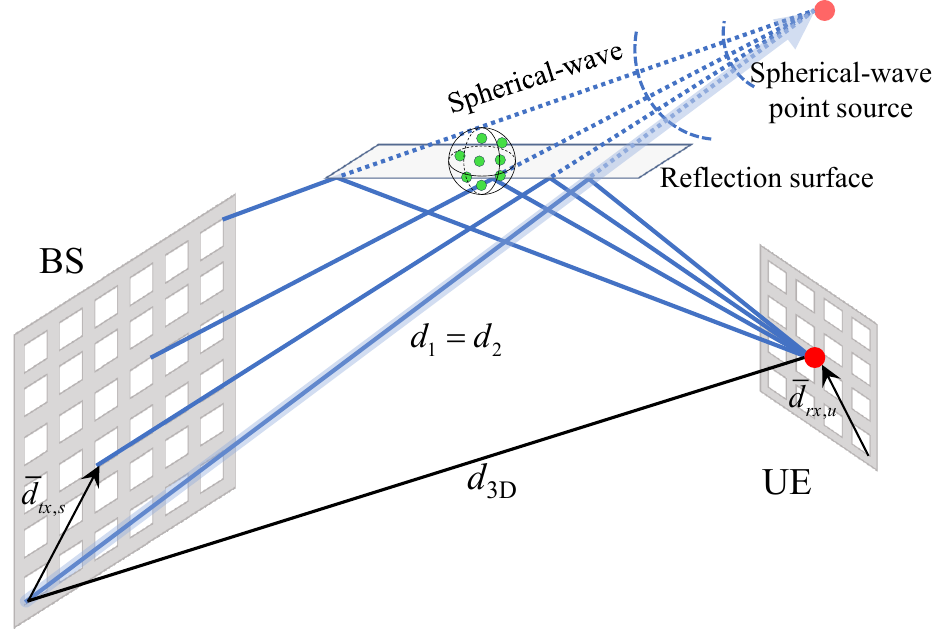} \label{fig_NF_model_b}}
\caption{Near-field propagation modeling: (a) Specular reflection and (b) Non-specular reflection.}
\label{fig_NF_model}
\end{figure*}

Following extensive discussions in 3GPP RAN1, a new near-field modeling methodology was proposed to bridge this gap. 
3GPP adopted a randomized near-field modeling approach that introduces the concept of a \emph{spherical-wave source distance} rather than a direct distance to a physical scatterer. This facilitates a precise description of wavefront curvature. Instead of generating absolute distances, the model defines a scaling factor: the ratio of the spherical-wave source distance to the total propagation path length. This normalization ensures consistent relative values across scenarios. As shown in Fig.~\ref{fig_NF_model}, clusters are classified into specular and non-specular types. For specular clusters, the spherical-wave source is the mirror image of the TX/RX relative to the reflection surface; thus, both BS and UE scaling factors are fixed to one. For non-specular clusters, the BS-side scaling factor is generated via a Beta distribution fitted to RT data:
\begin{equation}
 {d_{1,n,m}} = \left\{ {\begin{array}{*{20}{c}}
{{s_{{\rm{BS}},n}}\left( {{d_{{\rm{3D}}}} + {\tau _n} \cdot c + \Delta \tau  \cdot c} \right),}&{n = 1,2},\\
{{s_{{\rm{BS}},n}}\left( {{d_{{\rm{3D}}}} + {\tau _{n,i}} \cdot c + \Delta \tau  \cdot c} \right),}&{{\rm{others}}},
\end{array}} \right.
\label{equ_NF_d1}
\end{equation}
where $s_{\mathrm{BS},n}$ denotes the scaling factor for cluster $n$, and $\tau_{n,i}$ is the delay of the $i$-th sub-cluster. The scaling factor is defined as:
\begin{equation}
 {s_{{\rm{BS}},n}} = \left\{ {\begin{array}{*{20}{c}}
1&{n \le N_{\rm{spec}}},\\
{\rm{Beta}(\alpha,\beta)}&{n > N_{\rm{spec}}}.
\end{array}} \right.
\label{equ_s_BS}
\end{equation}

The UE-side scaling factor for non-specular clusters is the complement of the BS-side factor, ensuring the sum of scaled distances does not exceed the total path length. This mechanism achieves a statistically accurate and computationally efficient representation of near-field propagation, capturing non-linear phase evolution while maintaining compatibility with the 3GPP TR 38.901 framework.

\begin{figure}[h]
\centering
\subfloat[]
{\includegraphics[width=4.2cm]{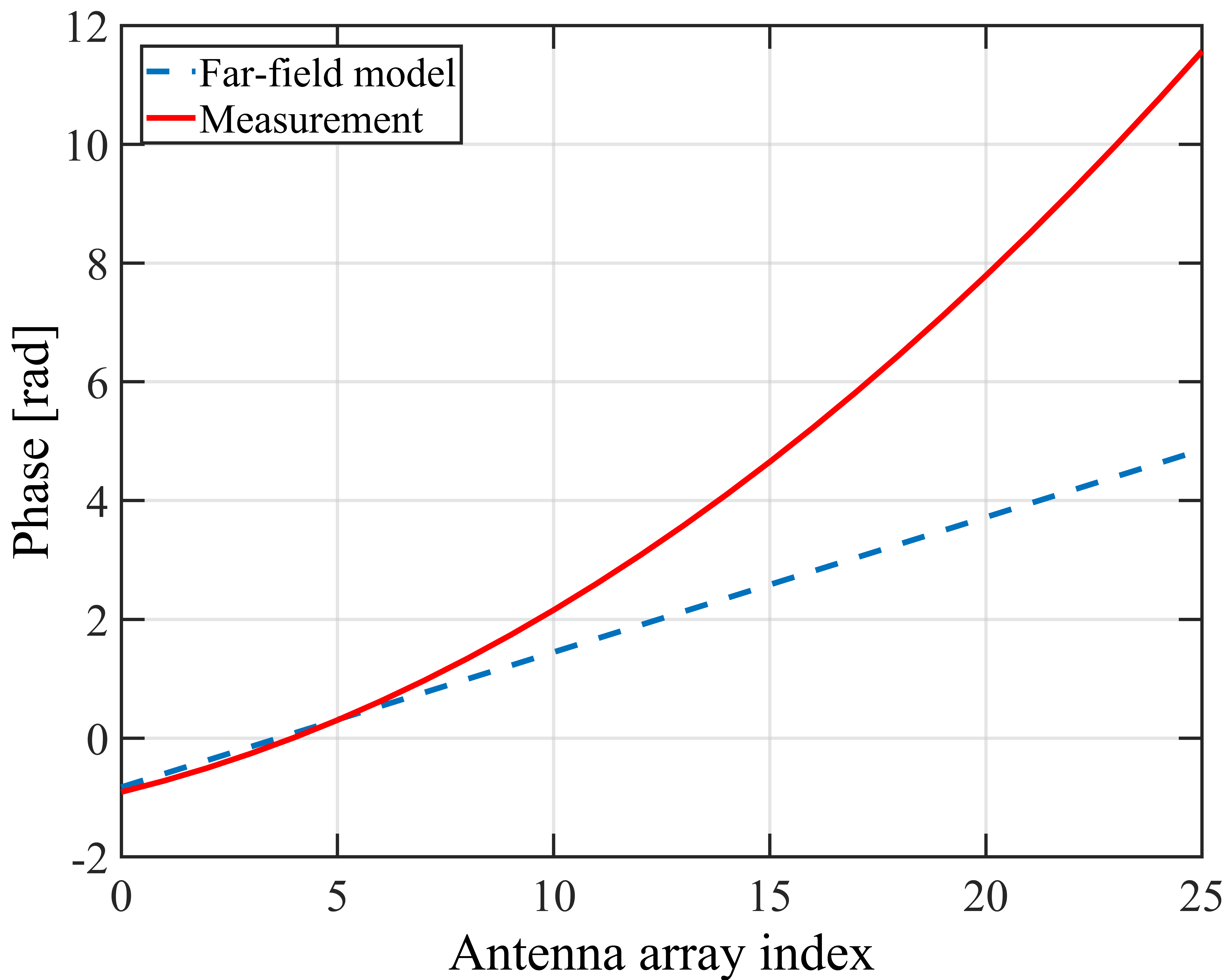}\label{fig_NF_phase}}
\hfill
\subfloat[]
{\includegraphics[width=4.2cm]{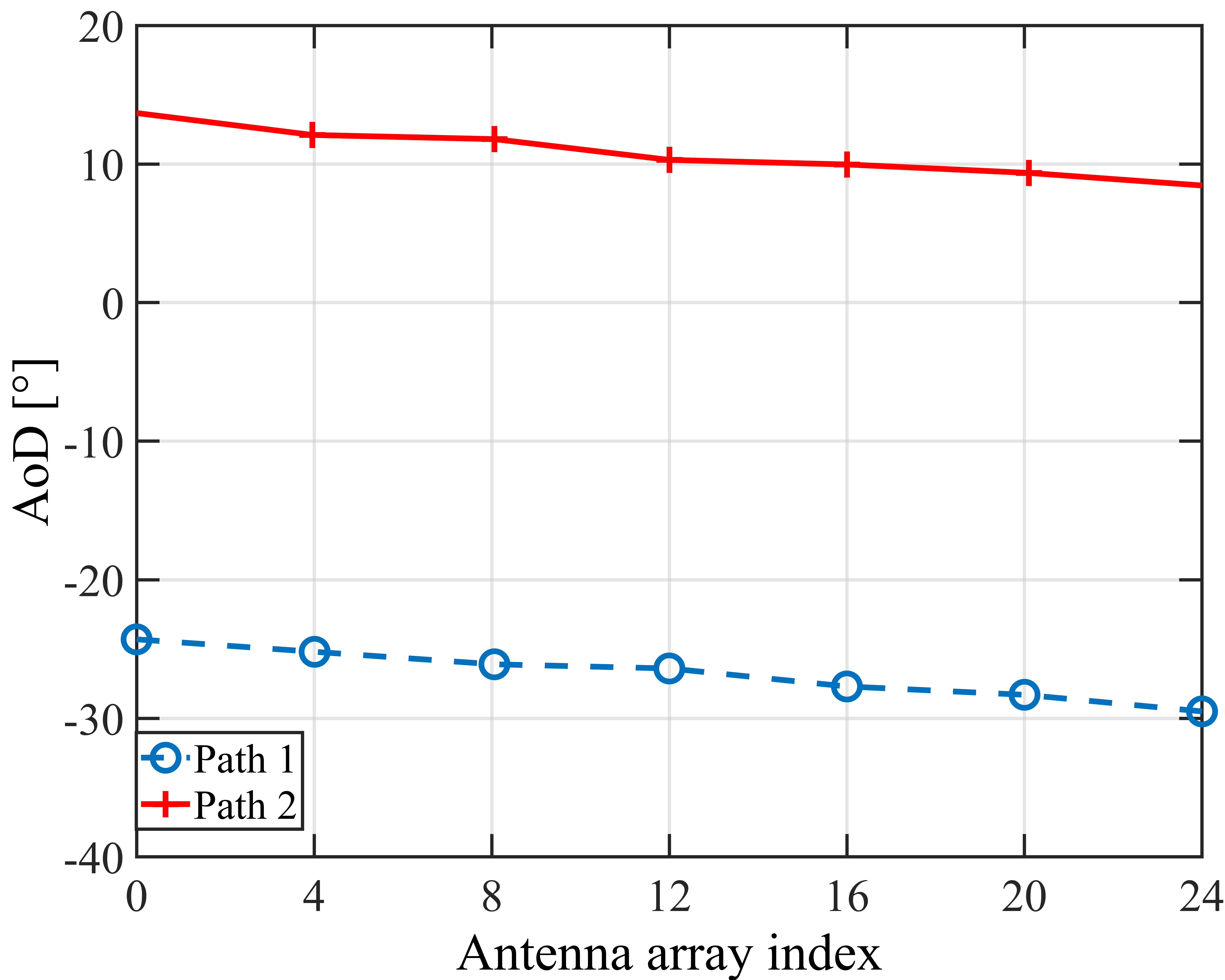} \label{fig_NF_angle}}
\caption{Evolution of element-wise (a) Phase and (b) AoD across the antenna array \cite{Miao_JSAC_NF}.}
\label{fig_NF_parameter}
\end{figure}

\paragraph{Antenna Element-Wise Channel Parameters}
In the adopted near-field modeling framework, the necessity of updating channel parameters at the element level depends on their sensitivity to the array aperture size. Modeling the BS-side array phase is designated as mandatory to faithfully capture wavefront curvature. In contrast, angular parameters are optional and may be included for scenarios requiring finer spatial resolution. For the UE side, given the typically smaller apertures, both phase and angle updates are optional. Other parameters—specifically amplitude, delay, Doppler, and polarization—generally do not require element-wise updates. The following subsections summarize the modeling principles for each category.

\textbf{Array Phase Parameters}: Accurate element-wise phase modeling is indispensable, as phase variation constitutes the most critical signature of near-field propagation. As illustrated in Fig.~\ref{fig_NF_phase}, measurements reveal a pronounced non-linear phase evolution across the array, a phenomenon that the conventional linear far-field model fails to capture \cite{Miao_JSAC_NF}. In the near-field model, the array phase is determined by spherical-wave propagation distance differences. For the direct path, the element-wise phase is derived from the exact geometric positions of the BS and UE elements:
\begin{equation}
 \exp\!\left(-j \frac{2\pi d_{\mathrm{3D}}}{\lambda_0}\right)
\exp\!\left(-j \frac{2\pi (\,|\vec r_{u,s}| - d_{\mathrm{3D}}\,)}{\lambda_0}\right).
\label{equ_NF_phase}
\end{equation}
For non-direct paths, the phase depends on the antenna element location relative to the spherical-wave source. For the BS side, the phase term is updated as:
\begin{equation}
\exp\!\left(
 j2\pi\,\frac{\bigl(d_{1,n,m} - d_{1,n,m}\, r_{\mathrm{tx},n,m} - d_{\mathrm{tx},s}\bigr)}{\lambda_0}
\right).
\label{equ_NF_angle_BS}
\end{equation}
For the UE side, the phase term is updated as:
\begin{equation}
 \exp\!\left(
 j2\pi\,\frac{\bigl(d_{2,n,m} - d_{2,n,m}\, r_{\mathrm{rx},n,m} - d_{\mathrm{rx},u}\bigr)}{\lambda_0}
\right).
\label{equ_NF_angle_UE}
\end{equation}

\textbf{Angular Parameters}: Modeling element-wise angular parameters is optional, as angular variation across the array is often limited. However, as evidenced in Fig.~\ref{fig_NF_angle}, noticeable inter-element variations in path angles do exist. Modeling these variations is beneficial for high-resolution use cases such as localization and sensing. For the direct path, the Angle of Arrival (AoA) and Angle of Departure (AoD) are determined directly by the element coordinates. For non-direct paths, they are derived geometrically based on the relationship between each element and the cluster’s spherical-wave source.

\begin{figure}[htbp]
	\centering
	\includegraphics[width=7.5cm]{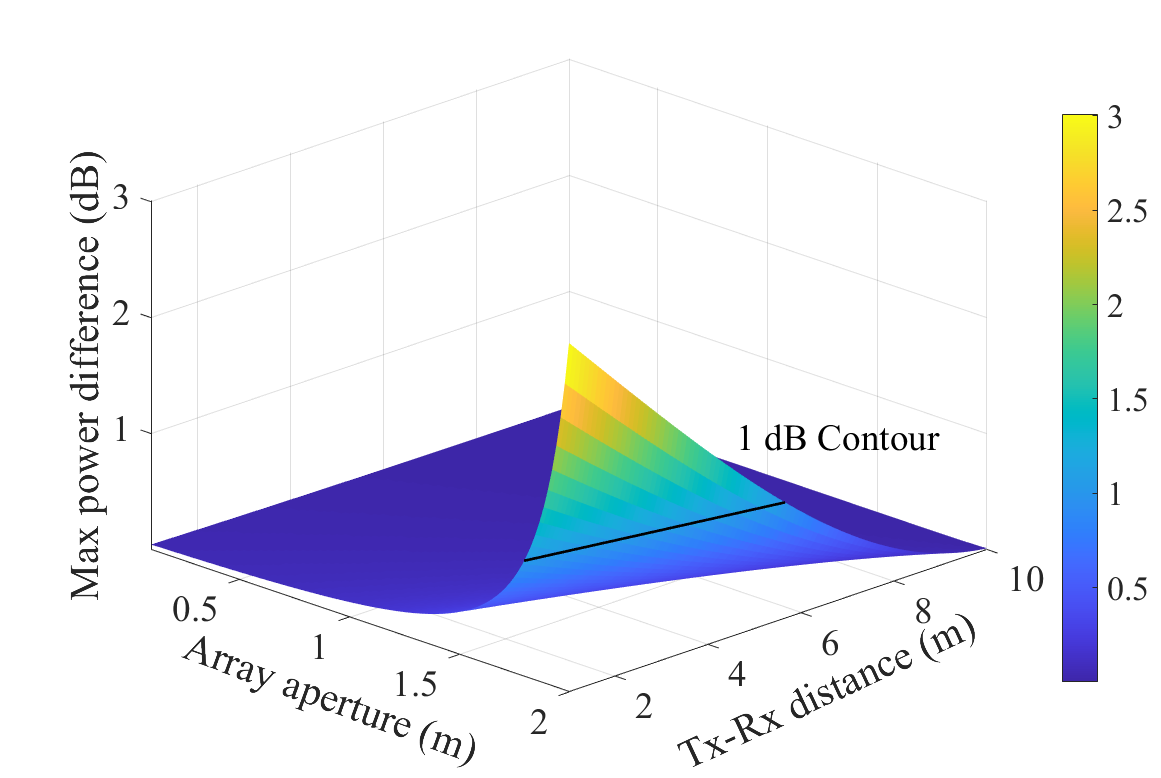}
	\caption{Maximum intra-array power difference versus array aperture and TX–RX distance, with the 1 dB contour indicating the boundary beyond which near-field amplitude variation becomes negligible.}
 \label{NF_amplitude}
\end{figure}

\textbf{Amplitude Parameters}: As depicted in Fig.~\ref{NF_amplitude}, significant power disparity among array elements (i.e., near-field gain variation) emerges only when the array aperture is comparable to the TX–RX distance. Theoretical studies define the ``uniform power distance"—where the minimum-to-maximum power ratio across the array is 0.9—as approximately 1.9 times the array aperture \cite{C2_UPD}. Since practical communication distances in cellular networks typically exceed this threshold, amplitude differences across the array are generally negligible and, therefore, do not require explicit element-wise modeling in the baseline.

\textbf{Delay Parameters}: Element-wise delay variations caused by propagation distance differences are typically negligible compared to the system sampling interval. For instance, considering a system operating at 15 GHz with a 400 MHz bandwidth, the maximum delay difference between adjacent elements is approximately 0.033 ns, which is far below the 1.25 ns sampling period. Consequently, explicit delay modeling is unnecessary unless for ultra-large arrays or high-precision sensing applications.

\textbf{Other Parameters}: Parameters such as Doppler shift, polarization matrix, and cross-polarization ratio (XPR) require no element-wise updates. Doppler variations across elements are negligible even in high-mobility scenarios. Similarly, polarization characteristics show no measurable element-wise variation. These parameters are therefore inherited directly from the far-field model, ensuring both consistency and computational simplicity.

\subsubsection{Spatial Non-Stationarity}
\begin{figure}[htbp]
	\centering
	\includegraphics[width=8cm]{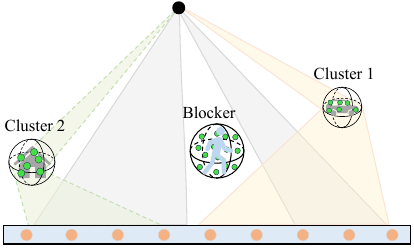}
	\caption{Illustration of SNS caused by partial blockage and incomplete scattering, where the LOS path and Cluster~1 are partially blocked and Cluster~2 interacts with a finite-size scatterer, making the clusters visible only to a subset of the array.}
 \label{fig_SNS_illustratiion}
\end{figure}

SNS is characterized by the spatial variation of received power across the antenna array, where different antenna elements observe varying energy contributions from the same cluster or ray. This phenomenon is modeled by applying element-wise power attenuation factors \cite{Meng2025EuCAPXL-MIMOSpatialNonStationary}. SNS can manifest at both the BS and UE sides. On the BS side, SNS typically arises from partial blockage and incomplete scattering, resulting in clusters being visible only to a subset of the array, as shown in Fig.~\ref{fig_SNS_illustratiion}. On the UE side, SNS is primarily induced by the handheld form factor, where the user's hand, head, and device body introduce element-dependent shadowing and attenuation. The following subsections detail the SNS modeling methodologies for the BS and UE sides, respectively.

\paragraph{Methodology for BS-Side SNS}


As summarized in Table~\ref{Tab_NF_SNS}, several representative approaches have been proposed to capture SNS effects. Ray-tracing-based models can explicitly resolve path visibility and interactions with the environment but are typically too complex for standardized system-level simulations \cite{Yuan_RT}. Fading channel models represent SNS by assigning non-identical fading statistics to different antenna elements \cite{SNS_fading}, which is mathematically tractable and suitable for link-level analysis, but lacks geometric interpretation and is incompatible with the 3GPP GBSM framework. VR models, such as COST 2100, associate each cluster with a spatial region along the array where it is observable 
\cite{C1_VR_COST,C1_VR_ljz}, offering intuitive physical interpretation and intra-cluster consistency; however, they require environment-specific parameterization, ignore inter-cluster dependencies, and introduce abrupt power discontinuities at VR boundaries. The birth–death process treats the appearance and disappearance of clusters as a stochastic evolution along the array \cite{C1_cluster_LR,C1_BD_CX}, enabling efficient spatial correlation modeling, but its fully random nature leads to poor continuity across antenna elements and makes the required birth–death probabilities difficult to determine. 




\begin{figure*}[h]
\centering
\subfloat[]
{\includegraphics[width=6cm]{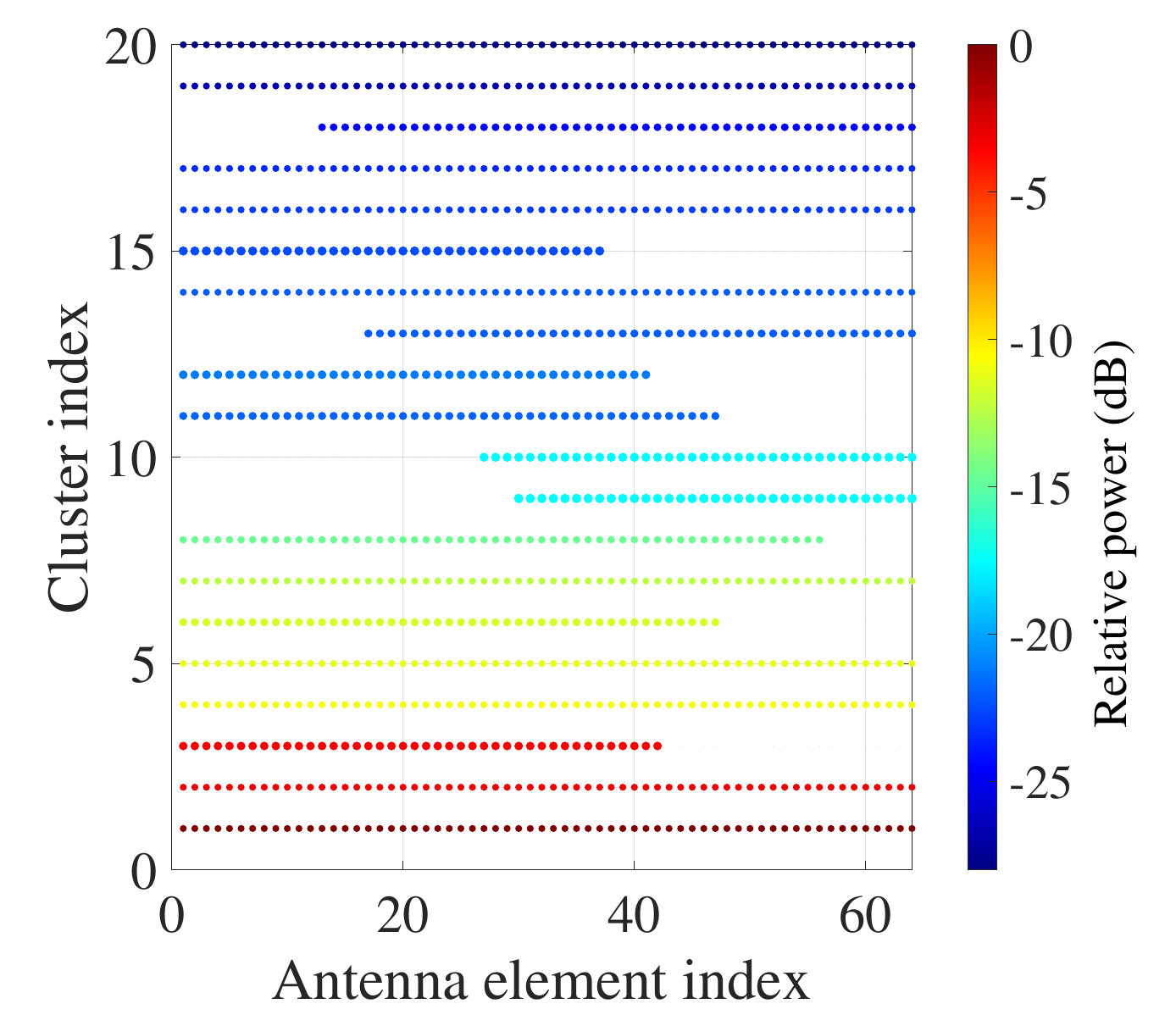}\label{fig_SNS_VR}}
\hfill
\subfloat[]
{\includegraphics[width=6cm]{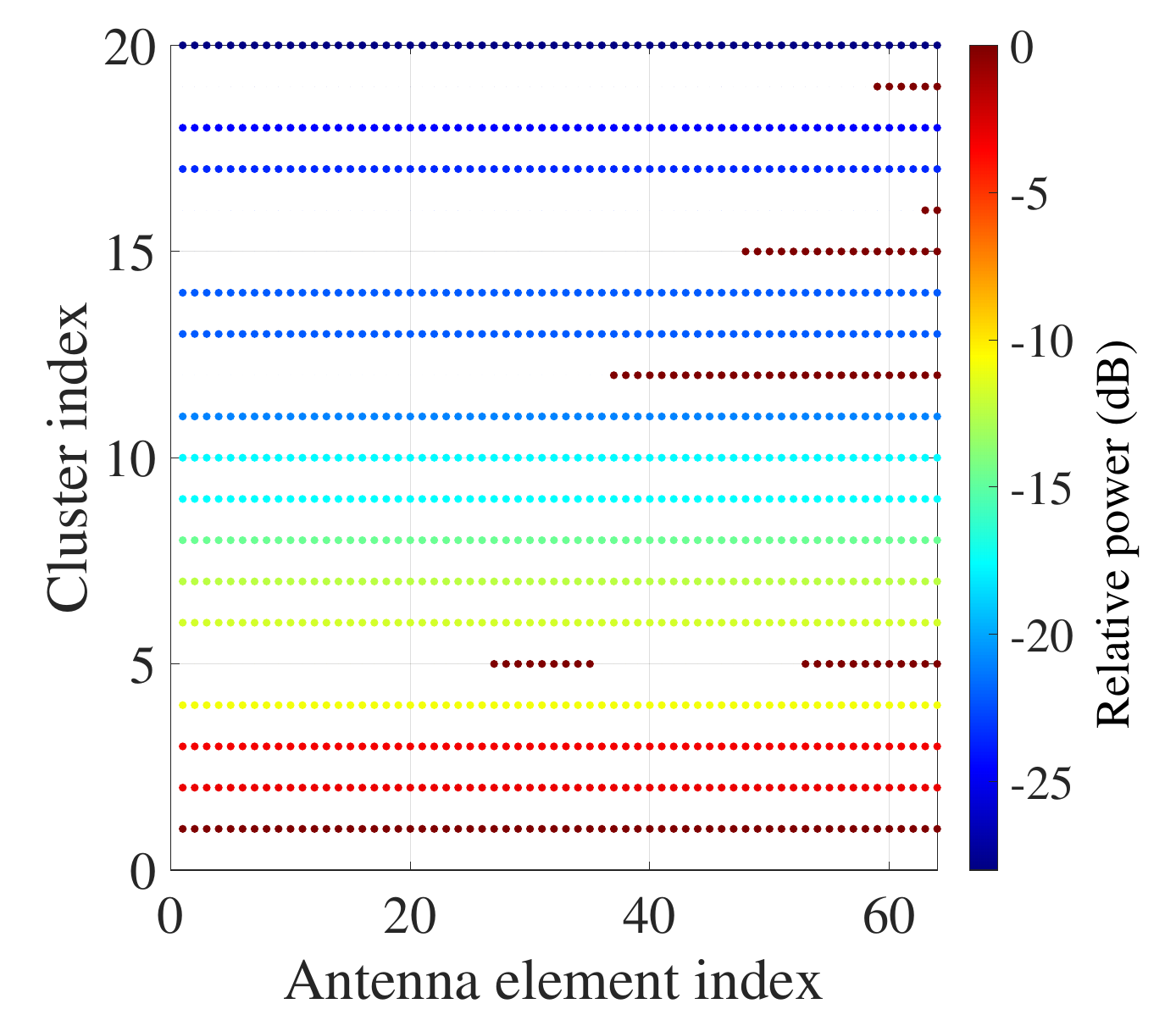}\label{fig_SNS_VP}}
\subfloat[]
{\includegraphics[width=6cm]{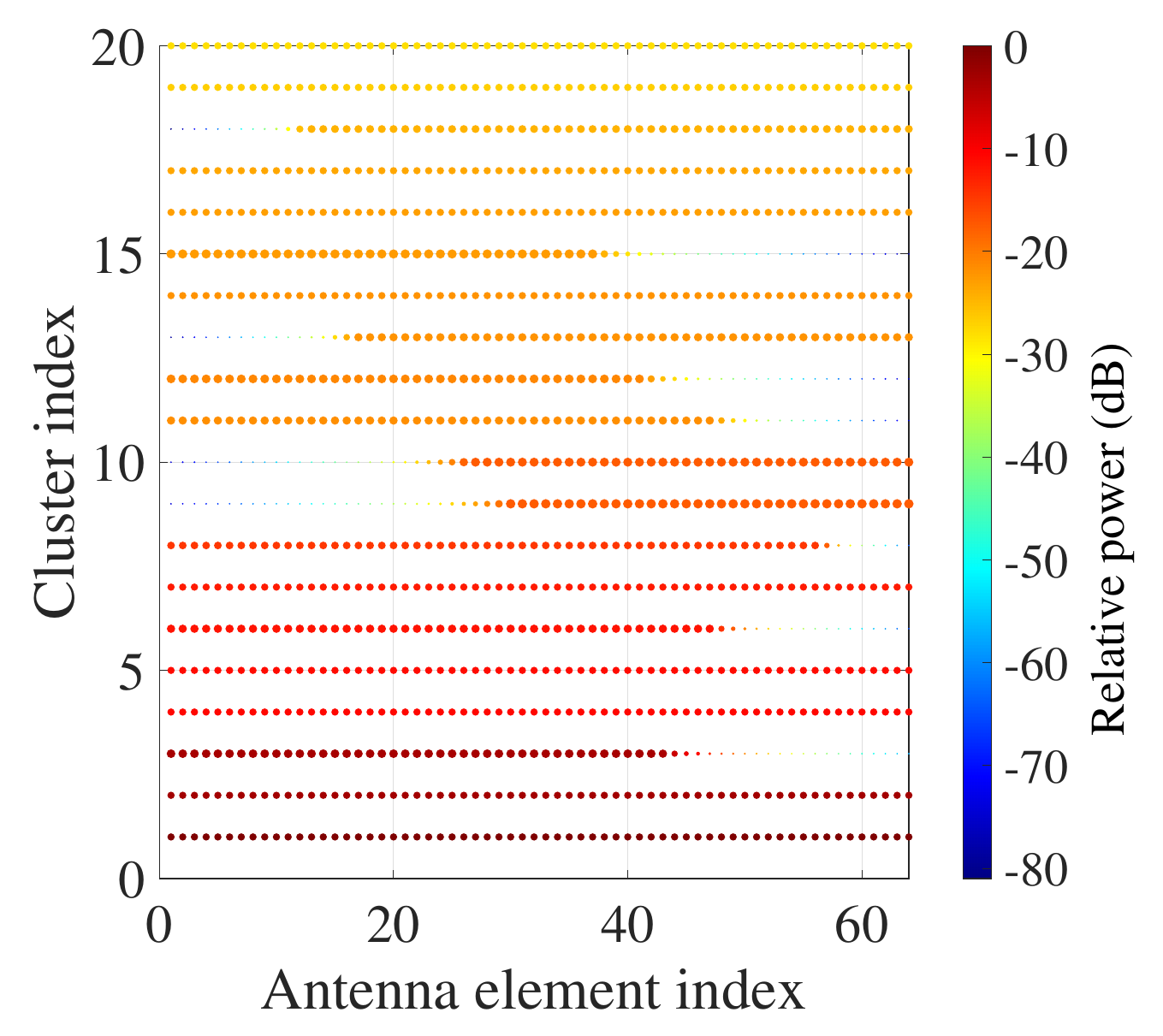}\label{fig_SNS_Proposed}}
\hfill
\caption{Evolution of cluster power across the antenna array for (a) the BS-VR model, (b) the birth–death process, and (c) the adopted stochastic model.}
\label{fig_SNS_Statistical}
\end{figure*}


\textbf{Adopted Stochastic-Based Model}: To overcome the aforementioned limitations, 3GPP adopted a hybrid stochastic framework that balances physical realism with computational efficiency \cite{3GPP_38901}. This method introduces an \emph{SNS probability} to quantify the likelihood of a cluster being stationary (visible to the entire array), effectively capturing the coexistence of stationary and non-stationary clusters.
First, for the \(n\)-th cluster, a visibility limitation is triggered if \(x_n < Pr_{sns}\), where \(x_n\) is a random variable. The SNS probability \(Pr_{sns}\) is generated per UE following a truncated normal distribution:
\begin{equation}
x_n \sim \mathrm{Unif}(0,1), \qquad
Pr_{sns} \sim \mathcal{N}(\mu, \sigma^{2}).
\label{equ_SNS_step1}
\end{equation}

Second, to ensure statistical consistency, a \textit{Visibility Probability} (VP), denoted as $V_n$, is defined for each cluster based on its power. The VP determines the fraction of the array over which the cluster is visible and is modeled as:
\begin{equation}
 {{V}}_n = A \cdot \exp \left(- \dfrac{ \max(P_n^{\textrm{dB}})-P_n^{\textrm{dB}}}{R}\right)+B+\xi,
 \label{equ_SNS_step2}
\end{equation}
where $A$ and $R$ control the exponential decay, $B$ sets the lower bound, and $\xi \sim \mathcal{N}(0, \sigma^2)$ introduces random variation. $P_n^{\textrm{dB}}$ is the cluster power in dB. Based on $V_n$, a rectangular VR of dimensions $a \times b$ is generated on the array plane. The width $a$ is drawn uniformly as $a \sim {\cal{U}}(V_n \cdot W,\, W)$, and the height is derived as $b = V_n \cdot H \cdot W / a$, where $H$ and $W$ are the array dimensions.

Finally, to overcome the unphysical power discontinuities observed in traditional VR and birth–death models, the adopted model applies a smooth power attenuation mechanism, yielding a continuous and physically consistent power evolution across the array, as illustrated in Fig.~\ref{fig_SNS_Statistical}. For stationary clusters, the attenuation is unity ($\alpha_{s,n} = 1$). For SNS clusters, the element-wise attenuation \(\alpha_{s,n}\) is computed as \cite{3GPP_huawei_wuhan}:
\begin{equation}
 {\alpha _{s,n}} = \left\{ {\begin{array}{*{20}{c}}
{1,}&(x_s, y_s) \in \mathcal{R} ,\\
{{\exp{\left( { - C \cdot \dfrac{{ {\tilde d_{s,n}}}}{{{D_n}}}} \right)}},}&(x_s, y_s) \notin \mathcal{R},\\
\end{array}} \right.
\label{equ_SNS_step3}
\end{equation}
where $\mathcal{R}$ denotes the rectangular VR region. $\tilde d_{s,n}$ represents the Euclidean distance from the antenna element $(x_s, y_s)$ to the nearest boundary of $\mathcal{R}$, and $D_n$ is the diagonal length of the VR. The roll-off factor $C$ controls the sharpness of the transition zone. This framework successfully models the ``soft'' visibility boundaries observed in reality.

\textbf{Adopted Physical Blocker-Based Model}: Recognizing partial blockage as a primary physical driver of SNS, 3GPP also incorporates a deterministic blocker-based model \cite{3GPP_38901}. Multiple blockers with configurable geometry are deployed between the BS and UE. The attenuation is computed via a simplified knife-edge diffraction model:
\begin{equation}
\begin{aligned}
 L_{{\rm{dB}},u,s,1}/&L_{{\rm{dB}},s,n,m} = \\
 &- 20{\log _{10}}\left( {1 - \left( {{F_{h1}} + {F_{h2}}} \right)\left( {{F_{w1}} + {F_{w2}}} \right)} \right),
\end{aligned}
\label{equ_SNS_Blockage}
\end{equation}
where \(F\) terms represent the Fresnel diffraction parameters at the four edges of the blocker. Typically, the model simplifies to consider only the dominant edge, assuming infinite projection for others. This approach offers high physical fidelity for scenarios dominated by explicit blockage objects.

\paragraph{Methodology for UE-Side SNS}

During 3GPP standardization, several strategies were evaluated for modeling UE-side SNS, primarily to capture the effects of the user's hand and head \cite{Nokia_sns}. The candidates ranged from high-fidelity geometric models (introducing explicit small-scale blockers around the UE) to stochastic angular blocking models. While geometric approaches offer precision, they incur high computational complexity and require detailed modeling of dynamic blocker orientation. Ultimately, a fixed attenuation value model was adopted due to its balance of simplicity and statistical accuracy \cite{3GPP_38901}. This measurement-driven approach classifies UE usage into three typical scenarios: (i) one-hand grip, (ii) dual-hand grip, and (iii) one-hand with head proximity. Based on the selected scenario and operating frequency (categorized into ranges: $<$1 GHz, 1-8.4 GHz, and 14.5-15.5 GHz), a pre-defined mask of attenuation values is applied to the antenna elements. For example, in the 1–8.4 GHz band, the element-wise attenuation can be as low as only a few decibels—down to 0.6 dB for certain antenna locations—while exceeding 10 dB for others, reflecting the highly spatially selective nature of self-blockage effects across the array. This implicitly captures the complex shadowing effects of the human body without requiring explicit geometric modeling of the user.

\section{Simulation Implementation of the Basic Model}

\begin{figure*}[htbp]
	\centering
	\includegraphics[width=6.4 in]{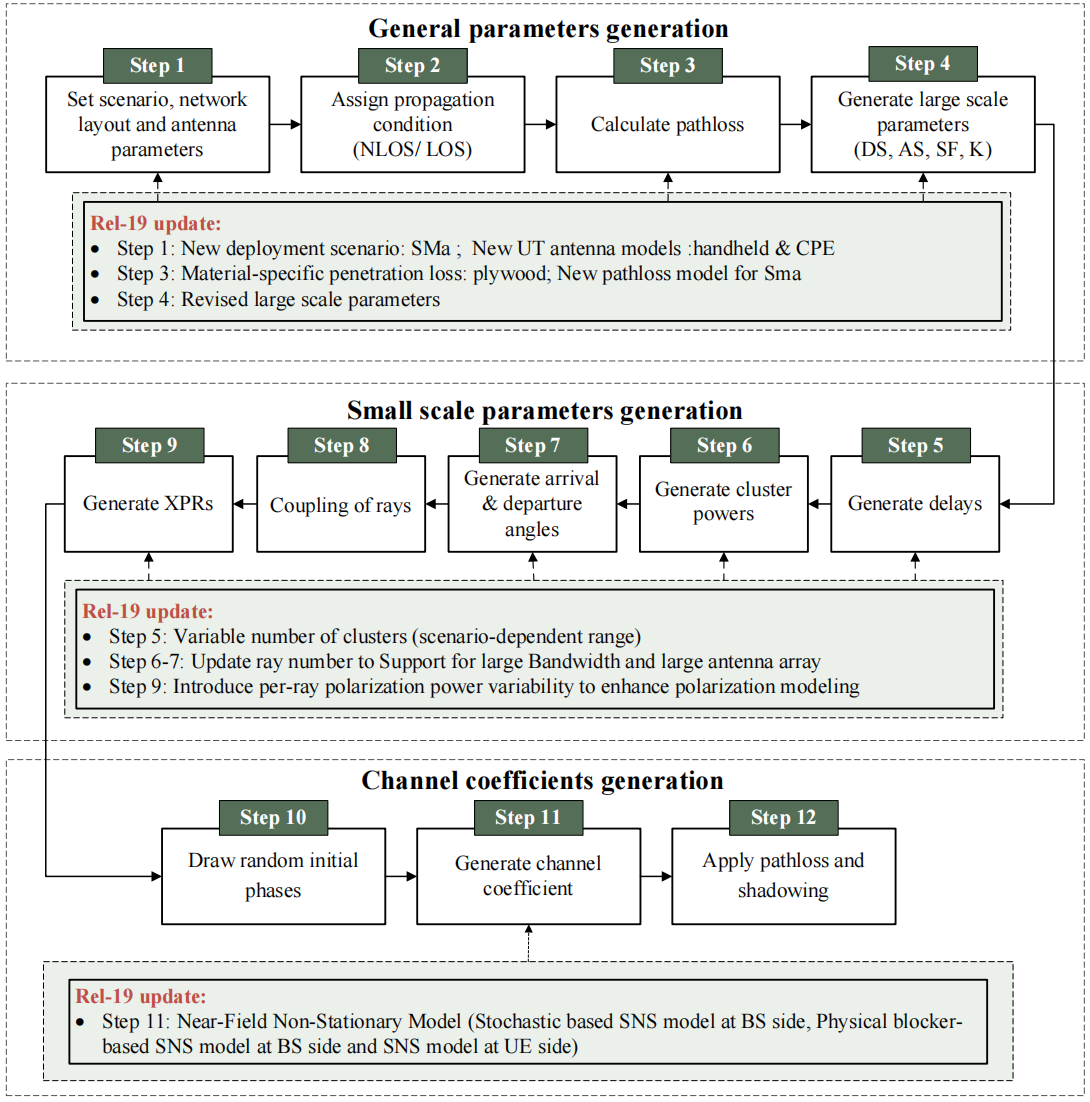}
	\caption{Algorithmic procedure for the 3GPP Rel-19 channel model generation.}\label{flowchart}
\end{figure*}

The 3GPP TR 38.901 specification \cite{38901} delineates a rigorous stochastic framework for channel generation. Based on this standard, we outline the simulation procedure, which is architecturally structured into 12 sequential steps grouped into three operational stages: the definition of general channel parameters, the generation of Small-Scale Parameters (SSPs), and the synthesis of channel coefficients. The comprehensive workflow is depicted in Fig. \ref{flowchart}, where standard procedures are denoted in black and Rel-19 specific updates are highlighted in red.

\subsection{General Parameters Generation}
The generation of general parameters constitutes the foundational phase of the simulation, establishing the macroscopic network topology and defining the large-scale fading environment. This stage encompasses the definition of scenario types, network layout geometries, antenna configurations, and propagation conditions. These parameters serve as the inputs for calculating path loss and generating spatially correlated LSPs, thereby determining the baseline signal-to-interference-plus-noise ratio (SINR) geometry for system-level evaluations.

\subsubsection{\textbf{Step 1}} Set scenarios, network layout, and antenna parameters. 

\textbf{Set scenarios:} Channel characteristics are intrinsically tied to the propagation environment. For instance, indoor environments typically exhibit richer multipath scattering due to enclosed structures compared to open outdoor settings. The 3GPP framework defines a suite of standardized test environments to represent diverse geographic and usage conditions:
\begin{itemize}
    \item Legacy scenarios: The model includes UMa for large-scale city coverage, UMi for dense street canyons, RMa for high-speed open terrain, and indoor scenarios like InH and Indoor Factory (InF).
    \item Rel-19 new scenario (SMa): To address the specific coverage challenges in the 7–24 GHz band, Rel-19 introduces the SMa scenario. This environment is characterized by continuous low-rise residential buildings (typically 1–2 floors) and prominent vegetation, which induces significant clutter loss. Unlike UMa where street canyons guide propagation, SMa represents a coverage-limited regime dominated by foliage attenuation and diffraction over rooftops.
\end{itemize}

It should be noted that the 3GPP channel model primarily comprises these standardized scenarios (InH, InF, UMa, UMi, RMa, and SMa), as defined in 3GPP TR 38.901. The model adopts a unified GBSM framework, while distinct parameter sets are provided for different frequency ranges to capture the variations in propagation characteristics over wide bandwidths. Table \ref{scenario} summarizes the mapping between the usage scenarios, test environments, and corresponding 3GPP channel models.
\begin{table*}[htbp]
    \centering
    \caption{The mapping among the usage scenarios, test environments, and channel models.}
    \label{scenario}
    \scriptsize
    \begin{threeparttable}
    \begin{tabular}{|p{2.2cm}|p{1.5cm}|p{1.2cm}|p{2.4cm}|p{1.2cm}|p{1.5cm}|p{3.2cm}|}
        \hline
        \textbf{Test environments} & Urban Micro-street canyon & Urban Macro & Indoor office & Rural Macro & Suburban Macro & Indoor Factory \\ \hline
        \textbf{Channel models} & UMi-street canyon & UMa & \begin{tabular}[c]{@{}l@{}}Indoor-office open office \\ Indoor-office mixed office\end{tabular} & RMa & SMa & \begin{tabular}[c]{@{}l@{}}InF-SL (sparse clutter, low BS) \\ InF-DL (dense clutter, low BS) \\ InF-SH (sparse clutter, high BS) \\ InF-DH (dense clutter, high BS) \\ InF-HH (high Tx, high Rx)\end{tabular} \\ \hline
    \end{tabular}
    \end{threeparttable}
\end{table*}

Each test environment is characterized by specific parameters, such as the ISD, BS antenna height, and UE speed. These parameters must be accurately configured in simulations to reflect realistic deployment conditions. Table \ref{scenario_parameter} summarizes these configuration parameters for five representative test environments. Among them, the SMa and RMa scenarios feature the largest ISD and BS antenna heights, reaching up to 1800 m and 35 m, respectively. In contrast, the Indoor Office (InH) environment utilizes a BS antenna height of only 3 m, typically representing ceiling-mounted deployments. 

\begin{figure}[htbp]
	\centering
	\includegraphics[width=3.1in]{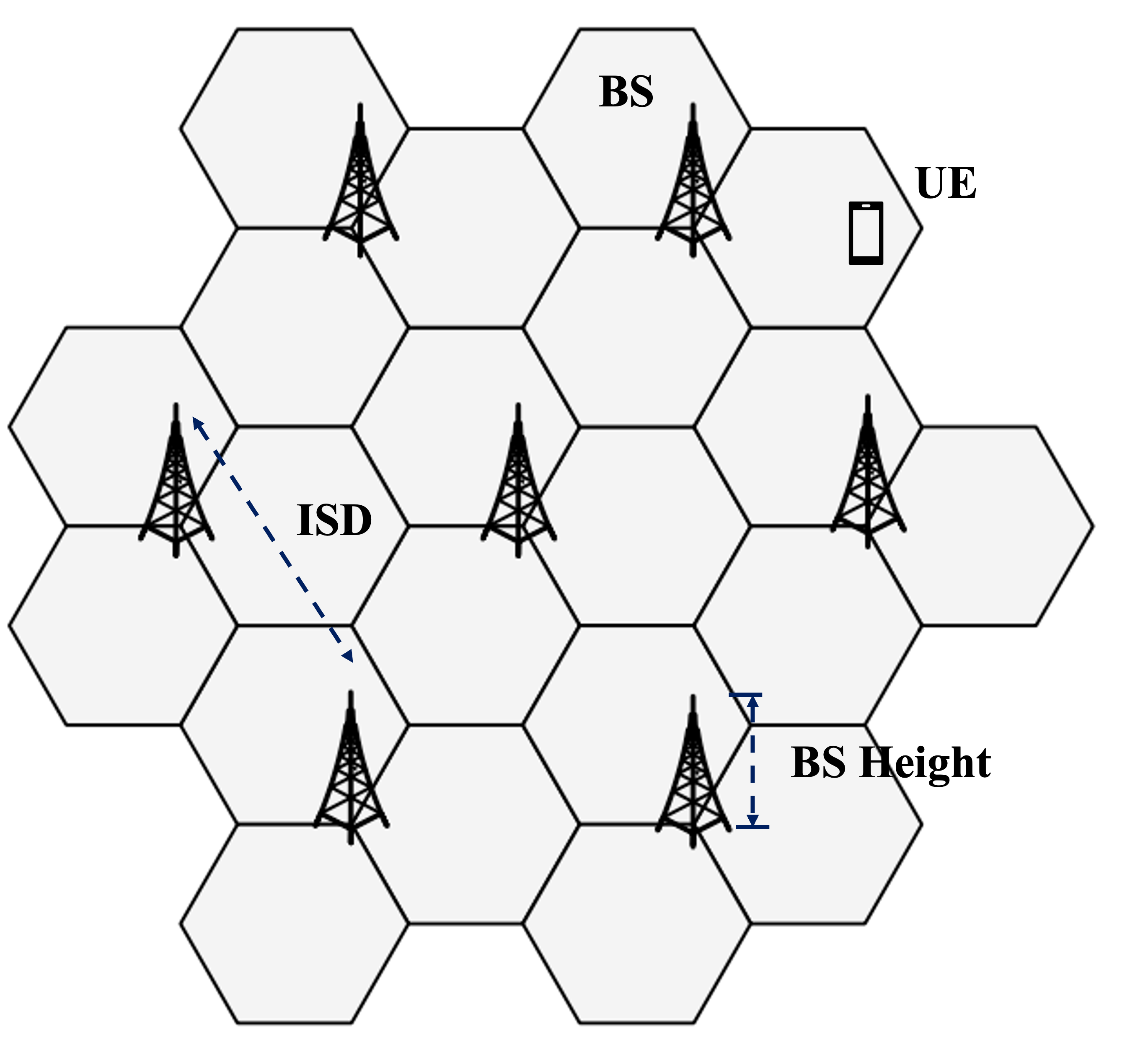}
	\caption{The network layout of outdoor test environments.}\label{indoor_layout}
\end{figure}

\begin{table*}[htbp]
    \centering
    \caption{Key configuration parameters for the five test environments.}
    \label{scenario_parameter}
    \scriptsize
    \begin{tabular}{|c|c|c|c|c|c|}
        \hline
        \textbf{Test Environments} & \textbf{UMi-street canyon} & \textbf{UMa} & \textbf{InH} & \textbf{RMa} & \textbf{SMa} \\ \hline
        BS antenna height (m) & 10 & 25 & 3 (ceiling) & 35 & 35 \\ \hline
        \begin{tabular}[c]{@{}l@{}}Min. distance\\BS to UE (m)\end{tabular} & 10 & 35 & 0 & 35 & 35 \\ \hline
        ISD (m) & 200--500 & 200--500 & -- & -- & 1200--1800 \\ \hline
        UE speed (km/h) & 3 & 3 & 3 & -- & \begin{tabular}[c]{@{}l@{}}Indoor: 3\\Outdoor (car): 30\end{tabular} \\ \hline
        UE antenna height (m) & \begin{tabular}[c]{@{}l@{}}Same as 3D-UMi\\in TR36.873\end{tabular} & \begin{tabular}[c]{@{}l@{}}Same as 3D-UMa\\in TR36.873\end{tabular} & 1 & 1.5 & \begin{tabular}[c]{@{}l@{}}outdoor: 1.5\\residential: 1.5/4.5\\commercial: 1.5/4.5/7.5\\/10.5/13.5\end{tabular} \\ \hline
        UE distribution & Uniform & Uniform & Uniform & Uniform & \begin{tabular}[c]{@{}l@{}}horizontal: Uniform\\vertical: Uniform across\\all floors\end{tabular} \\ \hline
        Indoor UE ratio & 80\% & 80\% & -- & \begin{tabular}[c]{@{}l@{}}50\% indoor,\\50\% in car\end{tabular} & 80\% \\ \hline
    \end{tabular}
\end{table*}
\textbf{Network layout:} The network topology is established to define the geometric relationship between Base Stations (BSs) and User Terminals (UEs). For indoor test environments, such as the InH scenario, the layout is modeled as a rectangular area representing a typical floor plan of an office or shopping mall. The area dimensions are 120 m $\times$ 50 m, with 12 BSs deployed at a height of 3 m and spaced approximately 20 m apart. UEs are uniformly distributed across the entire area. While internal walls are not explicitly modeled as physical obstacles, the propagation condition for each UE is determined by a stochastic Line-of-Sight (LOS) probability model, as specified in the 3GPP standard. For outdoor scenarios, as shown in Fig.~\ref{indoor_layout}, BSs are arranged in a regular hexagonal grid \cite{Chheda1999} employing a wrap-around technique to simulate an infinite network and eliminate boundary effects. Crucially, the configuration of the ISD for SMa differs from legacy models; it employs a geometrically derived ISD of 1299 m (corresponding to a 750 m cell radius) and a BS height of 35 m, reflecting the need for wider coverage in clutter-dominated suburban environments.



\begin{figure}[htbp]
	\centering
	\includegraphics[width=3.1in]{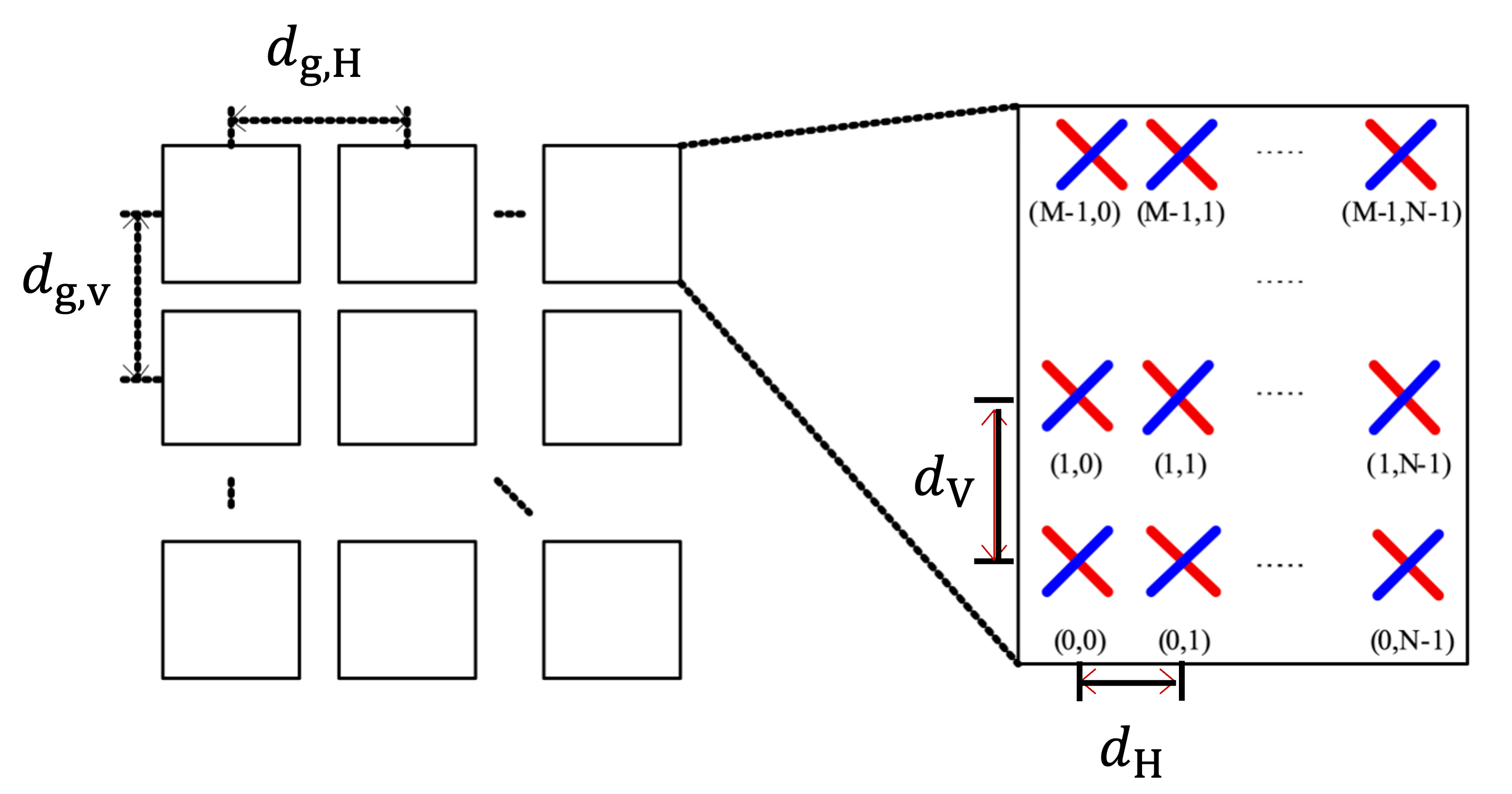}
	\caption{The structure of antenna arrays at the BS \cite{M2412}.}\label{antenna_structure}
\end{figure}

\textbf{Antenna configuration:} Simulation accuracy relies heavily on the precise modeling of antenna arrays. At the BS side, arrays are modeled as Uniform Planar Arrays (UPAs). As illustrated in Fig. \ref{antenna_structure}, the configuration typically comprises $M \times N$ antenna elements organized into $M_{g} \times N_{g}$ antenna panels. The arrangement of antenna elements is highly dependent on the polarization scheme \cite{Balanis2016}. In most practical 5G deployments, dual-polarized antennas with $\pm 45^{\circ}$ slant polarization are utilized. To minimize mutual coupling and avoid the formation of grating lobes, the spacing between adjacent antenna elements, denoted as $d_{H}$ and $d_{V}$, is typically set to $\lambda/2$ \cite{Balanis2016}. Accordingly, the horizontal and vertical distances between adjacent antenna panels, $d_{g,H}$ and $d_{g,V}$, are configured as $M\lambda/2$ and $N\lambda/2$, respectively.
  
For each antenna element, the expression of the horizontal radiation power pattern is specified as 
\begin{equation}
A_{dB}''(\theta''=90^{\circ},\phi'')=-\textrm{min}\{12(\frac{\phi''}{\phi_{3dB}})^2, A_{max}\},
\end{equation} 
where $-180^{\circ}\leq \phi \leq 180^{\circ}$, $\phi_{3dB}$ is the horizontal 3 dB beamwidth, $A_{max}$ is the maximum side lobe level attenuation. The expression of antenna element vertical radiation power pattern can be written as 
\begin{equation}
A_{dB}''(\theta'',\phi''=0^{\circ})=-\textrm{min}\{12(\frac{\theta''-90^{\circ}}{\theta_{3dB}})^2, SLA_{V}\},
\end{equation} 
where $0^{\circ}\leq \theta \leq 180^{\circ}$, $\theta_{3dB}$ is the vertical 3 dB beamwidth, and $\theta_{tilt}$ is the tilt angle in the vertical domain. The combined antenna element power pattern is 
\begin{equation}\label{antenna_pattern}
\begin{aligned}
A''_{dB}(\theta'',\phi'') 
&= -\min\Bigl\{ 
- \bigl(
A''_{dB}(\theta'',\phi''=0^{\circ}) \\
&\quad + A''_{dB}(\theta''=90^{\circ},\phi'')
\bigr),
A_{\max}
\Bigr\}.
\end{aligned}
\end{equation}
The relationship between the radiation field pattern and power pattern can be expressed as
\begin{equation}
A''(\theta'',\phi'')=|F''_{\theta''}(\theta'',\phi'')|^2+|F''_{\phi''}(\theta'',\phi'')|^2.
\end{equation} 

At the UE side, Rel-19 introduces a fundamental shift from the legacy point-source assumption to refined form-factor models. To accurately capture user interaction effects at higher frequencies, handheld and CPE devices are now modeled with specific candidate antenna locations (e.g., distributed along the edges or corners of a chassis) and specific polarization orientations. This geometric refinement is essential for evaluating the impact of body blockage and spatial non-stationarity in the 7–24 GHz band.

\subsubsection{\textbf{Step 2}} Assign Propagation condition.

\begin{figure}[htbp]
	\centering
	\includegraphics[width=3.1in]{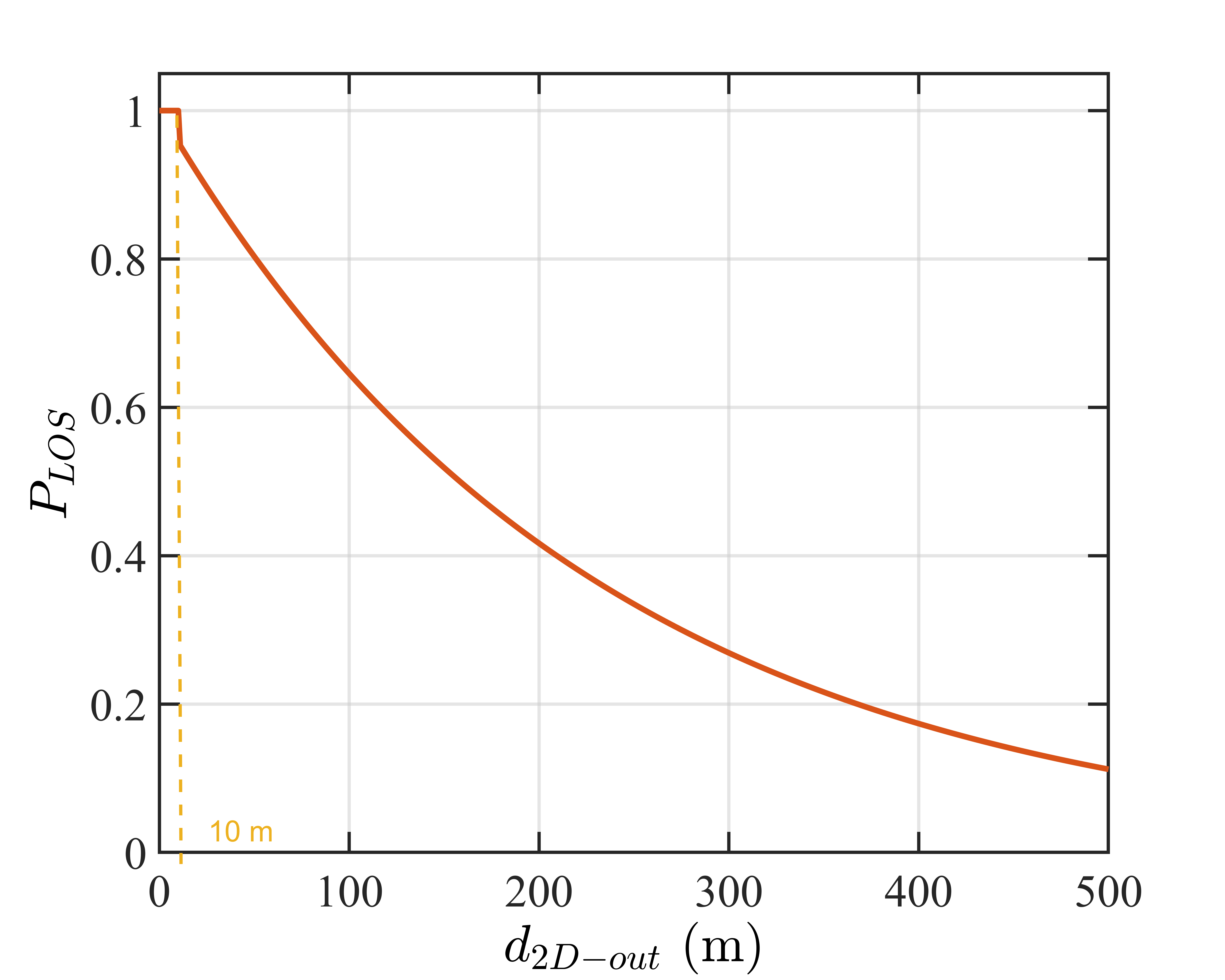}
	\caption{The LOS probability for SMa test environments.}\label{LOS_probability}
\end{figure}

\textbf{LOS/NLOS assignment:} The propagation condition for each link is stochastically classified as either Line-of-Sight (LOS), implying the presence of a dominant direct path, or Non-Line-of-Sight (NLOS), indicating obstruction by environmental clutter. In simulations, this state is determined by comparing a uniformly distributed random variable $(\xi \sim U[0,1])$ against a distance-dependent LOS probability function, $Pr_{LOS}(d_{2D})$. If $\xi <Pr_{LOS}(d_{2D})$, the link is assigned as LOS; otherwise, it is NLOS. Fig. \ref{LOS_probability} describes the LOS probability for SMa test environments. As observed, the probability is unity for short distances ($d_{2D}\leq 10$ m), reflecting a high likelihood of visual clearance. As the separation distance increases, $Pr_{LOS}$ exhibits a negative exponential decay.

\textbf{Indoor/outdoor state:} For outdoor deployment scenarios, an additional binary state—Indoor or Outdoor—must be assigned to each UE to determine whether building penetration loss is applicable \cite{Zheng2016,Yu2016,Kristem2017}. Unlike the distance-dependent LOS state, the indoor/outdoor status is typically determined by a scenario-specific constant probability. As specified in Table \ref{scenario_parameter}, the SMa scenario assumes a high indoor user ratio of 80\%, similar to UMa and UMi, reflecting a residential usage pattern. In contrast, the RMa scenario assumes only 50\% indoor users, with the remaining outdoor users potentially located inside vehicles. This distinction is critical for SMa simulations in the 7–24 GHz band, as the ``Indoor" state triggers the calculation of severe Outdoor-to-Indoor (O2I) penetration losses in the subsequent step.

\subsubsection{\textbf{Step 3}} Calculate Path loss.

\begin{figure}[htbp]
	\centering
	\includegraphics[width=3.1in]{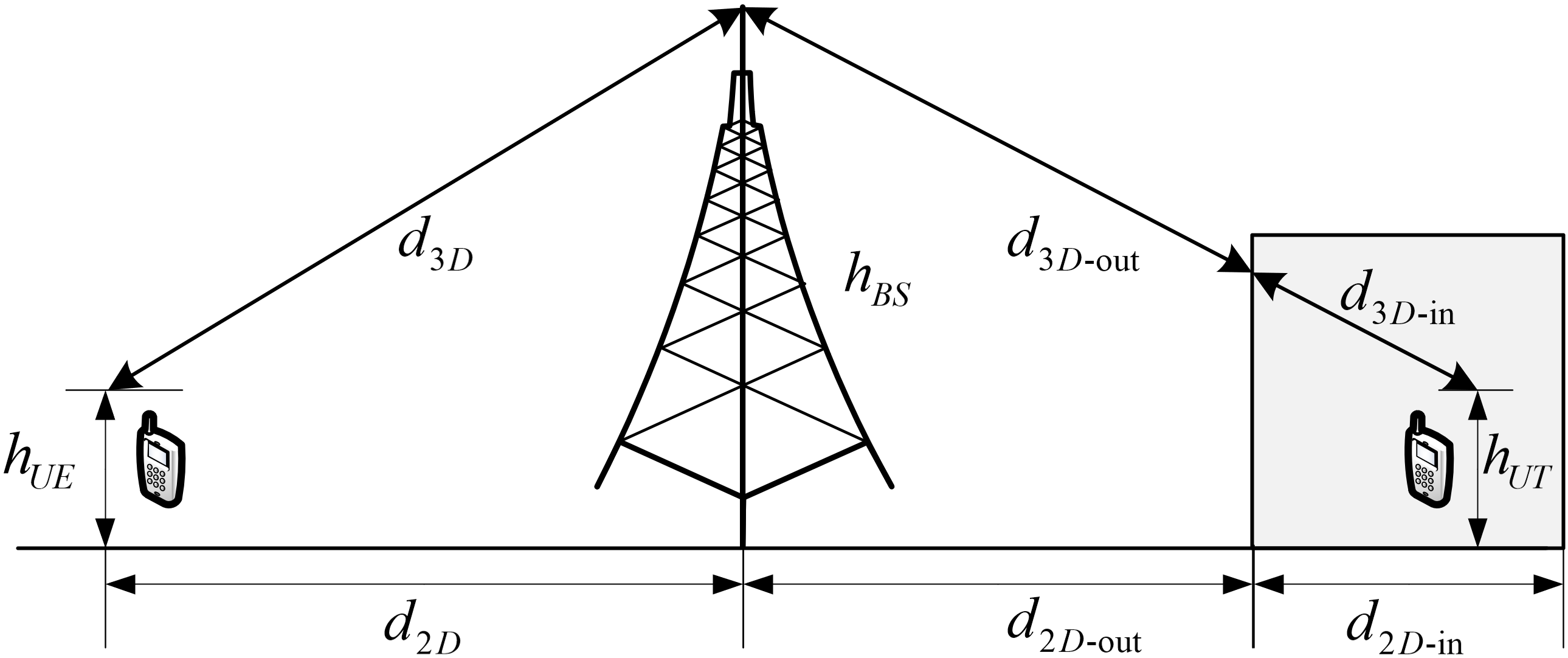}
	\caption{A sketch of one indoor UE and one outdoor UE \cite{M2412}.}\label{indoor_outdoor}
\end{figure}

This step is to calculate the path loss for each link. As discussed in Step 2, UEs may be outdoor or indoor. An example of that is shown in Fig. \ref{indoor_outdoor}. $h_{BS}$ is the height of the BS. $h_{UE}$ is the height of the antennas of the UE. $d_{3D}$ is the 3D distance while $d_{2D}$ is the 2D distance. It should be noted that if UEs are indoor and BSs are outdoor, both indoor and outdoor propagation distances ($d_{3D-out}$ and $d_{3D-in}$, respectively) should be considered. This indicates that in this case, the path loss should include outdoor loss, penetration loss, and inside loss as described in Equ. (\ref{PL_INDOOR_OUEDOOR}).

\textbf{Outdoor loss:} The outdoor path loss is modeled as a function of the 3D distance and carrier frequency. Tables \ref{outdoor_loss_model} show the models of path loss and shadow fading for SMa, respectively. $d_{BP}$ is the breaking point (BP). It indicates the distance at which the propagation exponent is changed. In indoor channel models, the 3GPP 5G channel model assumes that there is no BP. But in outdoor channel models, like UMi, UMa, RMa, and SMa, the path loss is modeled as a piecewise function of $d_{BP}$. The definition of $d_{BP}$ differs along with different channel models. For example, in RMa and SMa channel models, $d_{BP}$ is equal to $2\pi h_{BS}h_{UE}f_c/c$. In UMi and UMa channel models, $d_{BP}$ is equal to $4h'_{BS}h'_{UE}f_c/c$, where $h'_{BS}$ and $h'_{UE}$ are the effective antenna heights at the BS and the UE \footnote{For UMi and UMa, $h'_{BS}=h_{BS}-h_E$, $h'_{UE}=h_{UE}-h_E$, where $h_E$ is the effective environment height.}. 

\begin{table*}[htbp]
\centering
\caption{Outdoor path loss model for SMa street canyon scenario.}
\label{outdoor_loss_model}
\renewcommand{\arraystretch}{1.4}
\setlength{\tabcolsep}{10pt}

\begin{tabular}{|c|c|p{7.2cm}|c|p{3.2cm}|}
\hline
\multirow{20}{*}{SMa}
& \multirow{13}{*}{LOS}
& 
\begin{equation*}
PL_{\text{SMa-LOS}}=
\begin{cases}
PL_1, & 10\,\mathrm{m} \le d_{2D} \le d_{BP} \\
PL_2, & d_{BP} \le d_{2D} \le 5\,\mathrm{km}
\end{cases}
\end{equation*}
&
\multirow{13}{*}{%
\makecell[c]{%
$d_{2D}\!\in\![10\,\mathrm{m},\,d_{BP}]$: $\sigma_{SF}=4$\\[2pt]
$d_{2D}\!\in\![d_{BP},\,5000\,\mathrm{m}]$: $\sigma_{SF}=6$
}}
&
\multirow{13}{*}{%
\makecell[l]{%
$10\,\mathrm{m} \le d_{2D} \le 5000\,\mathrm{m}$\\
$25\,\mathrm{m} \le h_{BS} \le 35\,\mathrm{m}$\\
$h = 10\,\mathrm{m}$
}}
\\
& & 
\begin{equation*}
\begin{aligned}
PL_1 =\;&20\log_{10}\!\left(\frac{40\pi d_{3D} f_c}{3}\right)\\
&+ \min(0.03h^{1.72},10)\log_{10}(d_{3D})\\
&- \min(0.044h^{1.72},14.77)
+ 0.002\,\log_{10}(h)\,d_{3D}
\end{aligned}
\end{equation*}
& & \\
& &
\begin{equation*}
PL_2 = PL_1(d_{BP}) + 40\log_{10}\!\left(\frac{d_{3D}}{d_{BP}}\right)
\end{equation*}
& & \\
\cline{2-5}
& \multirow{7}{*}{NLOS}
&
\begin{equation*}
\begin{aligned}
PL_{\text{SMa-NLOS}} = \;&161.04
- 7.1\log_{10}(W)
+ 7.5\log_{10}(h) \\
&- \bigl(24.37 - 3.7(h/h_{BS})^2\bigr)\log_{10}(h_{BS}) \\
&+ \bigl(43.42 - 3.1\log_{10}(h_{BS})\bigr) \\
&\bigl(\log_{10}(d_{3D}) - 3\bigr) 
+ 20\log_{10}(f_c) \\
&- \bigl(3.2(\log_{10}(11.75h_{UE}))^2 - 4.97\bigr)
\end{aligned}
\end{equation*}
&
\multirow{7}{*}{$\sigma_{SF}=8$}
&
\multirow{7}{*}{
\makecell[l]{%
$10\,\mathrm{m} \le d_{2D} \le 5000\,\mathrm{m}$\\
$1\,\mathrm{m} \le h_{UE} \le 14\,\mathrm{m}$\\
$h = 10\,\mathrm{m}$, $W = 10\,\mathrm{m}$
}}
\\
\hline
\end{tabular}
\end{table*}

\begin{figure}[htbp]
	\centering
	\includegraphics[width=3.1in]{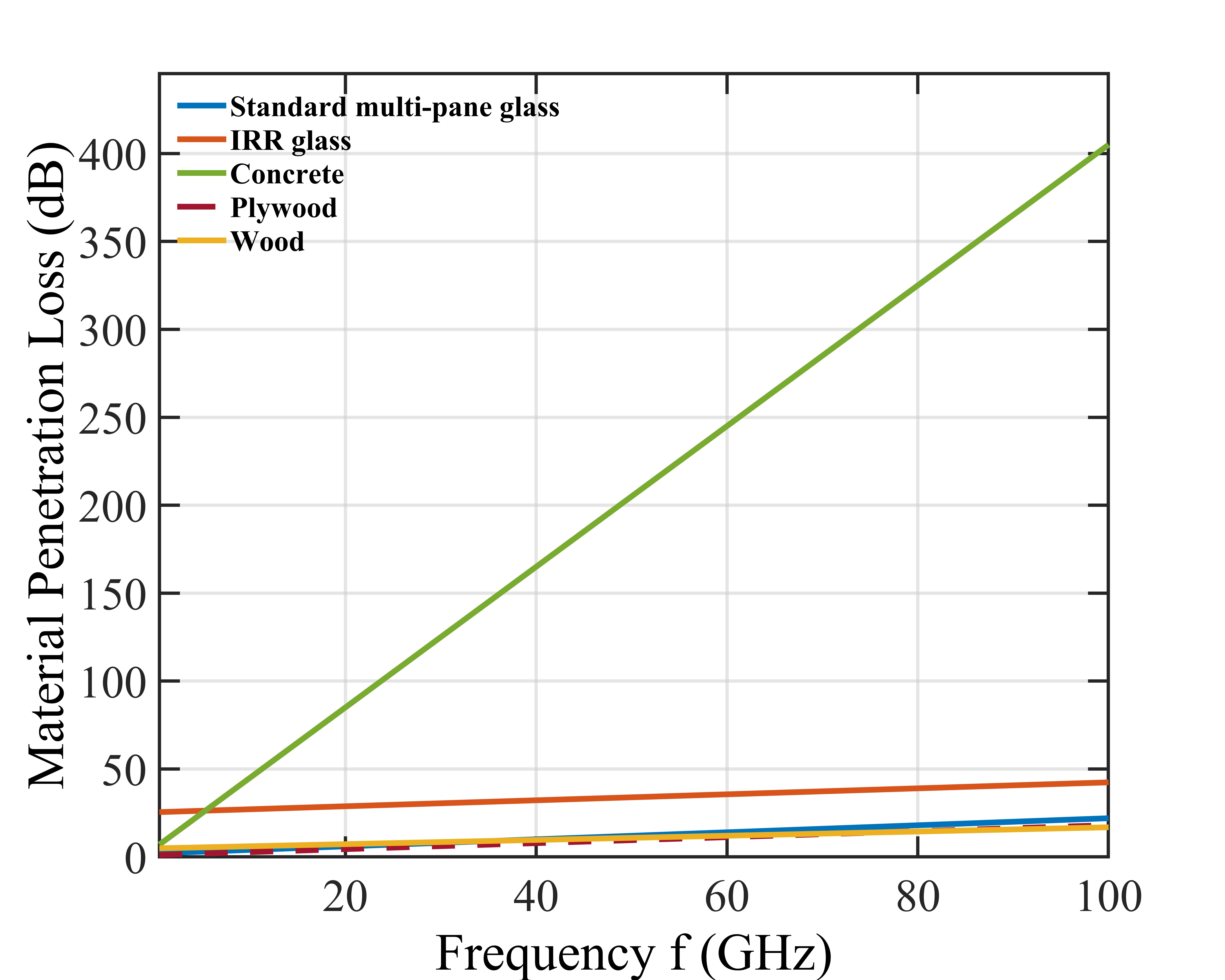}
	\caption{Material penetration losses \cite{M2412}.}\label{penetration_loss}
\end{figure}

\textbf{Material-dependent penetration loss:} Penetration loss represents the bottleneck for 7–24 GHz coverage. It is governed by the electromagnetic properties of the obstructing materials. Fig. \ref{penetration_loss} compares the frequency-dependent attenuation of four common materials: standard multi-pane glass, infrared reflective (IRR) glass, concrete, and wood. It is evident that concrete imposes the most severe attenuation with a steep frequency dependence, making Outdoor-to-Indoor (O2I) communications particularly challenging in the millimeter-wave bands. Conversely, wood exhibits relatively lower loss. This physical reality necessitates a material-specific modeling approach.

\begin{figure}[htbp]
	\centering
	\includegraphics[width=3.1in]{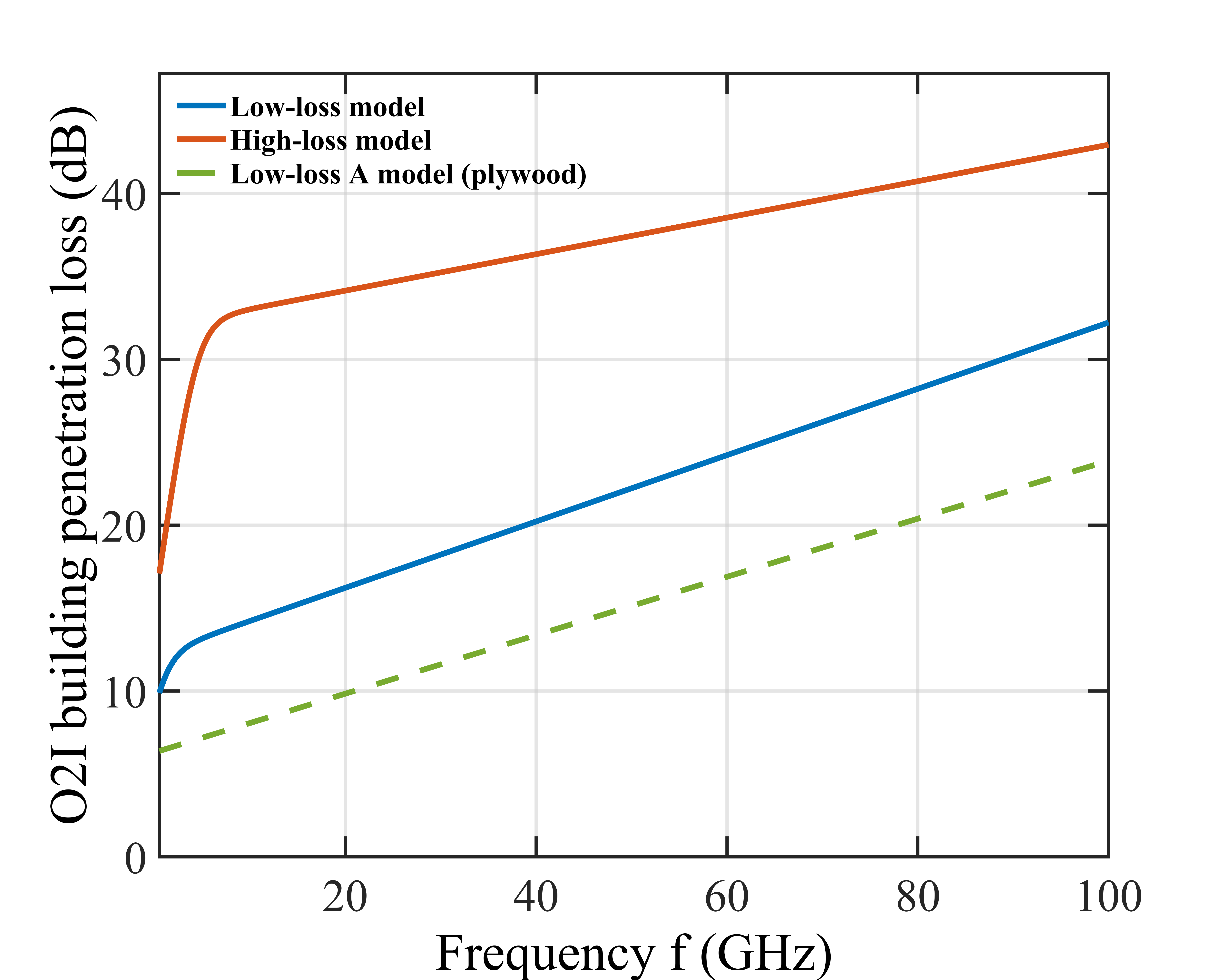}
	\caption{SMa O2I building penetration loss models.}\label{indoor_loss}
\end{figure}

\textbf{O2I building penetration loss:} To implement this, Rel-19 adopts a Low-loss/High-loss model framework, as summarized in Table \ref{tab:indoor_loss_model} and visualized in Fig. \ref{indoor_loss} \footnote{This figure shows the sum of the path loss through the external wall and the inside loss.}. The Low-loss model corresponds to buildings with standard multi-pane glass, while the High-loss model represents modern energy-efficient buildings using IRR glass. Commercial buildings are statistically more likely to follow the High-loss model. For the SMa scenario, the standard supports Low-loss, High-loss, and a specific ``Low-loss A" model (incorporating plywood characteristics typical of suburban housing). The indoor loss component $PL_{in}$ increases linearly with the depth into the building ($d_{2D-in}$). For SMa, $d_{2D-in}$ is generated as a UE-specific random variable, uniformly distributed between 0–25 m for commercial buildings and 0–10 m for residential buildings, reflecting the smaller footprint of suburban houses.

\begin{table*}[htbp]
\centering
\caption{Outdoor loss model for SMa.}\label{tab:indoor_loss_model}
\renewcommand{\arraystretch}{1.6}
\setlength{\tabcolsep}{18pt}
\begin{tabular}{|>{\centering\arraybackslash}p{3cm}|c|c|c|}
\hline
\multirow{2}{*}{\textbf{Model Type}} & \multicolumn{1}{c|}{\textbf{Path Loss through External Wall}} & \multicolumn{1}{c|}{\textbf{Indoor Loss}} & \multicolumn{1}{c|}{\textbf{Standard Deviation}} \\
& \multicolumn{1}{c|}{$PL_{\text{tw}}$ (dB)} & \multicolumn{1}{c|}{$PL_{\text{in}}$ (dB)} & \multicolumn{1}{c|}{$\sigma_{p}$ (dB)} \\
\hline

\textbf{Low-loss model} & 
$\begin{aligned}
5 &- 10\log_{10}\Big(0.3 \cdot 10^{\frac{-L_{\text{glass}}}{10}} \\
&+ 0.7 \cdot 10^{\frac{-L_{\text{concrete}}}{10}}\Big)
\end{aligned}$ &
$0.5d_{\text{2d-in}}$ & 4.4 \\
\hline

\textbf{High-loss model} & 
$\begin{aligned}
5 &- 10\log_{10}\Big(0.7 \cdot 10^{\frac{-L_{\text{IRR-glass}}}{10}} \\
&+ 0.7 \cdot 10^{\frac{-L_{\text{concrete}}}{10}}\Big)
\end{aligned}$ &
$0.5d_{\text{2d-in}}$ & 6.5 \\
\hline

\textbf{Low-loss A model} & 
$\begin{aligned}
5 &- 10\log_{10}\Big(0.3 \cdot 10^{\frac{-L_{\text{glass}}}{10}} \\
&+ 0.7 \cdot 10^{\frac{-L_{\text{plywood}}}{10}}\Big)
\end{aligned}$ &
$0.5d_{\text{2d-in}}$ & 4.4 \\
\hline
\end{tabular}
\end{table*}

\subsubsection{\textbf{Step 4}} Generate large-scale parameters.

\textbf{Statistical characterization:} This step generates the set of global LSPs, including the root-mean-square (RMS) delay spread (DS), angular spreads (ASA, ASD, ZSA, ZSD), Ricean K-factor (K), and shadow fading (SF). The 3GPP framework provides statistical distributions for these parameters derived from extensive measurement campaigns across scenarios (InH, UMi, UMa, RMa, and SMa). It is observed that the RMS DS and angular spreads follow a log-normal distribution, while SF and K-factor (in dB) follow a normal distribution. Notably, these statistics exhibit strong frequency dependency in the 7–24 GHz band: as frequency increases, the mean values of delay and angular spreads generally decrease due to higher path loss suppressing distant multipath components, while their variances tend to increase, reflecting greater channel variability.

\textbf{Spatial correlation modeling:} In system-level simulations, independently generating LSPs based solely on the above statistics would fail to capture the physical reality that users in close proximity experience similar channel conditions \cite{Zhao2007}. When one UE is connected to multiple BSs, correlations of these links mostly probably exists in many scenarios, at least between BSs near each other \cite{Baum2005}. Some measurements indicate that these correlations are a function of the angle between BSs directions being seen from the UE, i.e, $\theta_{BS}$ \cite{Gudmundson1991,Perahia2001}. However, the amount of measurement data is limited; these correlations of links from the same UE to multiple BSs are usually modeled as zero. In addition, correlations of links from multiple UEs to the same BS are proportional to their relative distance $d_{UE}$. The closer the relative distance, the more similar the propagation environments. Thus, LSPs are correlated between links. This kind of correlation is called auto-correlation. Also, LSPs of intra-site links are correlated according to the distance between UE positions. The cross-correlation distance of $d_{cor}$ is usually used to quantify correlation. 
\begin{figure}[htbp]
	\centering
	\includegraphics[width=3.1in]{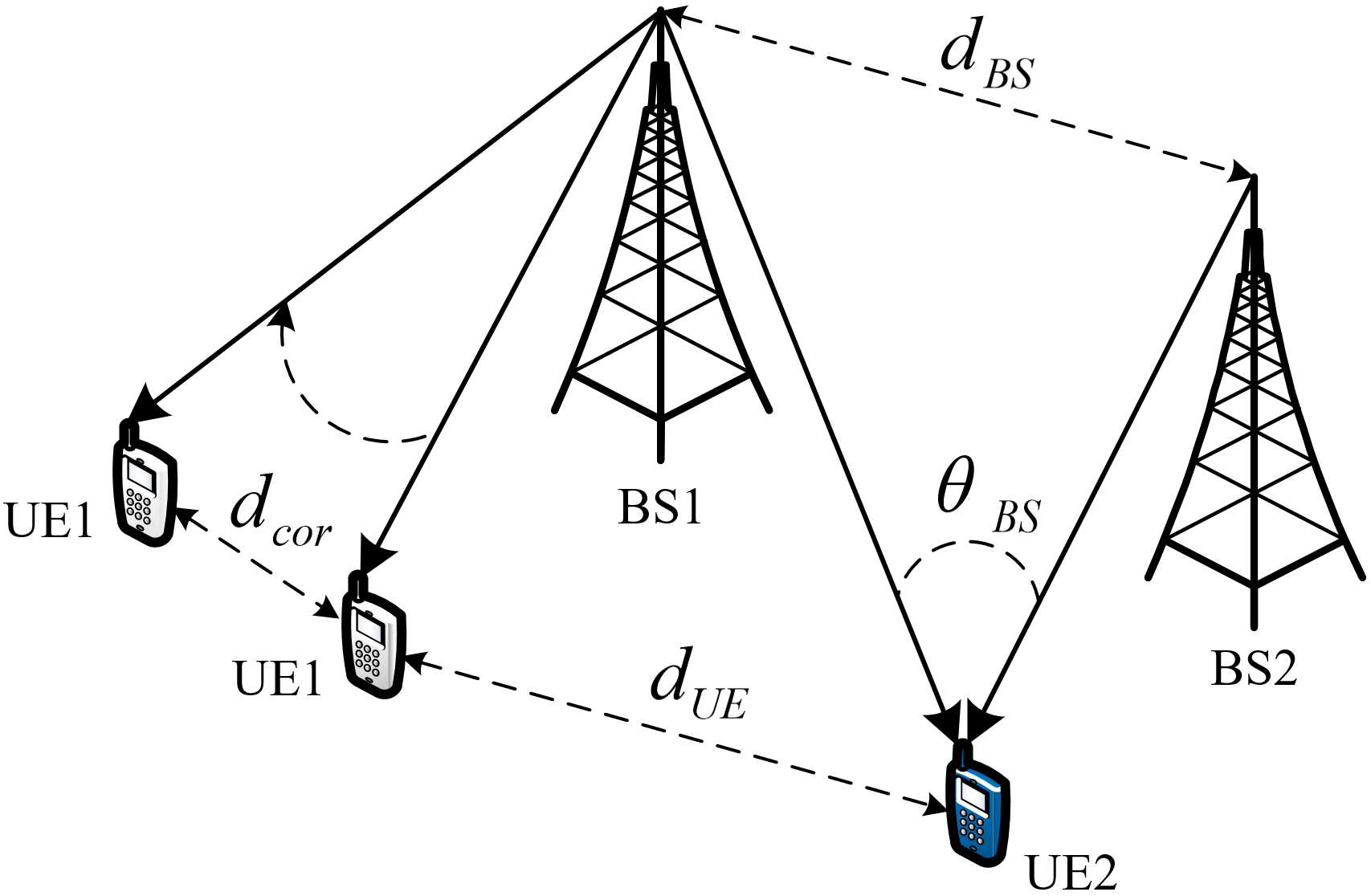}
	\caption{The illustration of correlations between different links.}\label{different_correlation}
\end{figure}

\begin{figure}[htbp]
	\centering
	\includegraphics[width=3.1in]{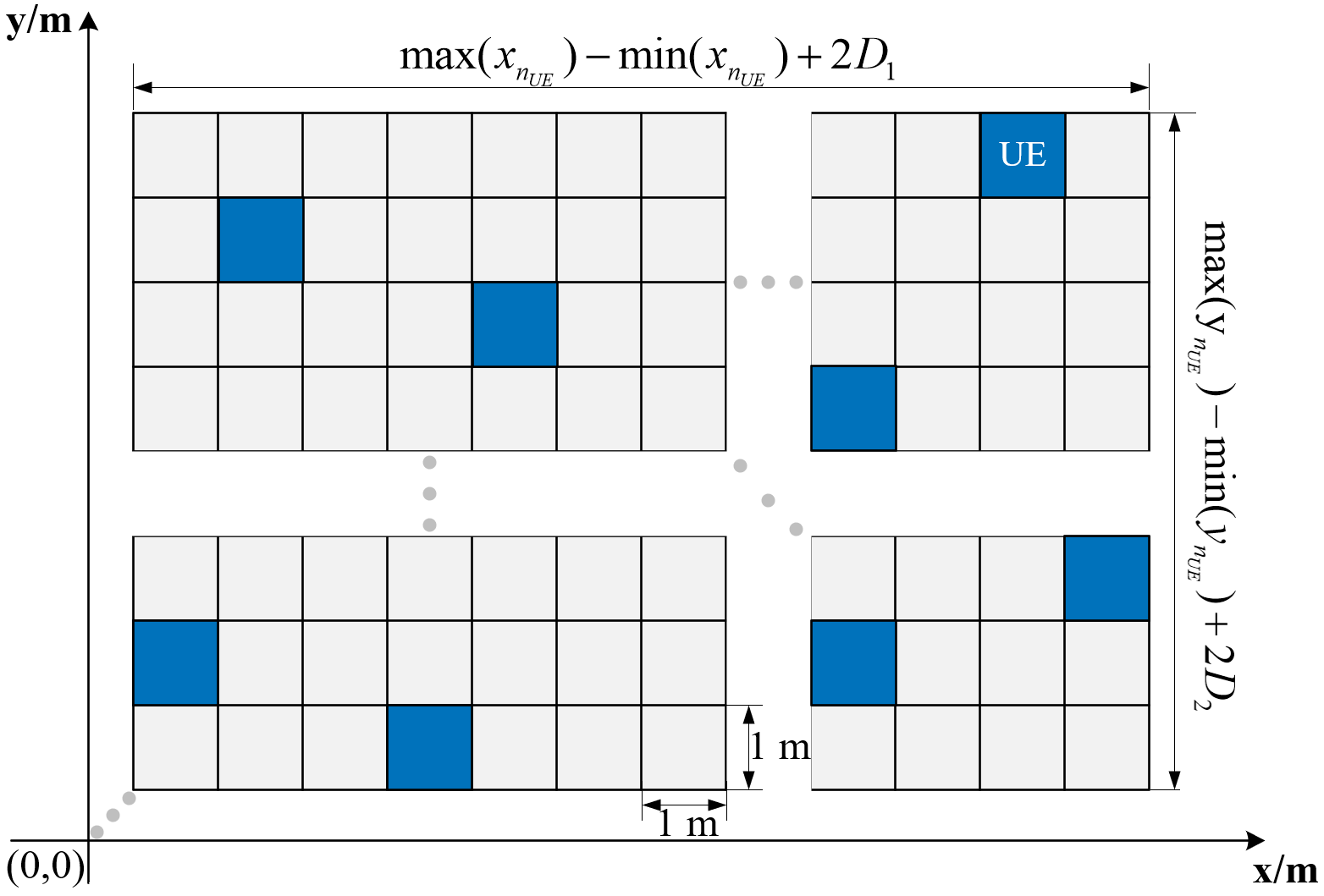}
	\caption{The sketch of a uniform grid.}\label{grid}
\end{figure}

The influence of LSPs exponential auto-correlation and cross-correlation is generated separately. Assuming there are $M_{LSP}$ LSPs per link and $N_{UE}$ UEs linked to the same BS at positions $(x_{n_{UE}},y_{n_{UE}})$, where $n_{UE}=1,...,N_{UE}$. Then, a uniform grid of positions that covers $N_{UE}$ UEs is generated. As shown in Fig. \ref{grid}, size of the grid is set based on UEs' positions, i.e., $(\textrm{max}(x_{n_{UE}})-\textrm{min}(x_{n_{UE}})+2\textrm{D}_1)\times(\textrm{max}(y_{n_{UE}})-\textrm{min}(y_{n_{UE}})+2\textrm{D}_2)$. $D_1$ and $D_2$ are constant values that can be defined personally. The basic principle is to make all UEs be on the grid. Empirically, the size of the grid node is set to $1 \textrm{m}\times 1 \textrm{m}$, which is helpful to map onto UEs. To each grid node, $M_{LSP}$ Gaussian iid$\sim$N(0,1) random numbers are assigned for representing $M_{LSP}$ LSPs, separately. Then, we use a two dimensional finite impulse response (FIR) filter to filter the grid of random numbers, and get exponential auto-correlation. Impulse response of the filter for the $m_{LSP}$th is
\begin{equation}
f_{m_{LSP}}(d)=\textrm{exp}(\frac{-d_{UE}}{d_{cor}}),
\end{equation} 
where $d_{UE}$ is the distance between UEs and $d_{cor}$ is the correlation distance, see Fig. \ref{different_correlation}. It is worth noting that $d_{cor}$ is special for each LSP. After filtering, we can get the correlated random numbers, $\xi(x_{n_{UE}},y_{n_{UE}})$, at $N_{UE}$ grid nodes.

For the cross-correlation, it is generated independently of the LSPs of $N_{UE}$ UEs by linear transformation
\begin{equation}
\widetilde{\textbf{S}}(x_{n_{UE}},y_{n_{UE}})=\sqrt{\textbf{C}_{M_{LSP}\times M_{LSP}}}\xi(x_{n_{UE}},y_{n_{UE}}),
\end{equation} 
where $\textbf{C}_{M_{LSP}\times M_{LSP}}$ is the cross-correlation matrix \footnote{Under LOS conditions, $\textbf{C}_{M_{LSP}\times M_{LSP}}$ is a $7\times7$ matrix. But under NLOS and O2I conditions, it is a $6\times6$ matrix. This is because the K-factor is not modeled in the latter case.}. Elements of this matrix are the cross-correlation values between LSPs. Some cross-correlation values are not available as the K-factor is not modeled under NLOS and O2I conditions. It is worth noting that there is a mapping between elements of $\textbf{C}_{M_{LSP}\times M_{LSP}}$ and elements of $\xi(x_{n_{UE}},y_{n_{UE}})$. Finally, we can get each LSP by

\begin{equation}
s_{n_{UE}}=\begin{cases}
10^{\tilde{s}_{n_{UE}}\cdot\sigma_{n_{UE}}+\mu_{n_{UE}}},\ s_{n_{UE}}\sim N(\mu,\sigma),\\
\tilde{s}\cdot\sigma_{n_{UE}}+\mu_{n_{UE}},\ \textrm{log}10(s_{n_{UE}})\sim N(\mu,\sigma),\\
\end{cases}
\end{equation} 
where $\tilde{s}_{n_{UE}}$ is a element of $\widetilde{\textbf{S}}$, $\sigma_{n_{UE}}$ and $\mu_{n_{UE}}$ are statistics of LSPs. These LSPs for different BS-UE links are uncorrelated, but the LSPs for links from co-sited TRxPs to a UE are the same. In addition, these LSPs for the links to UE on different floors are uncorrelated. For RMS angular spreads, they have limitation on values, i.e., $\sigma_{ASA}=\textrm{min}(\sigma_{ASA},104)$, $\sigma_{ASD}=\textrm{min}(\sigma_{ASD},104)$, $\sigma_{ESA}=\textrm{min}(\sigma_{ESA},52)$, and $\sigma_{ESD}=\textrm{min}(\sigma_{ESD},52)$. 
\subsection{Small Scale Parameters Generation}
Predicated on the spatially correlated LSPs derived in Step 4, this stage involves the instantiation of Small-Scale Parameters (SSPs)—including delays, cluster powers, angles, and cross-polarization power ratios (XPRs)—to explicitly describe the Multipath Components (MPCs). Table \ref{small_scale_parameters_UMi} provides the reference statistical parameters and distributions specifically calibrated for the SMa scenario.
\begin{table*}[htbp]
	\centering
	\caption{Reference values or statistics to generate SSPs for SMa.}\label{small_scale_parameters_UMi}
	\renewcommand\arraystretch{1.25}
	\begin{tabular}{|c|c|c|c|c|}
		\hline
		\multicolumn{2}{|c|}{\multirow{2}{*}{\textbf{Parameters}}}                    & \multicolumn{3}{c|}{\textbf{SMa}} \\ \cline{3-5} 
		\multicolumn{2}{|c|}{}                                                        & LOS                    & NLOS                  & O2I                   \\ \hline
		\multicolumn{2}{|c|}{Delay distribution}                                      & Exp                    & Exp                   & Exp                   \\ \hline
		\multicolumn{2}{|c|}{\begin{tabular}[c]{@{}c@{}}AOD and AOA \\ distribution\end{tabular}} & \begin{tabular}[c]{@{}c@{}}Wrapped\\  Gaussian\end{tabular} & \begin{tabular}[c]{@{}c@{}}Wrapped\\  Gaussian\end{tabular} & \begin{tabular}[c]{@{}c@{}}Wrapped \\ Gaussian\end{tabular} \\ \hline
		\multicolumn{2}{|c|}{\begin{tabular}[c]{@{}c@{}}EOD and EOA \\ distribution\end{tabular}} & Laplacian & Laplacian& Laplacian\\ \hline
		\multicolumn{2}{|c|}{\begin{tabular}[c]{@{}c@{}}Delay scaling\\ parameters $r_{\tau}$\end{tabular} }     & 2.4                    & 1.5                     & 1.5                   \\ \hline
		\multirow{2}{*}{XPR [dB]}              & $\mu_{XPR}$      & 8                      & 4                     & 4                    \\ \cline{2-5} 
		& $\sigma_{XPR}$    & 4                      & 3                     & 3                    \\ \hline
		\multicolumn{2}{|c|}{Number of clusters}                                      & 15                     & 14                    & 14                   \\ \hline
		\multicolumn{2}{|c|}{\begin{tabular}[c]{@{}c@{}}Number of rays\\ per cluster $r_{\tau}$\end{tabular} }                              & 20                     & 20                    & 20                    \\ \hline
		\multicolumn{2}{|c|}{Cluster DS ($C_{DS}$)}                                  & max(0.25,6.5622-3.4084log$_{10}(f_c)$)                     & max(0.25,6.5622-3.4084log$_{10}(f_c)$)                           & max(0.25,6.5622-3.4084log$_{10}(f_c)$)  \\ \hline
		\multicolumn{2}{|c|}{Cluster ASD ($C_{ASD}$)}                                & 2.08                      & 1.33                    & 1.33                    \\ \hline
		\multicolumn{2}{|c|}{Cluster ASA ($C_{ASA}$)}                                 & 5                     & 10                    & 10                     \\ \hline
		\multicolumn{2}{|c|}{Cluster ESA ($C_{ESA}$)}                                & 7                      & 7                     & 7                     \\ \hline
		\multicolumn{2}{|c|}{\begin{tabular}[c]{@{}c@{}}Per cluster shadowing\\ std $\zeta$ [dB]\end{tabular}} & 3                      & 3                     & 3                     \\ \hline
	\end{tabular}
    \end{table*}
\subsubsection{\textbf{Step 5}:} 
\begin{figure}[htbp]
	\centering
	\includegraphics[width=3.1in]{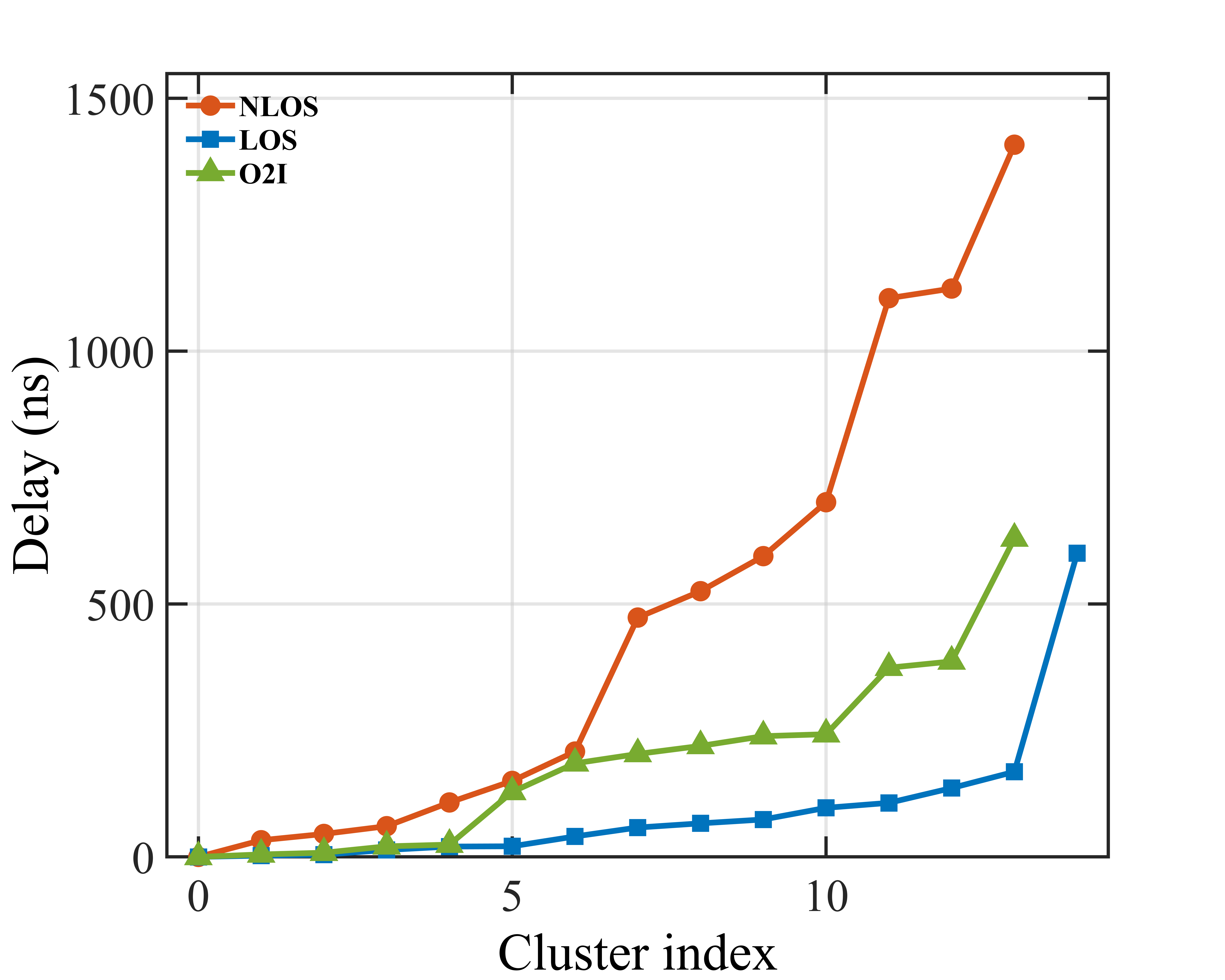}
	\caption{The delays of clusters under LOS, NLOS, and O2I conditions in SMa.}\label{delays_UMi}
\end{figure}

Generate delays. Electromagnetic waves propagate through diverse paths, arriving at the UE with varying excess delays. In the simulation, delays are modeled stochastically using an exponential distribution. The unscaled delays $\tau_n'$ are generated as
\begin{equation}
{\tau_n'} = -{r_\tau}DS\textrm{ln}({X_n}),
\end{equation}
where ${r}_{\tau}$ is the delay scaling parameter, ${X}_{n}\sim$ uniform (0, 1), and cluster index $n = 1 , \cdots, N$. With ${X}_{n}$ following a uniform distribution, the delay values ${\tau}_{n}'$ are drawn from this range. Then, normalize the delays by subtracting the minimum delay and sort the normalized delays to ascending order. It's worth noting that additional scaling of delays is required to compensate for the effect of LOS peak addition to the delay spread in the case of LOS condition, and the scaling constant depends on the K-factor, i.e.,
\begin{equation}
C_{\tau}=0.7705-0.0433K+0.0002K^2+0.000017K^3,
\end{equation}
where $K$ [dB] is the Ricean K-factor that is the LSP generated in Step 4. Thus, the scaled delays is $\tau_n^{LOS}=\tau_n/C_{\tau}$. Fig \ref{delays_UMi} shows an example of delays in SMa. $\tau_{LOS}$, $\tau_{NLOS}$. $\tau_{O2I}$ represents delays of clusters under LOS, NLOS, and O2I conditions, respectively.  

\subsubsection{\textbf{Step 6}} Generate cluster powers. Cluster powers ${P}_{n}$are modeled as a decaying exponential function of the generated delays, superimposed with a stochastic shadowing term:
\begin{equation}
{P_n}' = \textrm{exp}(-{\tau_n}\frac{{r_\tau}-1}{{r_\tau}DS})\cdot{10^{-\frac{{Z_n}}{10}}},
\end{equation}
where ${Z}_{n}\sim N(0,{\zeta}^{2})$  represents the per-cluster shadowing variability in dB. The powers are then normalized such that their sum equals unity. Also, we need to normalize the cluster powers so that the sum power of all cluster powers is equal to one. Similarly to generating delays, we need to consider the LOS path especially. The LOS path is the first path that arrived at UEs. Thus, an additional specular component should be added to the first cluster under LOS conditions. The power ratio of the LOS path to other stochastic paths is equal to $K_R$ (in linear). Thus, the power of the LOS path is 
\begin{equation}
P_{1,LOS}=\frac{K_R}{1+K_R}.
\end{equation}
Accordingly, the powers of the scattering clusters are re-scaled to maintain the total energy balance:
\begin{equation}
P_{n}=\frac{1}{1+K_R}\frac{P'_{n}}{\sum_{n=1}^NP'_{n}}+\delta(n-1)P_{1,LOS}.
\end{equation}

\begin{figure}[htbp]
	\centering
	\includegraphics[width=3.1in]{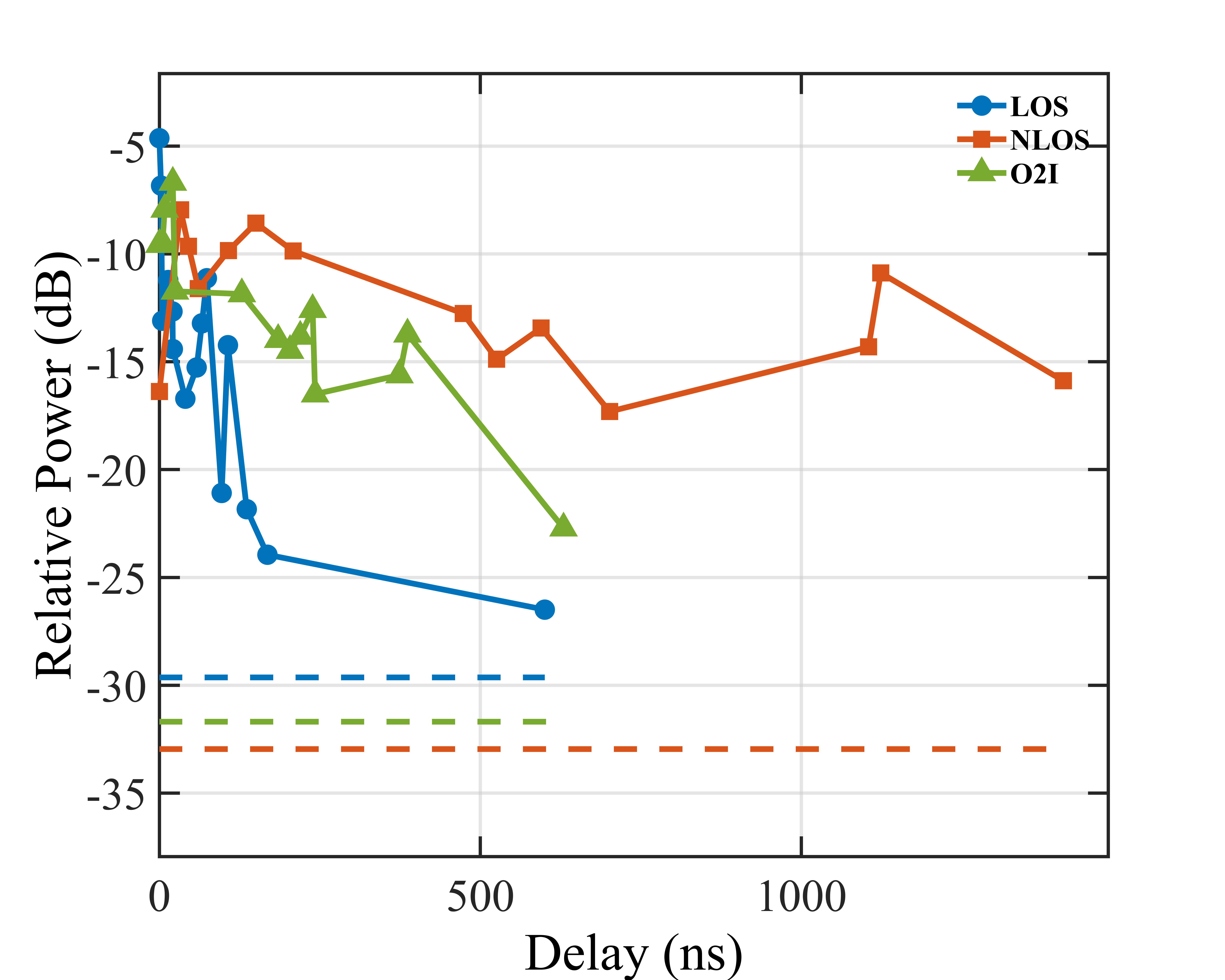}
	\caption{The power delay profiles under LOS, NLOS, and O2I conditions in SMa.}\label{power_UMi}
\end{figure}

Each cluster is assumed to comprise $M=20$ rays, with power uniformly distributed as ${P}_{n}/M_{n}$. Fig. \ref{power_UMi} depicts representative power delay profiles for the SMa scenario. To analyze the cluster distribution, a dynamic range threshold is applied: a dashed line is plotted 25 dB below the maximum cluster power to indicate the relevant power range.

\subsubsection{\textbf{Step 7}} Generate arrival and departure angles. 
The generation of angles—AoA, AoD, ZoA, and ZoD—employs the Inverse CDF method. The composite angular power spectrum is modeled as a Wrapped Gaussian distribution for azimuth angles and a Laplacian distribution for zenith angles. Taking AoA as an example, the cluster angles are derived as:
\begin{equation}
{\phi_{n, AOA}'} =
\frac{2(\textrm{ASA}/1.4)\sqrt{-\textrm{ln}(P_n/\textrm{max}(P_n))}}{C_\phi},\ \textrm{Gaussian},
\end{equation}
Similarly, for zenith angles (ZoA):
\begin{equation}
{\theta_{n, ZOA}'} = -\frac{\textrm{ZSA}\textrm{ln}(P_n/\textrm{max}(P_n))}{C_\phi},\ \textrm{Laplacian},
\end{equation}
where ``ASA" and ``ZSA'' represent the RMS ASA and ZSA generated in Step 4, ${C}_{\phi}$ and $C_\theta$ are defined as scaling factors. Similar to the scaling factor $C_{\tau}$, they are also a function of the K-factor under the LOS condition. This is to compensate for the effect of LOS peak addition to the angular spreads. $C_{\phi}^{NLOS}$ and $C_{\theta}^{NLOS}$ are scaling factors related to the total number of clusters \footnote{The scaling factors are set to 0 when there is no reference value for the number of clusters.}. 

It is noted that to introduce random variation, we need to assign positive or negative sign to the angles by multiplying with a random variable $X_n$ with uniform distribution to the discrete set of $\{1,\ -1\}$, and add a Gaussian variable $Y_{n}\sim$N(0,(ASA/7)$^2$). This process can be written as
\begin{equation}
\phi_{n,AOA}=X_n\phi'_{n,AOA}+Y_n+\phi_{LOS,AOA},
\end{equation} 
where $\phi_{LOS,AOA}$ is the AOA of the LOS ray that can be calculated according to positions of UEs and BSs. This equation is to generate AOAs of clusters under NLOS and O2I conditions. For the LOS condition, we need to make the first cluster to the LOS direction by
\begin{equation}
\phi_{n,AOA}=X_n\phi'_{n,AOA}+Y_n-(X_1\phi'_{1,AOA}+Y_1-\phi_{LOS,AOA}).
\end{equation} 

Until now, we have got AOAs of clusters. Next, AOAs of each ray in a cluster can be got by adding an offset angle $\alpha_m$
\begin{equation}
\phi_{n,m,AOA}=\phi_{n,m,AOA}+c_{ASA}\alpha_m,
\end{equation} 
where $c_{ASA}$ is the cluster ASA see Table \ref{small_scale_parameters_UMi}, $\alpha_m$ is a normalized constant value. Table \ref{angle_offset} gives the exact values of ray offset angles, the sum of all $\alpha_m$ is equal to 1. 
\begin{table}[htbp]
	\centering
	\caption{Ray offset angles for AOA.}\label{angle_offset}
	\begin{tabular}{|c|c|}
		\hline
		\textbf{Ray number $m$} & \textbf{Basis vector of offset angles $\alpha_m$} \\ \hline
		1,2                   & $\pm$0.0447                                    \\ \hline
		3,4                   & $\pm$0.1413                                    \\ \hline
		5,6                   & $\pm$0.2492                                    \\ \hline
		7,8                   & $\pm$0.3715                                    \\ \hline
		9,10                  & $\pm$0.5129                                    \\ \hline
		11,12                 & $\pm$0.6797                                    \\ \hline
		13,14                 & $\pm$0.8844                                    \\ \hline
		15,16                 & $\pm$1.1481                                    \\ \hline
		17,18                 & $\pm$1.5195                                    \\ \hline
		19,20                 & $\pm$2.1551                                    \\ \hline
	\end{tabular}
\end{table}


The generation of other angles follows a procedure similar to AOA as described above. It is worth noting that azimuth angles and elevation angles should be wrapped within $[-180^{\circ},180^{\circ}]$ and $[0,180^{\circ}]$, respectively. For example, if we get a  $\theta_{n,m,ZOD}\in[180^{\circ},360^{\circ}]$, it should be set to $(360^{\circ}-\theta_{n,m,ZOA})$. Details can be found in \cite{M2412}. 
%
%
%
%

\subsubsection{\textbf{Step 8}} 
\begin{figure}[htbp]
	\centering
	\includegraphics[width=3.7in]{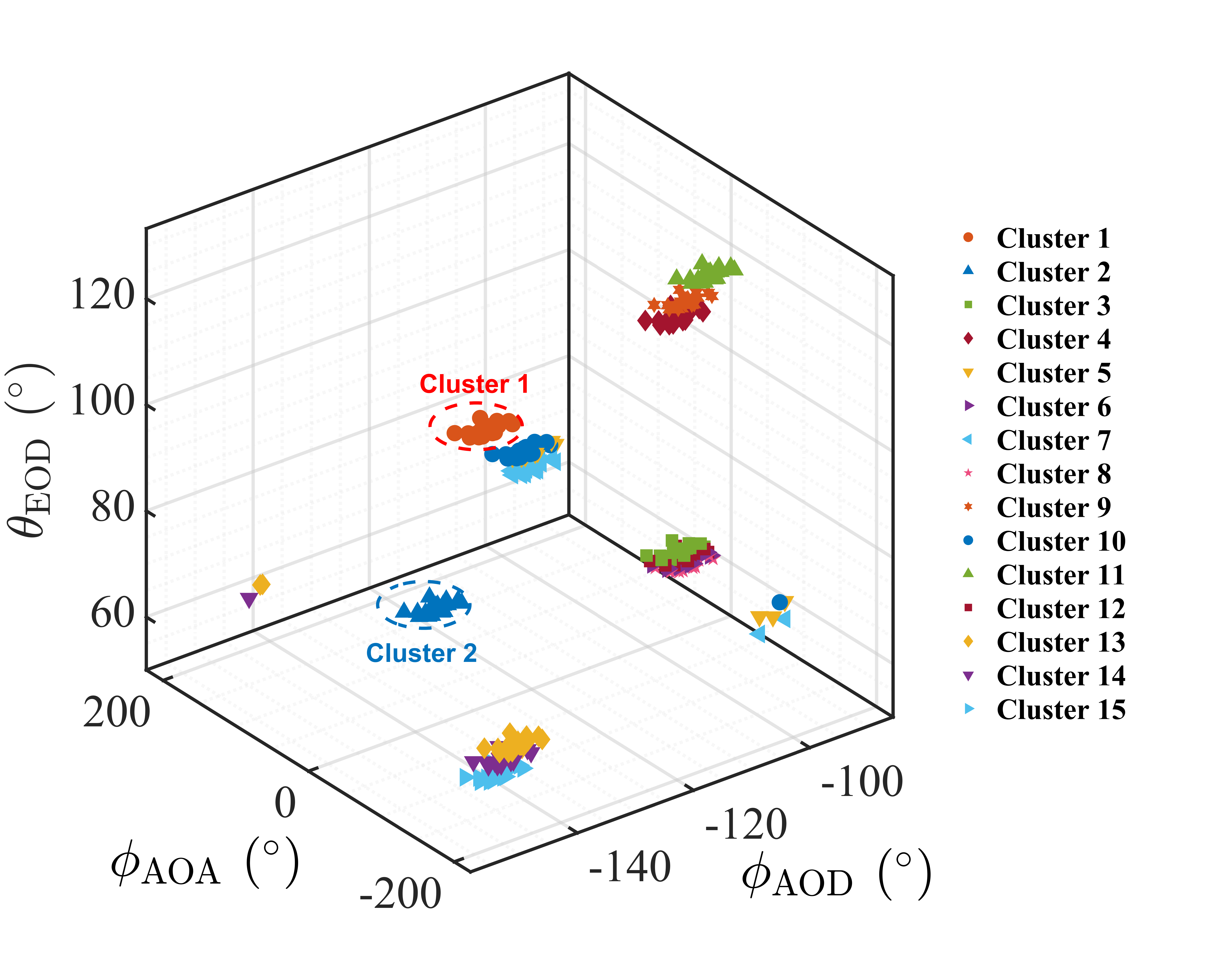}
	\caption{An example of clusters under the LOS condition for SMa.}\label{clusters}
\end{figure}

Coupling angles. The independently generated azimuth and zenith angles for both transmission and reception must be coupled to form physically meaningful 3D paths. Thus, we need to couple randomly $\phi_{n,m,AOA}$, $\phi_{n,m,AOD}$, $\theta_{n,m,ZOA}$, and $\theta_{n,m,ZOD}$ within a cluster $n$. Firstly, couple randomly $\phi_{n,m,AOA}$ to $\phi_{n,m,AOD}$ within a cluster $n$, or within a sub-cluster in the case of the two strongest clusters. Secondly, couple randomly $\theta_{n,m,ZOA}$ to $\theta_{n,m,ZOD}$ using the same procedure. Thirdly, couple randomly $\phi_{n,m,AOD}$ with $\theta_{n,m,ZOD}$ with a cluster $n$, or within a sub-cluster in the case of the two strongest clusters. Fig. \ref{clusters} shows an example of clusters after coupling. As shown in the figure, each colored group represents a distinct cluster of multipath components. The coupling ensures that rays within the same cluster (such as the highlighted Cluster 1 and Cluster 2) are spatially correlated. This reflects the physical reality where paths reflecting off the same object or group of objects arrive and depart within a concentrated angular range.
\subsubsection{\textbf{Step 9}} XPR generation. To model polarization effects, the Cross-Polarization Power Ratio (XPR) $\kappa_{n, m}$ for each ray is generated following a log-normal distribution:
\begin{equation}
\kappa_{n, m} = 10^{X_{n,m}/10},
\end{equation}
where $X_{n,m}\sim N({\mu}_{XPR}, {\sigma}_{XPR}^{2})$. Example values of ${\mu}_{XPR}$ and ${\sigma}_{XPR}$ can be found in Table \ref{small_scale_parameters_UMi}. 

\subsection{Channel Coefficients Generation}
The final stage synthesizes the time-variant Channel Impulse Response (CIR) by superimposing the generated spatial and temporal components.
\subsubsection{\textbf{Step 10}} Draw initial random phases.
To model the random phase shifts induced by scattering, initial phases are generated stochastically. For each ray $m$ of each cluster $n$ and for each of the four polarization combinations ($\theta\theta$, $\theta\phi$, $\phi\theta$, $\phi\phi$), random phases $\Phi$ are drawn uniformly from the interval ($-\pi, \pi$). For LOS conditions, if the LOS phase is not deterministically calculated based on geometry, it follows the same random distribution.

\subsubsection{\textbf{Step 11}} Generate channel coefficients. 
This step computes the complex channel coefficients for each cluster and each Transmit-Receive (Tx-Rx) element pair. The Doppler frequency component is applied based on the arrival angles (AoA, ZoA) and the UE velocity vector $\bar{v}$. A critical feature of the 3GPP model is the modeling of Intra-Cluster Delay Spread. While weaker clusters ($n=3,4,...,N$), are modeled as single delay taps, the two strongest clusters ($n=1$ and $n=2$) are decomposed into three sub-clusters to capture fine-grained temporal dispersion. The delays of these thre sub-clusters are
\begin{equation}
\begin{split}
\tau_{n,1}&=\tau_n,\\
\tau_{n,2}&=\tau_n+1.28C_{DS},\\
\tau_{n,3}&=\tau_n+2.56C_{DS},\\
\end{split}
\end{equation}
where $C_{DS}$ is cluster delay spread. Also, 20 rays of a cluster are mapped to sub-clusters as shown in Table \ref{sub_cluster}.
\begin{table}[htbp]
	\centering
	\caption{Sub-cluster information for intra cluster delay spread clusters.}\label{sub_cluster}
	\begin{threeparttable}
	\begin{tabular}{|c|c|c|c|}
		\hline
		\textbf{\begin{tabular}[c]{@{}c@{}}Sub-\\ cluster\\ i\end{tabular}} & \textbf{\begin{tabular}[c]{@{}c@{}}Mapping to \\ rays $R_i$\end{tabular}} & \begin{tabular}[c]{@{}c@{}}\textbf{Power}\\ \textbf{$|R_i|$/$M$}\end{tabular} & \textbf{\begin{tabular}[c]{@{}c@{}}delay  offset\\ $\tau_{n,i}-\tau_n$\end{tabular}} \\ \hline
		$i$=1                                                                   & \begin{tabular}[c]{@{}c@{}}$R_1$=\{1,2,3,4,5,6,\\ 7,8,19,20\}\end{tabular}   & 10/20          & 0                                                                     \\ \hline
		$i$=2                                                                   & \begin{tabular}[c]{@{}c@{}}$R_2$=\{9,10,11,\\ 12,17,18\}\end{tabular}        & 6/20           & 1.28$c_{DS}$\tnote{1}                                                                 \\ \hline
		$i$=3                                                                   & $R_3$={13,14,15,16}                                                        & 4/20           & 2.56$c_{DS}$                                                                 \\ \hline
	\end{tabular}
\begin{tablenotes}
	\item[1] $c_{DS}$ is the cluster delay spread.
\end{tablenotes}
\end{threeparttable}
\end{table}
The NLOS channel impulse response is constructed by summing the contributions of the sub-clusters (for $n=1,2$) and the standard clusters (for $n\geq3$), as expressed:
\begin{equation}
\centering
\begin{split}
H_{u,s}^{NLOS}(\tau,t)=&\sum_{n=1}^2\sum_{i=1}^3\sum_{m\in R_i}h_{n,m}(t)\delta(\tau-\tau_{n,i})+\\
&\sum_{n=3}^Nh_{n,m}(t)(t)\delta(\tau-\tau_n),
\end{split}
\end{equation}
where $h_{n,m}(t)$ represents the complex coefficient of ray $m$ in cluster $n$ (including antenna patterns, polarization, and Doppler terms). 
%

\subsubsection{\textbf{Step 12}}
The normalized small-scale coefficients are scaled by the large-scale attenuation factors. Specifically, the Path Loss (calculated in Step 3) and Shadow Fading (generated in Step 4) are applied to the impulse response to produce the final system-level channel realization suitable for performance evaluation.


\section{Simulation Implementation of Additional Modeling Components}
To quantify the impact of the new modeling features—specifically near-field propagation and SNS—we conducted system-level simulations in two representative scenarios: UMi and indoor hotspot (InH). The simulation parameters and assumptions are summarized in Table~\ref{Tab_Performance}, where 1000 UEs were randomly deployed for each simulation run to ensure statistical stability.

\begin{table}
\renewcommand\arraystretch{1.2}
    \centering
    \caption{Simulation assumptions for performance evaluation}
        \begin{tabular}{m{2.5cm}<{\centering}|m{2.2cm}<{\centering}|m{2cm}<{\centering}}
    \hline
      Parameters & \multicolumn{2}{c}{Values}\\
          \hline
        Scenarios & UMi & InH  \\
        \hline
        Carrier frequency & \multicolumn{2}{c}{7 GHz }    \\
        \hline
       BS antenna height  & 10 m &3 m  \\
       \hline
       SNR & \multicolumn{2}{c}{10 dB}  \\
       \hline
       BS antenna configuration & \multicolumn{2}{c}{\makecell*[c]{64×16 dual-polarized UPA, \\half-wavelength spacing, directional pattern}}   \\
       \hline
       UE antenna configuration &\multicolumn{2}{c}{\makecell*[c]{8 antennas distributed on a handheld device,\\ dual-polarization \cite{3GPP_38901}}}   \\
       \hline
       UE antenna pattern & \multicolumn{2}{c}{Isotropic}   \\
       \hline
       UE location &LOS scenario with outdoor UE at 1.5~m height &LOS scenario with indoor UE at 1 m height  \\
       \hline
       \multicolumn{3}{c}{Radius for uniform UE deployment centered at the BS} \\
       \hline
       Near-field simulation &20, 50, and 100 m & 2, 5, and 10 m \\
       \hline
       SNS simulation & 100 m & 10 m \\
        \hline
    \end{tabular}
    \vspace{-0.5cm}
    \label{Tab_Performance}
\end{table}

\subsection{Near-Field Propagation Simulation}

To incorporate near-field propagation characteristics, the standard channel generation procedure (specifically Step 11 in TR 38.901) is augmented with three additional sub-steps:
\begin{itemize}
\item \textbf{Absolute Time of Arrival Generation}: The absolute propagation delay is derived by combining the relative cluster delays generated in Step 5, the geometric 3D LOS propagation delay, and the specific excess delay of NLOS components.
\item \textbf{Spherical-Wave Source Distance Generation}: Using \eqref{equ_NF_d1} and \eqref{equ_s_BS}, the distances from the BS and UE to the spherical-wave source are computed. For both specular and non-specular clusters, scaling factors are first generated and then applied to the total path length (derived from the absolute delay) to determine the effective source distances.
\item \textbf{Element-Wise Parameter Update}: Based on the computed source distances, the specific direction vectors for each ray at every TX/RX antenna element are determined. Consequently, the element-wise array phase and angular parameters are calculated according to \eqref{equ_NF_phase}, \eqref{equ_NF_angle_BS}, and \eqref{equ_NF_angle_UE} to update the final channel coefficients.
\end{itemize}

\begin{figure}[htbp]
\centering
\subfloat[]
{\includegraphics[width=8.3cm]{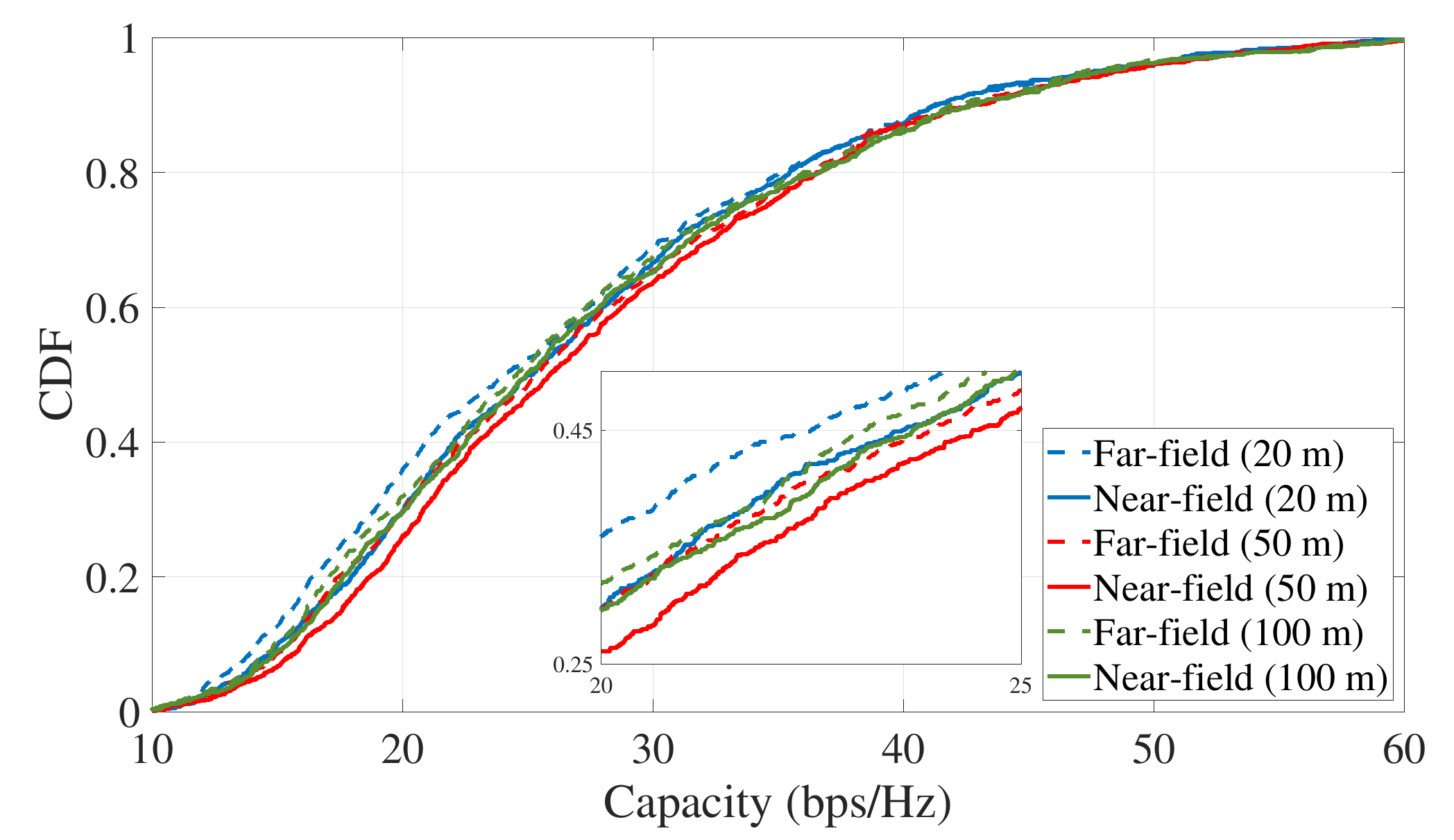}
\label{fig_NF_Capacity_a}}
\hfill
\subfloat[]
{\includegraphics[width=8.3cm]{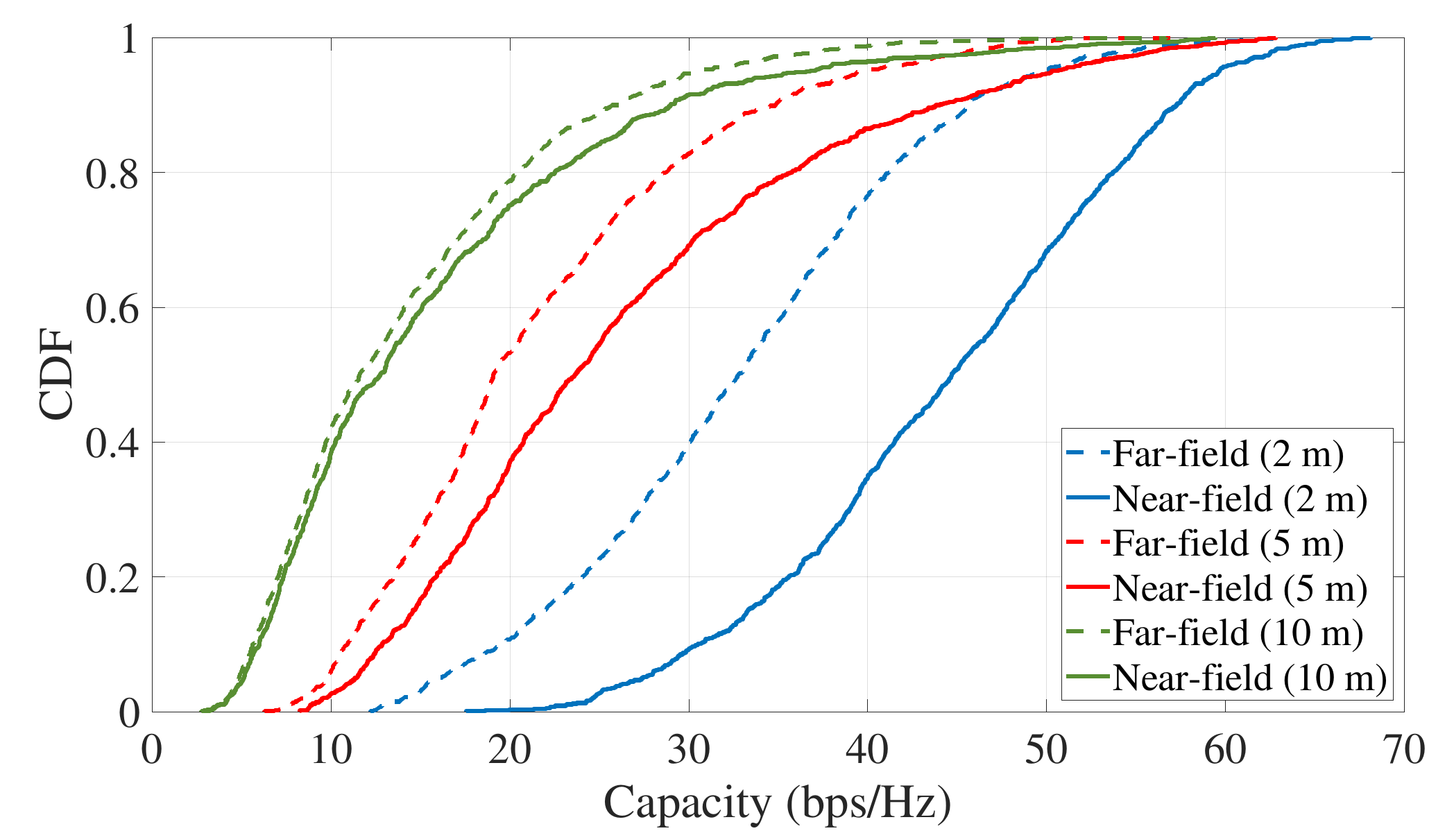}
\label{fig_NF_Capacity_b}}
\caption{Channel capacity comparison between near-field and far-field models in (a) UMi and (b) InH scenarios.}
\label{fig_NF_Capacity}
\end{figure}

Fig.~\ref{fig_NF_Capacity} presents the channel capacity comparison between the near-field and far-field models. As illustrated, the near-field model yields consistent capacity improvements due to the utilization of spherical wavefronts. In the UMi scenario, the gains are relatively modest—averaging 0.70, 0.59, and 0.44 bps/Hz for deployment radii of 20, 50, and 100 m, respectively. This is attributed to the larger BS–UE separations and the minimum horizontal distance constraint (10 m), which limits the dominance of near-field effects. Conversely, the InH scenario, characterized by shorter link distances, exhibits substantial capacity gains: 11.60, 4.75, and 1.46 bps/Hz for radii of 2, 5, and 10 m, respectively. These results confirm that near-field capacity benefits are most pronounced in short-range deployments where wavefront curvature is significant.

\subsection{Spatial Non-Stationarity Simulation}

The implementation of SNS modeling necessitates distinct procedures for the BS and UE to account for their differing blockage mechanisms. On the BS side, two alternative modeling approaches are supported. The recommended \textit{stochastic-based SNS model} is integrated into the coefficient generation step (Step 11) as follows:
\begin{itemize}
\item \textbf{Status Determination}: The SNS status of each cluster is determined using \eqref{equ_SNS_step1} by comparing a cluster-specific random variable against the UE's SNS probability threshold.
\item \textbf{Visibility Region Generation}: For clusters designated as SNS, a VP is computed based on cluster power via \eqref{equ_SNS_step2}. A rectangular VR satisfying this VP is then randomly generated on the array plane.
\item \textbf{Attenuation Calculation}: An element-wise power attenuation factor is calculated using \eqref{equ_SNS_step3}. For spatial stationary clusters, the factor is unity across the array. For SNS clusters, the factor remains unity within the VR but decays gradually outside it, ensuring smooth power transitions.
\end{itemize}
Alternatively, for the \textit{physical blocker-based model}, blockers are deployed after Step 1. The element-wise attenuation for each ray is then computed using the knife-edge diffraction model described in \eqref{equ_SNS_Blockage}.

On the UE side, SNS modeling accounts for device usage effects:
\begin{itemize}
\item \textbf{Scenario Selection}: One of four usage scenarios (one-hand grip, dual-hand grip, head-and-hand, or free space) is randomly selected based on predefined occurrence probabilities.
\item \textbf{Attenuation Calculation}: Element-wise power attenuation values are retrieved from look-up tables corresponding to the selected scenario and operating frequency, and are applied during coefficient generation.
\end{itemize}

\begin{figure}[htbp]
\centering
{\includegraphics[width=8.3cm]{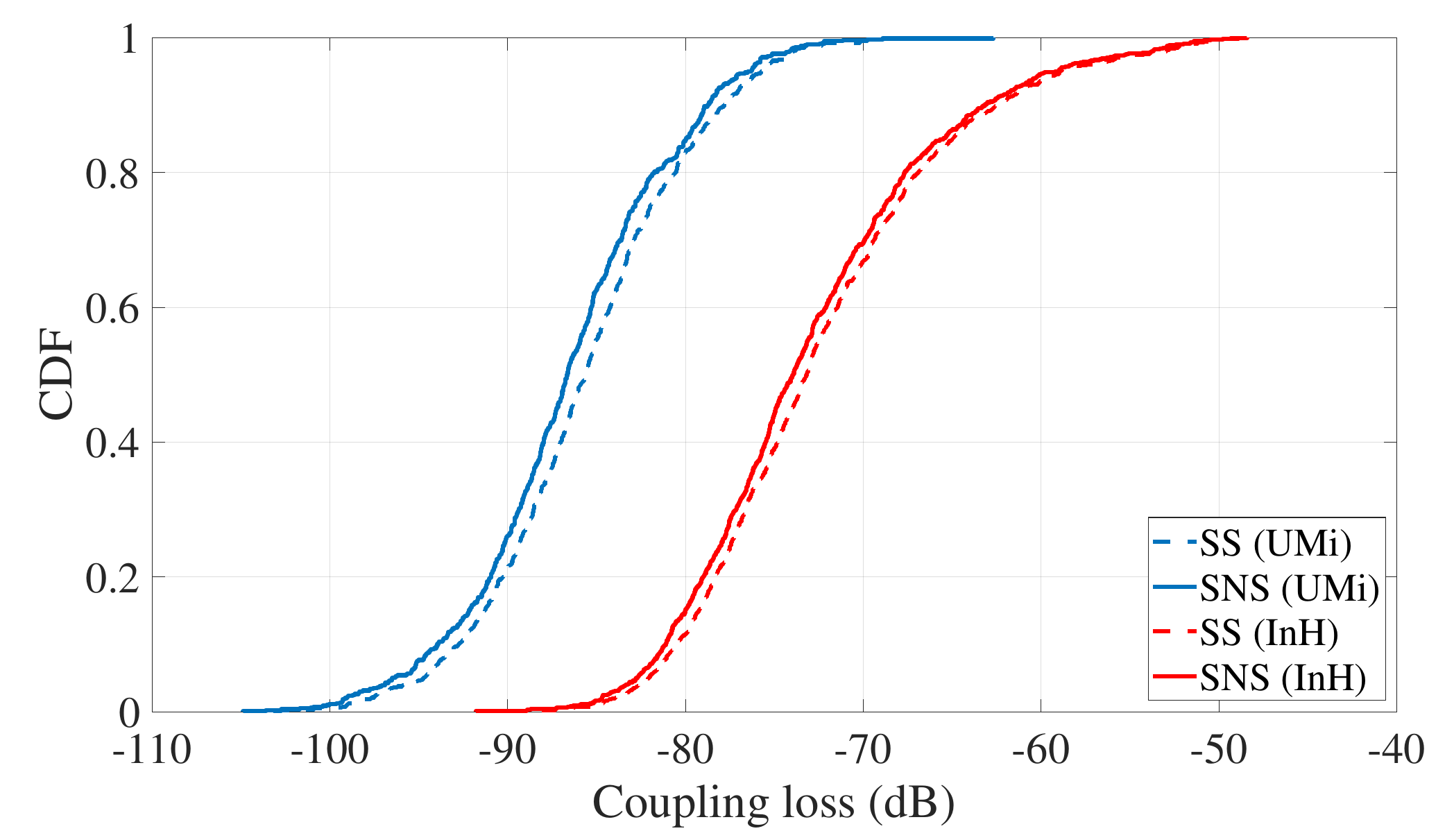}}
\caption{Impact of SNS on coupling loss in UMi and InH scenarios.}
\label{fig_SNS_ACG}
\end{figure}

The recommended stochastic-based model was applied to evaluate the impact of BS-side SNS. The coupling loss distributions for UMi and InH scenarios are shown in Fig.~\ref{fig_SNS_ACG}. The results indicate that enabling the SNS feature introduces additional attenuation (reflecting stronger spatial fading), resulting in an increase in coupling loss of approximately 0.91 dB in the UMi scenario and 0.67 dB in the InH scenario. The impact is less pronounced in the InH scenario due to a statistically lower SNS probability and higher VP compared to UMi.

\subsection{Large Bandwidth and Large Antenna Array}

To accurately characterize frequency-dependent channel characteristics, model parameters must be allowed to vary with frequency throughout the channel modeling procedure. 
The simulation experiments are conducted to identify the reasons for the inability to characterize frequency-dependent behavior. Simulation settings are summarized in Table \ref{sim_con}.

\begin{table}[t]
\centering
\caption{Simulation experiments configuration}
\label{sim_con}
\begin{tabular}{l|l}
\hline
\textbf{Parameters} & \textbf{Value / Type} \\ \hline
Scenario & UMa \\ \hline
Carrier frequency & 6, 9, 24 GHz \\ \hline
Antenna aperture size ($D$) & 1.5 m \\ \hline
Sample number & 10000 \\ \hline
Bandwidth & $c/D$, where $c$ is the speed of light \\ \hline
Horizontal array size ($D_h$) & 0.13 m \\ \hline
Vertical array size ($D_v$) & 1.49 m \\ \hline
$d_{2D}$ & 35 m \\ \hline
$h_{UE}$ & 1.5 m \\ \hline
\end{tabular}
\end{table}



\begin{figure}[htbp]
	\centering
	\includegraphics[scale=0.3]{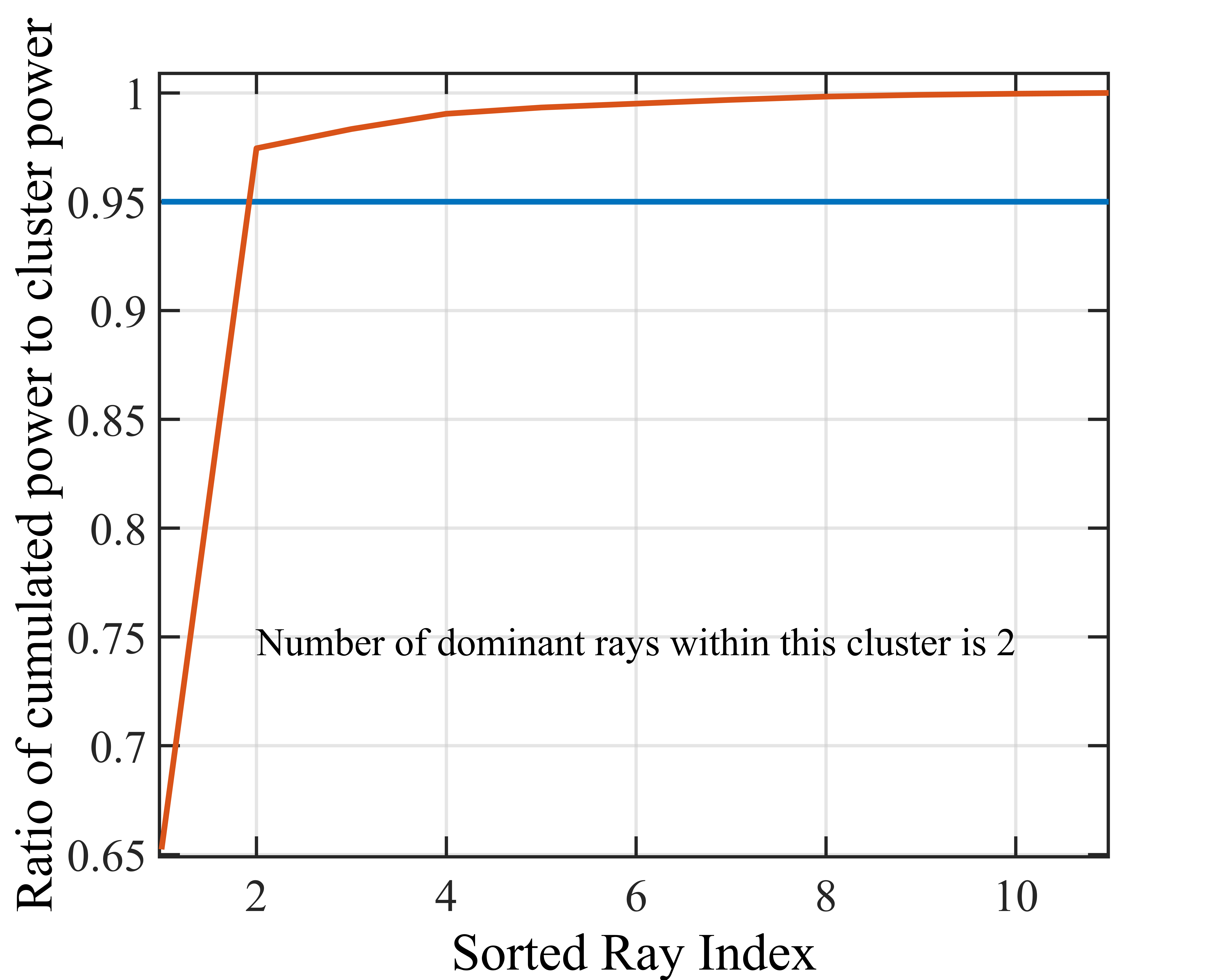}
	\caption{The Ratio of cumulated power to cluster power within example cluster in UMa scenarios at 13 GHz.}
    \label{dominant}
\end{figure}

\begin{figure}[htbp]
	\centering
	\includegraphics[scale=0.3]{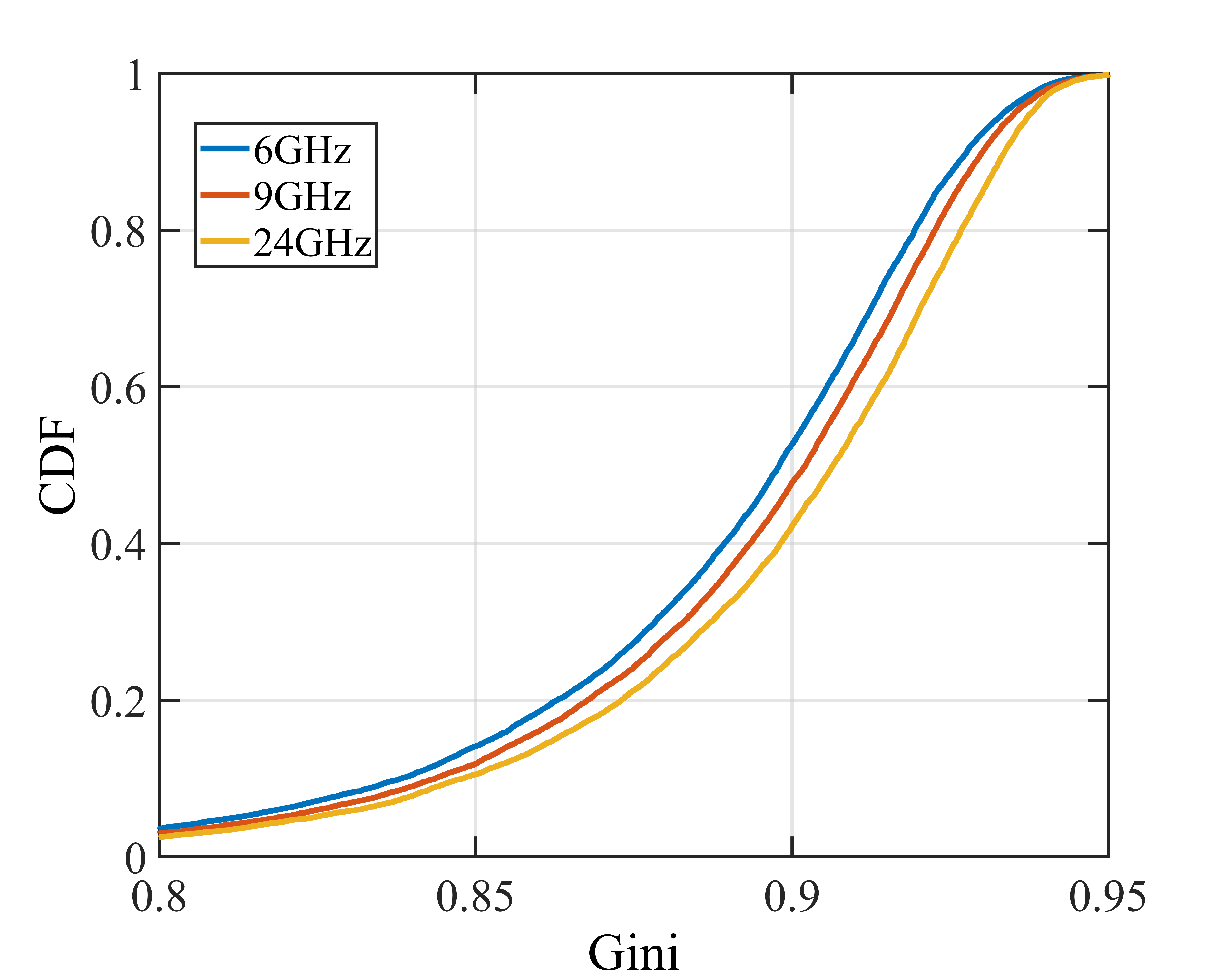}
	\caption{The CDF of the Gini index obtained by modified model in UMa scenarios.}
    \label{Gini_3}
\end{figure}

We revise the minimum number of multipath components to capture frequency-dependent channel behavior. Using the 13 GHz measurements in the UMa scenario as an example \cite{r1-2408095}, we first determine the number of dominant rays per cluster based on the 95\% cumulative power threshold relative to the total cluster power, as illustrated in Fig. \ref{dominant}. The average number of dominant rays per cluster is considered as $M_min$, which means that $M_{min}=3$ in the LOS scenario at 13 GHz. After updating the model for the number of multipath components, the channel model can characterize sparsity across frequencies, as shown in Fig. \ref{Gini_3}. When setting $M_{min}=3$ for the UMa LOS scenario, the actual number of rays per cluster $M$ is 4, 6, and 16 at 6, 9, and 24 GHz, respectively. This leads to a frequency-dependent channel impulse response, thereby capturing variations in channel sparsity across frequencies.

\section{Open Issues}
While 3GPP Release 19 represents a monumental stride in standardizing channel models for the 7–24 GHz band and ELAA systems, it serves as a foundational baseline rather than a definitive conclusion. The introduction of near-field propagation and spatial non-stationarity fundamentally disrupts the simplistic assumptions of legacy models, introducing complexities that current stochastic frameworks address through approximations. As the industry advances towards 6G deployment, several critical challenges regarding model fidelity, parameter acquisition, and system-level applicability remain unresolved. This section critically analyzes these open issues to illuminate the path for future research and standardization evolution.

\subsection{Measurement-Based Parameter Acquisition and Model Validation}
Currently, the parameterization of near-field and SNS channel models relies heavily on Ray-Tracing (RT) simulations due to the paucity of comprehensive empirical datasets. Although RT facilitates flexible scenario generation, its reliance on simplified environmental abstractions (e.g., idealized material properties and canonical geometries) inevitably limits its ability to capture the stochastic nuances of real-world propagation. Consequently, large-scale measurement campaigns encompassing diverse frequency bands, array configurations, and dynamic environments are imperative to derive realistic parameter tables and validate the physical consistency of the models. Furthermore, a clear distinction must be made between implementation consistency and model accuracy. While recent cross-company calibration efforts have successfully demonstrated that different simulators yield consistent statistical outputs (e.g., aligned coupling loss and eigenvalue distributions), this does not strictly equate to physical accuracy against ground-truth reality. Future research must therefore prioritize measurement-driven validation, ensuring that the mathematical abstractions of near-field and SNS effects faithfully reproduce the propagation characteristics observed in actual deployments.



\subsection{Impact on System Design and Performance Evaluation}
The incorporation of near-field propagation and SNS fundamentally alters wireless channel behavior, introducing significant complexity to system modeling and performance evaluation. Regarding near-field propagation, the acquisition of channel state information (CSI) becomes substantially more challenging \cite{Zhao2024NearFieldCommEng,Xing2025XL-ArrayWaveNumberDomain,Ding2025XL-MIMOUCANearFieldCodebook}. Conventional estimation methods exploiting angular sparsity—such as compressed sensing—become ineffective due to the ``energy spreading" effect observed in the joint angle–distance domain. This necessitates the development of novel beam training strategies that jointly resolve both angular and range dimensions. Conversely, the near-field regime unlocks new degrees of freedom for spatial multiplexing through emerging beamforming paradigms, such as beam focusing and orbital angular momentum (OAM).

Concurrently, the phenomenon of SNS invalidates the spatial stationarity assumption underpinning most existing channel estimation algorithms. This mismatch leads to significant estimation errors when antenna elements are exposed to heterogeneous scattering environments. Furthermore, SNS introduces distortions in array beam patterns—manifesting as main-lobe broadening and side-lobe elevation—since clusters are visible only to a subset of the array aperture. These effects collectively reshape the spatial structure of the channel, necessitating signal processing algorithms specifically optimized for spatially non-stationary conditions.

\subsection{Hybrid Channel Modeling for Near-Field and SNS}
A fundamental tension exists in current modeling methodologies: conventional stochastic models (GBSM) struggle to capture the intricate spatial determinism of ELAA channels, while deterministic methods (RT) suffer from prohibitive computational complexity. A promising evolutionary direction is hybrid channel modeling, which synergizes deterministic ray tracing with measurement-based stochastic calibration to balance fidelity and efficiency \cite{Wang2025XL-MovableArrayCrossField}. This approach enables physically consistent modeling of SNS features (e.g., deterministic blockage by specific obstacles) while retaining the statistical generalization of GBSM. The potential of hybrid modeling is particularly pronounced in higher frequency bands, such as terahertz (THz), where SNS effects are amplified by short wavelengths and dense arrays \cite{HybridTHz}. Future research should focus on developing scalable hybrid frameworks that enable accurate channel modeling under practical computational constraints.

\subsection{Refinement of Penetration Loss Parameters}
Empirical evidence reveals notable discrepancies between measurement results and standard 3GPP values for specific construction materials, particularly glass and concrete. These deviations, stemming from variations in material composition, thickness, and multi-layer structures, indicate that the current penetration loss parameters may not fully capture the complex propagation behavior of modern building materials in the FR3 band. Therefore, it is essential to revise and enrich the penetration loss database. Updating these parameters to account for a wider range of material characteristics and incidence angles will significantly enhance the accuracy of system-level simulations, ensuring they faithfully reflect the severe indoor-outdoor propagation losses characteristic of the 7–24 GHz band.

\section{Conclusion}
This tutorial has provided a comprehensive technical guide to the 3GPP Release 19 channel modeling standard, specifically tailored for the 6G FR3 (7–24 GHz) band. By bridging the gap between standardization specifications and practical simulation implementation, we have addressed the critical need for an accurate evaluation methodology in the era of ELAA.
We began by elucidating the standardization rationale, highlighting how the scarcity of empirical data and the physical paradigm shift induced by ELAA rendered legacy interpolation-based models insufficient. We then deconstructed the core enhancements in Rel-19, encompassing the recalibration of legacy scenarios (UMa/UMi), the definition of the new SMa scenario, and the mathematical formulation of near-field spherical wavefronts and SNS. Crucially, this article transitioned from theory to practice by presenting a step-by-step implementation guideline, demonstrating how to integrate these advanced features—such as distance-dependent phase corrections and visibility masks—into the established GBSM framework. As the industry moves towards 6G, the channel model presented herein serves as the foundational bedrock for assessing key enabling technologies, particularly ultra-massive multiple input multiple output (MIMO) and high-capacity mid-band transmissions. While Rel-19 represents a monumental stride, the evolution of channel modeling is far from complete. Future research must address the computational challenges of high-fidelity simulations, potentially through hybrid deterministic-stochastic frameworks or AI-native channel generation. It is our hope that this tutorial equips researchers and engineers with the necessary tools to navigate these complexities, fostering the accurate design and rigorous performance evaluation of future 6G networks.


\bibliographystyle{IEEEtran}
\bibliography{ref}
\vfill

\end{document}